\documentclass[11pt]{article}

\usepackage[utf8]{inputenc}
\usepackage{graphicx}           
\usepackage{amsmath,amssymb}    
\usepackage{cite}               
\usepackage{url}                
\newcommand{\degree}{\ensuremath{^{\circ}}}
\usepackage{bbm}
\usepackage{xcolor}             
\usepackage{hyperref}
\hypersetup{
    colorlinks=true,
    linkcolor=blue!75!black,    
    citecolor=magenta,          
    urlcolor=teal               
}
\usepackage{xurl}
\usepackage{pdfpages}
\usepackage{tcolorbox}

\title{\Large \bfseries \color{darkgray} Weather Emulators at the Frontier of Heat Extremes Predictability}

\author{
    Cas Decancq$^{1,*}$, 
    Thomas Mortier$^{1}$,
    Jessica Keune$^{2}$,
    and Diego G. Miralles$^{1}$,
}

\date{} 

\begin{document}

\maketitle

\noindent
$^{1}$Hydro-Climate Extremes Lab (H-CEL), Department of Environment, Ghent University, Ghent, Belgium.\\
$^{2}$European Centre for Medium-Range Weather Forecasts (ECMWF), Reading, United Kingdom.\\
$^{*}$e-mail: cas.decancq@ugent.be

\begin{abstract} 
Atmospheric predictability declines rapidly beyond the next ten days, such that forecasts at longer lead times primarily convey large-scale trends rather than specific states. Yet in a warming world, improving early warnings of extreme heat is an increasingly critical challenge. Here we evaluate six state-of-the-art deep learning weather emulators---Pangu-Weather, FuXi, ArchesWeather, AIFS, GraphCast and Aurora---alongside leading dynamical systems and statistical baselines in forecasting global near-surface temperature and extreme heat at lead times of 10--15 days. We find that several emulators rival or even surpass physics-based forecasts in deterministic temperature skill, but do so at the cost of reduced spectral fidelity, in a process widely known as \textit{blurring}. While all models show some degree of predictive skill for extreme heat, most emulators under-represent peak intensities, and IFS recall is greater than that of any of the emulators. These results highlight both the emerging potential of AI to enhance extended range temperature prediction, and the remaining challenges in delivering reliable, actionable early warnings in a changing climate.
\end{abstract}

\section{Introduction}

Extreme heat is one of the deadliest climate hazards \cite{zhao2021global, ballester2023heat, EEA2025_ClimateStateOutlook_3_2}. Nearly ten million heat-related excess deaths occurred in the first two decades of the new millennium \cite{zhao2021global}. Accurate forecasting of heat extremes is therefore crucial to reduce mortality and economic losses \cite{WMO2025_COP30_ClimateUpdate, rogers2011implementing, WMO2015Heatwaves}. By enabling timely public health responses, agricultural planning, and energy management, early warnings can mitigate societal and infrastructural impacts and enhance resilience under a warming climate \cite{WMO2015Heatwaves, rogers2011implementing, vsakic2022early}. Heatwaves often emerge from a complex interplay of processes \cite{miralles2019land, domeisen2023prediction, barriopedro2023heat, scholz2024atmospheric, fix2024detection, luo2024anthropogenic}, and while dynamical models can capture such mechanisms at short range, their predictive skill decays rapidly in time \cite{Haiden2021_ECMWF_Evaluation, rasp2024weatherbench}.

A long-standing paradigm has shaped weather forecasting: the state of the atmosphere becomes inherently unpredictable beyond roughly two weeks. At these lead times, forecasts are indicative of week-by-week changes rather than individual states \cite{ECMWF_FUG_ExtendedENS_8_2}. This limit, grounded in the chaotic dynamics of the atmosphere and famously captured by Lorenz’s butterfly effect \cite{Lorenz1969}, defines the upper bound of medium-range weather forecasting \cite{judt2018insights, zhang2019predictability, selz2022transition}. Because it is impossible to perfectly observe and model Earth’s vast nonlinear system, errors are unavoidable and rapidly amplify. This leads to divergent trajectories over time, with atmospheric predictability decreasing rapidly after about ten days \cite{white2017potential}. In recent years, however, advances in data-driven modeling and computational capacity have begun to challenge the notion that this limit is immovable \cite{vonich2025testing}. A new generation of deep learning models trained to emulate atmospheric dynamics has emerged, which rival the medium-range forecasts from the most sophisticated Numerical Weather Prediction (NWP) systems, such as the European Centre for Medium-Range Weather Forecasts (ECMWF) Integrated Forecasting System (IFS) \cite{ecmwf_ifs_cy49r1_2025}, or the systems developed by the China Meteorological Administration (CMA) \cite{cma_grapes_tigge_2018} and the USA National Centers for Environmental Prediction (NCEP) \cite{ncep_gefs_v12_2020}. Notable examples of AI emulators include Huawei’s Pangu-Weather \cite{bi2022pangu}, Google DeepMind’s GraphCast \cite{lam2023learning}, Fudan University's FuXi \cite{chen2023fuxi}, INRIA's ArchesWeather \cite{couairon2024archesweather}, ECMWF’s own Artificial Intelligence Forecasting System (AIFS) \cite{lang2024aifs}, and Microsoft’s Aurora \cite{bodnar2025foundation}. Collectively, these models have achieved in a few years what once required decades of incremental progress, prompting a “second revolution in weather forecasting” \cite{pasche2025validating}. Their skill for variables like temperature and precipitation is already comparable to or even surpasses operational NWP \cite{bi2022pangu, lam2023learning, chen2023fuxi, couairon2024archesweather, lang2024aifs, bodnar2025foundation}. However, it remains unclear whether these gains extend to extreme heat at the critical two-week horizon.

The subseasonal window, traditionally spanning lead times from two to six weeks \cite{robertson2018sub}, occupies a gap between classical weather and climate prediction. It is an inherently difficult regime, where predictability increasingly depends on slower-varying boundary conditions (such as sea and land surface states) that retain memory but provide weaker and more indirect constraints on atmospheric variability \cite{robertson2015improving, white2017potential}. Although historically the subseasonal range has received limited attention, forecasts here are critical for proactive disaster mitigation and planning \cite{white2017potential, ecmwf_aiweatherquest_2025}---an impactful example is the Forecast-based Financing programme by the Red Cross \cite{GermanRedCross_FbF}. In recognition of this, major initiatives have been launched to advance subseasonal prediction, including the joint WWRP--WCRP S2S Project \cite{vitart2018sub}, the U.S. Subseasonal Climate Forecast Rodeo \cite{hwang2019improving}, and more recently ECMWF’s AI Weather Quest competition \cite{ecmwf_aiweatherquest_2025}.

Heatwaves are among the most consequential targets for accurate forecasting \cite{donaldson2003cardiovascular, mcmichael2011climate, coumou2012decade, loughnan2014heatwaves, horton2016review, campbell2018heatwave, ballester2023heat}. Following the emergence of AI emulators, a key question remains: Can these models improve the accuracy of extreme heat forecasts at longer lead times? There exist studies on the applicability of emulators for medium-range extremes forecasting (e.g. \cite{magnusson2023_ml_extremes_ecmwf, ben2024rise, pasche2025validating}), whereas the evaluation of extremes in the subseasonal forecasting range has been oriented towards dynamical systems \cite{lin20222021, emerton2022predicting}. Until recently, most subseasonal machine learning methods targeted aggregated metrics for specific regions \cite{vijverberg2020subseasonal, benson2023soil, van2022using, weirich2023subseasonal, he2021sub, weyn2021sub}; and although the use of emulators for subseasonal forecasting is rapidly gaining attention \cite{nathaniel2024chaosbench, ecmwf_aiweatherquest_2025}, a multi-model benchmark for extreme heat is absent still. Although early warning systems beyond 10 days are ultimately probabilistic, few probabilistic emulators are both open source and computationally accessible. Moreover, many probabilistic approaches build upon architectural foundations shared with deterministic models. Evaluating deterministic emulators can therefore provide valuable insight into the ability of these architectures to forecast extreme heat beyond 10 days, helping to assess whether AI-based forecasting systems can extend the predictability horizon of heat extremes.

Here, we assess whether AI weather emulators outperform dynamical models in predicting extreme heat beyond the 10-day limit. Specifically, we systematically benchmark six deterministic state-of-the-art weather emulators and three dynamical forecasting systems on their ability to predict global land surface temperature and heat extremes at lead times of 10--15 days. Our findings reveal that while all models struggle to provide significant value over climatology for global near-surface temperatures, AI emulators are pushing the boundaries of predictability, outperforming traditional dynamical systems on several standard metrics. Despite their general skill, a critical disconnect remains regarding extreme heat. While current emulators improve upon baselines, they generally fall short of the IFS in extreme heat recall. This gap largely stems from a sacrifice in spectral fidelity; many emulators produce overly smoothed fields that fail to capture peak intensity. However, recent developments demonstrate that pipeline improvements can alleviate these deficiencies, delivering strong point-evaluation metrics without losing structural detail. The inherent advantages of emulators---their computational speed, accessibility, and ensemble scalability---offer an untapped potential for task-specific tailoring that dynamical models cannot match.

\section{Results}
\subsection{Extreme Heat Events and Model Selection}

\begin{figure}[h] 
    \centering
    \includegraphics[width=\linewidth]{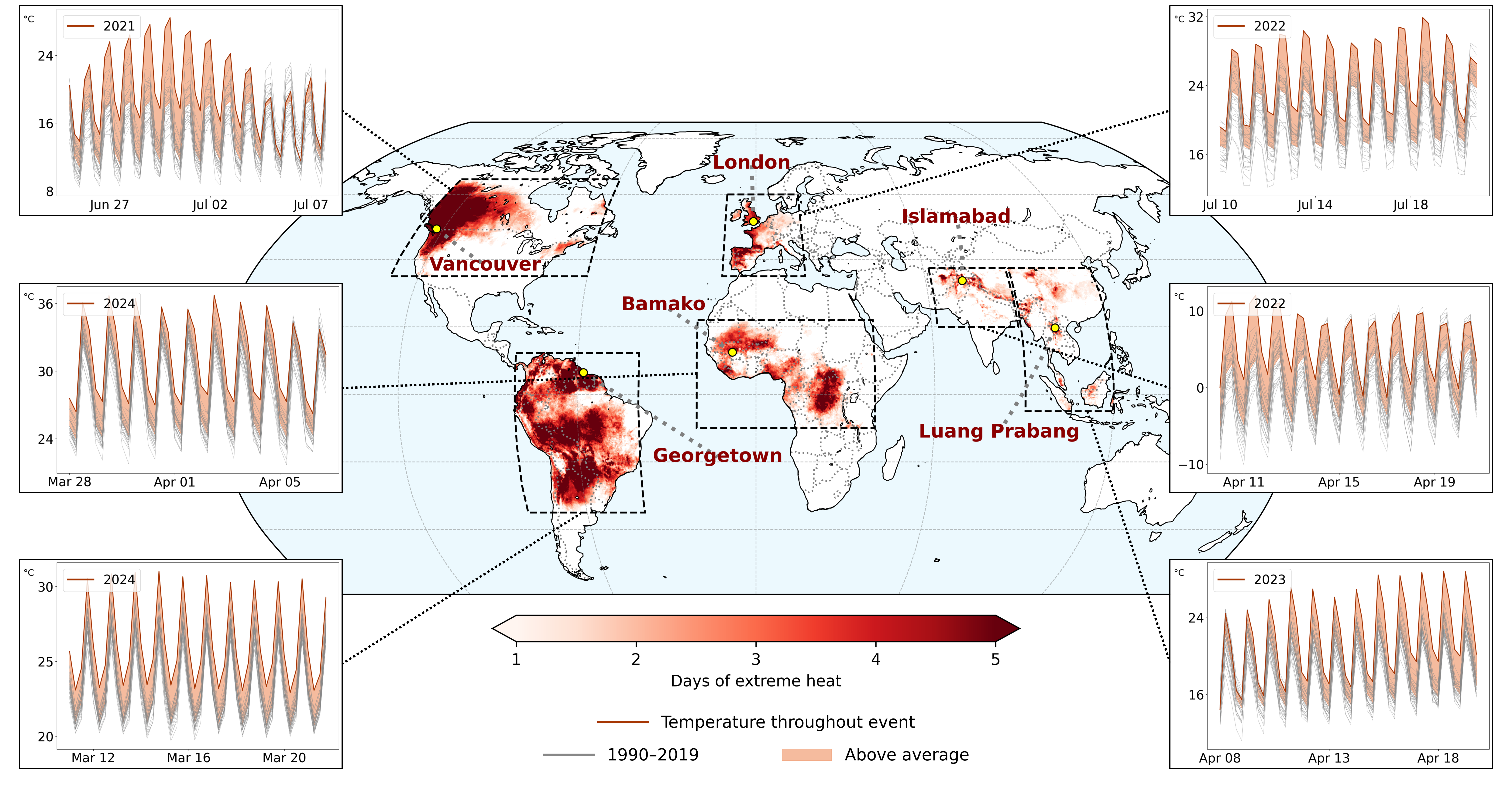}
    \caption{\textbf{Composite view of six extreme temperature events in the past five years.} The main panel shows the total number of days with ERA5 surface temperature \cite{hersbach2020era5} exceeding two standard deviations above its 1990--2019 climatological mean (Section \ref{M-Baselines}) during each event. Insets specify the studied period and display the evolution of near-surface temperature in the affected regions by means of a latitude-weighted average across locations experiencing at least three days of extreme heat. Gray lines show the same time series observed in years 1990-2019; orange shades indicate deviations from the region's climatology.}
    \label{fig:FIG1}
\end{figure}

Recent years have witnessed an alarming increase in both the frequency and severity of heat extremes. The 2021 Pacific Northwest “heat dome” shattered records by unprecedented margins \cite{emerton2022predicting}, triggering a 440\% surge in mortality in Vancouver \cite{henderson2022analysis}. In 2022, prolonged heat and drought across Europe killed tens of thousands and caused billions of euros in losses \cite{faranda2023persistent, ballester2023heat}, while India and Pakistan witnessed record-breaking temperatures from March through May \cite{zachariah2023attribution}. More recently, record-breaking heat has continued to affect South Asia, Africa, and Latin America \cite{aadhar20232022, zachariah2023attribution, WWA_Sahel_Heatwave_2024, AlJazeera_RioHeatIndex_2024, WMO_LAC_ClimateImpacts_2025, ClimateCentral_Brazil_Feb2025, Guardian_Heatwave_Brazil_2025}. Figure \ref{fig:FIG1} provides a composite view of the extent, duration, and intensity of several of these recent episodes. As global mean temperatures continue to rise \cite{masson2021climate, simolo2022quantifying}, events of this magnitude are projected to become far more common \cite{perkins2012increasing, beck2024increasing, rousi2022accelerated, dosio2025fast}. Recent estimates, for example, suggest a threefold increase in the likelihood of 2022-like mortality events for Europe, and up to a thirtyfold increase for southern Europe \cite{beck2024increasing}. These trends increase the need for forecasting systems that can provide reliable early warnings weeks in advance.

\begin{figure} 
    \centering
    \includegraphics[width=\linewidth]{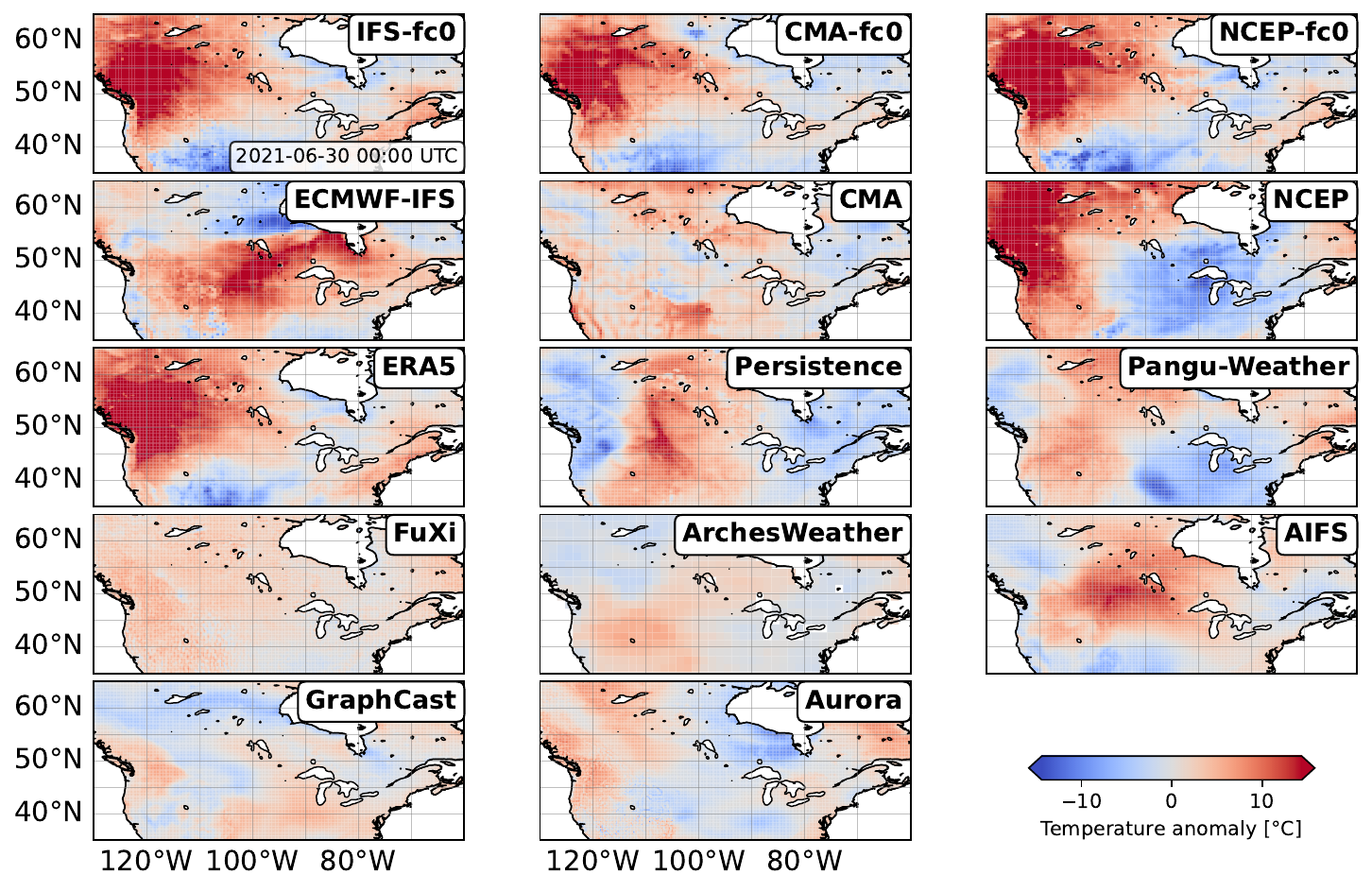}
    \caption{\textbf{Example anomaly maps.}
    Near-surface temperature anomaly at the peak extent of the 2021 heat dome over the Northwest Pacific (2021-06-30 00:00 UTC) in references (IFS-fc0, CMA-fc0, NCEP-fc0 and ERA5) and corresponding forecasts initialized 14 days prior (2021-06-16 00:00 UTC). Similar visualization is provided for all events in Figures S7--S12.}
    \label{fig:FIG2}
\end{figure}

We evaluate three numerical weather prediction systems (ECMWF-IFS, NCEP, CMA) and six deep-learning emulators (Pangu-Weather, FuXi, ArchesWeather, AIFS, GraphCast, Aurora) for their ability to forecast two-meter temperature and specifically extremes, defined as the exceedance of plus-two standard deviations (\(T_{surface} \geq \mu + 2\sigma\); with $\mu$ the ERA5 1990-2019 climatological mean and $\sigma$ its standard deviation---Figure S1), at lead times of 10--15 days (Section S4), and compare them to two standard baselines (climatology and persistence---Section \ref{M-Baselines}). The emulators were trained on ERA5 and are therefore evaluated against ERA5, whereas dynamical models are evaluated against their respective initial conditions (referred to as \textit{fc0}) at the forecast valid time, following established practice \cite{WMO2023Manual, lam2023learning, price2023gencast}. To compute anomalies, all forecasts are compared to climatology derived from ERA5, as neither emulator nor dynamical model climatologies are publicly available or easily replicated. Figure \ref{fig:FIG2} provides an example of forecast anomalies restricted to the bounds of the 2021 "heat dome" event (Table S2). The evaluation distinguishes global from event-level scales, the full near-surface temperature field from its heat extremes, and considers a suite of metrics; all details can be found in Sections \ref{M-ProblemStatement}--\ref{M-Categorical-Evaluation}.

\subsection{Global Near-Surface Temperature Forecasting}\label{section2.2}

\begin{figure} 
    \centering
    \includegraphics[width=\linewidth]{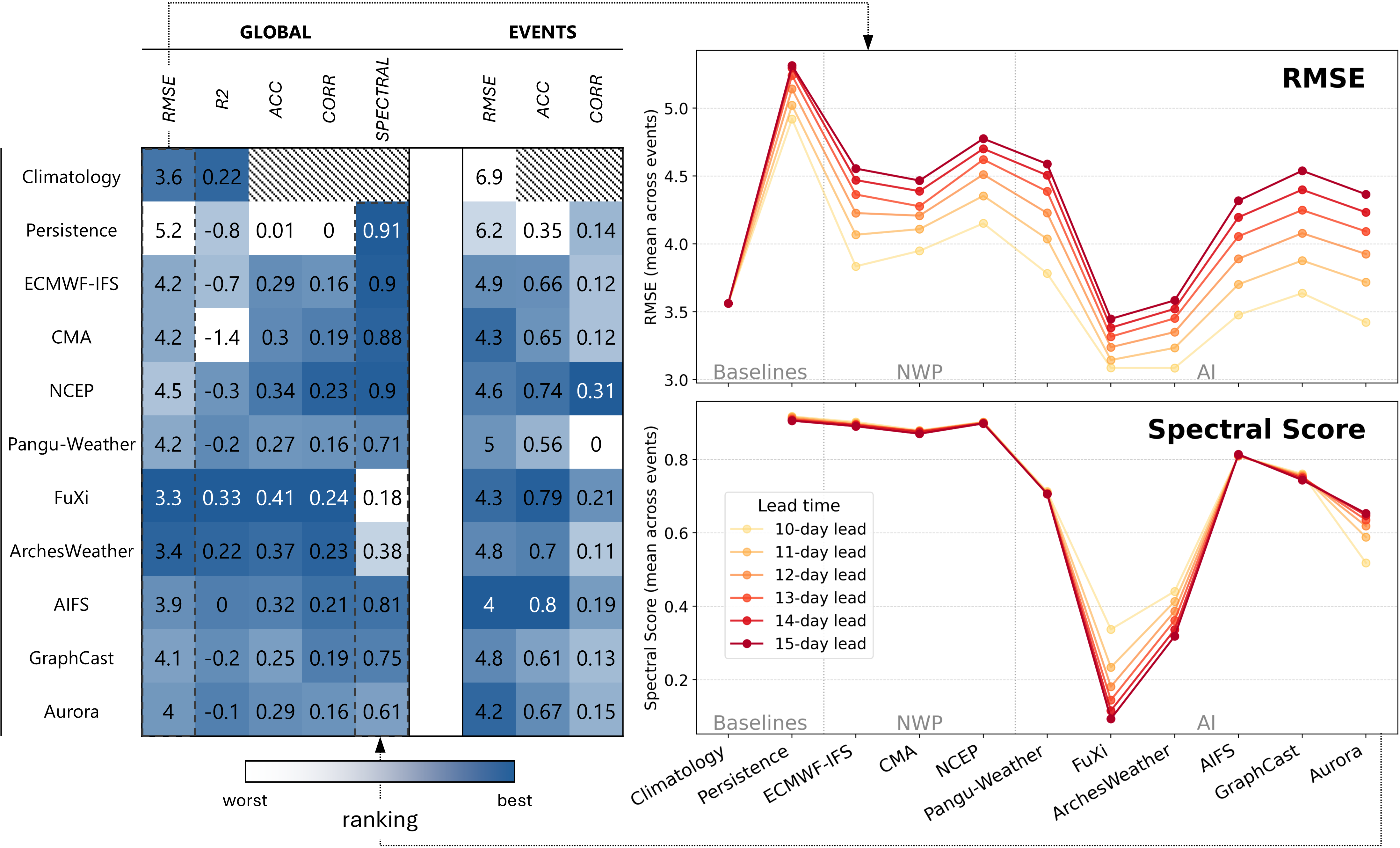}
    \caption{\textbf{Model benchmarking results.} Left panels: fixed-lead-time model forecast scores across events and averaged over lead days 10 to 15, per metric (Section \ref{M-Continuous-Evaluation}). A dark blue fill denotes a model performing best for that metric, whereas a white fill denotes the opposite. For each metric, the best score is reported in white font. Metrics considered are $RMSE$, $R^2$, $ACC$, temporal correlation and spectral score (Section \ref{M-Continuous-Evaluation}). Global subpanel: evaluation for all surface temperature data over global land area. Events subpanel: evaluation bounded to the event boxes, and restricted to the affected (red-hued) regions in Figure \ref{fig:FIG1} at times of recorded heat extremes. The spectral metric is global-only (Section \ref{M-Continuous-Evaluation}), and extreme heat time series are too short to get a stable $R^2$ estimate, which is why these metrics are reported for the global evaluation only. $ACC$, temporal correlation and spectral score are based on deviations from ERA5 climatology, and therefore not available for the climatological baseline (Section \ref{M-Continuous-Evaluation}). Right panels: $RMSE$ and spectral score for each model and dependency on forecast lead time. Averaging over these lead times yields the summary scores in the left panel.}  
    \label{fig:FIG3}
\end{figure}

For evaluating global forecasts at different lead times, root mean square error ($RMSE$), coefficient of determination ($R^2$), anomaly correlation coefficient ($ACC$), temporal correlation, and spectral score are considered (see Methods). The spectral score represents the model’s ability to correctly distribute energy across all spatial scales, ensuring that both small- and large-scale climate temperature anomaly patterns have the right intensity (Section \ref{M-Continuous-Evaluation}, Figure S12). When all five metrics are equally weighted, FuXi emerges as the top-performing model for global near-surface temperature forecasts, followed by ArchesWeather. FuXi averages the best scores across all metrics except the spectral score over the 10--15 day lead-time range (Figure \ref{fig:FIG3}, left). AIFS, Aurora, GraphCast and Pangu-Weather perform similarly, with AIFS showing slightly better results. As ACC, temporal correlation, and spectral score are calculated based on deviations from ERA5 climatology (Section \ref{M-Continuous-Evaluation}), the climatological forecast cannot be evaluated in this manner. However, considering only $RMSE$ and $R^2$, climatology is a top-performing predictor of temperature. This is expected because global near-surface temperature is dominated by seasonal and daily cycles and persistent large-scale spatial gradients, which are represented by climatology. Consequently, a forecast can achieve relatively low errors without accurately predicting temperature anomalies. These results indicate that models struggle to add skill beyond the expected seasonal and daily cycle. For each lead time separately, metric values and their standard deviation between events can be found in the Supplementary Tables S5--S6.

Notably, only FuXi and ArchesWeather manage to surpass global climatological $RMSE$ (Figure \ref{fig:FIG3}, left). On average throughout the 10- to 15-day range, their $RMSE$ is $8.0\%$ and $5.3\%$ lower than climatology, respectively. Even with a 15-day lead, FuXi $RMSE$ is still $3.0\%$ lower, whereas that of ArchesWeather becomes marginally higher (Figure \ref{fig:FIG3}, right). At higher latitudes, the two models show strong $RMSE$ and $R^2$ performance compared to other emulators and NWPs (Figures S14--S15), with patterns that closely resemble those of climatology. Among the other models, AIFS and Aurora perform better than GraphCast and Pangu-Weather but overall scores are modest. They only improve on climatological $RMSE$ at a 10-day lead, with their $RMSE$ increasing by over $20\%$ at a 15-day lead. Persistence is the worst estimator of global near-surface temperature at two-week lead time by a margin, followed by NCEP. ECMWF-IFS and CMA $RMSE$ are comparable to those of GraphCast and Pangu-Weather.

A pattern emerges in global evaluation: the data-driven models that excel at $RMSE$, $R^2$, $ACC$ and temporal correlation struggle to balance this skill with spectral fidelity (Figure \ref{fig:FIG3}, right; Table S5). Spectral analysis shows that, among emulators, AIFS more realistically captures variability in temperature anomalies across spatial scales. In contrast, FuXi and ArchesWeather tend to produce oversmoothed fields (Figure \ref{fig:FIG2}, right), a phenomenon called \textit{blurring} \cite{lam2023learning, lang2024aifs, rasp2024weatherbench, mathieu2015deep, isola2017image}. By optimizing for standard $L_p$ (e.g., MAE or MSE) norms, these models learned to prioritize mean-state accuracy over realistic weather patterns, resulting in fields that resemble climatology. With increasing lead time, prediction becomes increasingly difficult, and this effect strengthens (Figure \ref{fig:FIG3}, right). More generally, deep networks have a learning bias towards low frequency functions, meaning that global fluctuations are learned more quickly than local ones \cite{rahaman2019spectral}. FuXi in particular is a "three-stage" model, where three separate models are trained, one for each of the 0--5, 5--10 and 10--15-day ranges \cite{chen2023fuxi}. Interestingly, a sharp drop in spectral fidelity is visible from lead day 10 to lead day 11 (Figure \ref{fig:FIG3}, right; Table S5), indicating the model tailored to longer lead times learned that stronger smoothing optimizes MAE. The microensemble version of ArchesWeather evaluated here, on the other hand, consists of four iterations of the same model trained with different random seeds \cite{couairon2024archesweather} (Table S3). The deterministic forecast from ArchesWeather considered here is therefore a small ensemble average, which inherently filters out unpredictable features and averages out extremes \cite{bonavita2026forecast}.

Both FuXi and Aurora exhibit limited spatial coherence in their forecasts at local scales, with temperature anomaly fields showing high-frequency spatial noise (Figures \ref{fig:FIG2} and S7--12). This behavior indicates that local coherence is not well captured, likely as a consequence of the large spatial attention windows employed by these models, which may promote overfitting. In general, attention-based emulators rely on two-dimensional (FuXi) or three-dimensional (Pangu-Weather, ArchesWeather, Aurora) Swin Transformer architectures \cite{bi2022pangu,couairon2024archesweather,chen2023fuxi,bodnar2025foundation}, while graph-based methods such as GraphCast and AIFS more effectively capture spatial dependency through explicit locality, an inductive bias that assumes nearby grid points are more strongly related than distant ones \cite{lam2023learning, lang2024aifs}. This inductive bias reduces the number of trainable parameters, thereby lowering the risk of overfitting. Figures S7--S12 highlight the distinct spatial characteristics of the different model classes. The three physics-based models stand out for their spectral fidelity, despite their slightly higher errors and lower correlations (Figure \ref{fig:FIG3}, Table S5). Their spectral score all but matches that of persistence, indicating that the spatial variability of temperature anomalies in their forecasts is nearly indistinguishable from a reference state. It becomes clear that forecasting near-surface temperatures at the intersection of medium- and subseasonal ranges is a formidable challenge to any model, and there is a trade-off between spectral faithfulness and global skill. NWPs solve explicit physical differential equations, inherently producing more realistic temperature patterns and anomaly ranges. Emulators are typically trained to optimize $L_p$ loss functions, inherently favoring gradual \textit{blurring} with increasing lead time.

The atmosphere is a chaotic deterministic system, meaning that unobservable differences in initial conditions grow exponentially over time, leading to vastly different trajectories \cite{Lorenz1969, judt2018insights, zhang2019predictability, selz2022transition}. In the case of NWP systems, such diverging trajectories are observed in the fixed-lead time series. The sequential forecasts concatenated from different initializations are, for most regions on Earth, far more variable than actual temperature variability, since each time step originates from a dynamically independent forecast trajectory (Figure S19). However, the forecast activity---its standard deviation in time \cite{benbouallegue2024accuracy}---from a regular NWP rollout is more in line with that of the reference (Figure S18). Emulators, which tend towards climatological states at extended lead times, show less divergent trajectories. FuXi's variability is the lowest of all. For most of the globe, a 10--15 day FuXi rollout is 25--75\% less variable than the ERA5 reference, although it is better calibrated over land than oceans (Figure S18). Of all models, the variability in Pangu-Weather and ArchesWeather rollouts best matches that of their reference. Most emulators show an interesting pattern with too little variability over tropical ocean regions, and distortions at minimum one of the poles (Figures S18--19). Although latitude-weighting is typically done to account for area distortion caused by grids, temperature variability is far greater at higher and lower latitudes. As a result, errors in the tropical belt, especially over oceans where variability is lower, contribute less to $L_p$-norm minimization. The same latitude weighting also attributes a (near-)zero importance to the Earth's poles, which may contribute to the observed distortions.

\subsection{Forecasts in Extreme Heat Events}

For global temperature forecasting over land, most models provide limited added value relative to the climatological forecast. For extreme heat events, however, the balance shifts. In affected regions (Section \ref{M-Continuous-Evaluation}), all models improve greatly on climatological $RMSE$ at all lead times (Figure S13), where AIFS now is the best performing model for extreme heat (Figure \ref{fig:FIG3}, left panel, events). Under these extreme conditions, the climatological $RMSE$ almost doubles relative to global evaluation, whereas that of AIFS remains remarkably stable. FuXi's $RMSE$, in contrast, increases by $30\%$, and that of ArchesWeather by roughly $40\%$. $ACC$ quantifies the degree of similarity between the spatial patterns in forecast and observed anomalies; if a forecast's $ACC$ value is 0.6 or higher it is generally considered skillful, and below 0.6 it is considered that the positioning of synoptic scale features ceases to have value for forecasting purposes \cite{ecmwf_fug_2018}. Using this definition, most models produce skillful forecasts for extreme heat events. With the top ranked $ACC$ and $RMSE$ across events and lead times, AIFS best reproduces the combined spatial pattern and intensity of extreme heat in the affected regions.

In heat extremes, the models are more competitive relative to baselines like climatology or persistence. This indicates that even at the edge of the medium range window, some degree of predictability of extremes is recoverable. Persistence scoring better than climatology suggests a slow-varying memory of land--atmospheric extremes, with upcoming hot states generally preceded, even two weeks before, by states of already elevated surface temperature. Regardless of evaluating for the full temperature spectrum or heat extremes only, correlation scores are weak or even negligible. The highest correlation is that of NCEP for the hot events, at $0.31$ (Figure \ref{fig:FIG3}); the lowest is that of Pangu-Weather under the same circumstances, at zero. No model accurately predicts the evolution of temperature in time with high skill. Nevertheless, AIFS stands out because it achieves consistent scores for global temperature forecasting and top scores for event-scale extreme heat forecasting, without showing obvious weaknesses. FuXi and ArchesWeather---which in the global analysis showed the best $RMSE$ and $R^2$ results by a margin---exhibit lower performance for extreme heat events. Rather than mesoscale structures, they forecast large-scale smoothed anomalies (Figures S6--S11), and as a consequence, severely underestimate the extent of both regional and global heat extremes (Figure S5). FuXi's time series, in particular, show little temperature variability (Figure S18--S19).

\subsection{Heat Extremes Classification Skill}\label{section2.4}

\begin{figure} 
    \centering
    \includegraphics[width=\linewidth]{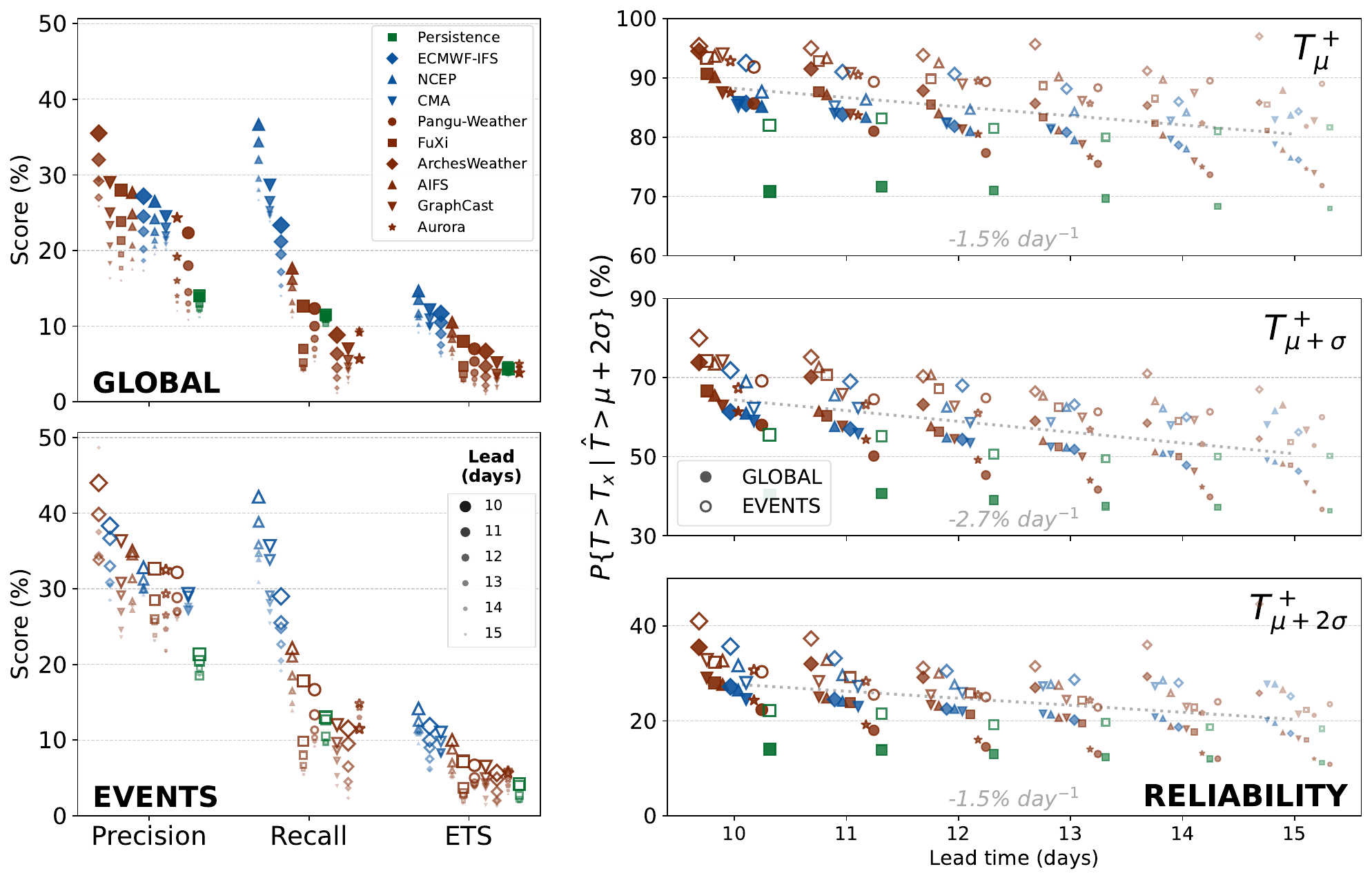}
    \caption{\textbf{Categorical forecast skill and reliability of extreme heat forecasts.} Near-surface temperature and forecasts are cast into binary format indicating extreme heat occurrence. Left panels: precision, recall, and the $ETS$ (Section \ref{M-Categorical-Evaluation}) as a function of lead time at global (top) and event-level (bottom) scales. Right panels: given an extreme heat forecast, probability of experiencing above-normal ($T\geq\mu$, top), above-hot ($T\geq\mu + \sigma$, middle), and extreme ($T\geq\mu + 2\sigma$, bottom) temperature (Sections \ref{M-Categorical-Evaluation}). Again, a distinction is made between global and event scales, and across lead times. These panels summarize the reliability of extreme heat forecasts.}
    \label{fig:FIG4}
\end{figure}

To evaluate categorical forecast skill, the left panels in Figure \ref{fig:FIG4} summarize each model's ability to identify heat extremes using precision, recall, and the Equitable Threat Score ($ETS$; Section \ref{M-Categorical-Evaluation}). We note that accuracy is not an adequate metric on its own, as by definition and despite sampling bias towards hot events, extremes in the temperature data are rare (Figure S3). As a result, a climatological forecast that contains no hot extremes scores a $95\%$ accuracy. FuXi and ArchesWeather, in comparison, respectively average about $95\%$ and $96\%$ accuracy with recall under $10\%$ for most lead times---lower than persistence (Figure \ref{fig:FIG4}, top left). NCEP shows the opposite pattern, combining the highest recall with the lowest accuracy, falling below $90\%$ at longer lead times. At a 10-day lead, it correctly retrieves about $45\%$ of event-level and $37\%$ of global heat extremes. However, NCEP forecasts turn out to be hot-biased. Compared to its reference, NCEP at 10- to 15-day leads globally produces $\sim40\%$ more heat extremes ($11.5\times10^6 \ km^2$) than actually observed, as well as $\sim20-25\%$ less cold extremes (Tables S4--S5). Similarly, CMA shows an hot extreme excess of about $12\%$. Evaluating for hot extremes recall does not account for frequency bias, thereby rewarding models that overpredict heat events. ECMWF-IFS forecasts contain $15\%$ fewer hot extremes than observed at 10 days lead, increasing further to $20\%$ less at 15 days lead. Despite this, its recall is greater than that of any of the emulators (Figure \ref{fig:FIG4}). At the event-level scale, IFS recall 15 days ahead is comparable to that of AIFS 10 days ahead. IFS precision increases notably when evaluating its forecasts for the events only (Figure \ref{fig:FIG4}, left), indicating enhanced skill for large-scale, coherent events.

The accuracy--recall tradeoff is apparent in most deep learning models, consistent with the \textit{blurring} effect seen in deterministic rollouts \cite{lam2023learning, lang2024aifs, rasp2024weatherbench}. Dynamical models generally show slightly lower precision, although the difference with emulators is less pronounced than in terms of recall (Figure \ref{fig:FIG4}, left). Most emulators produce only a fraction of the total observed extent of extreme heat (Table S4). Their behaviour can also change with lead time: where ArchesWeather and Aurora both predict about $1.5$ million km² of heat extremes at a 10 day lead, this decreases to $0.28$ million $km^2$ for ArchesWeather while it increases to $5.3$ million $km^2$ for Aurora (Figure S5, Table S4). Among emulators, AIFS has the best recall; from  $17.7\%$ at lead day 10 to $11.2\%$ at lead day 15, it consistently produces about $65\%$ of global extreme heat area (Table S4).

The dynamical models score highest on the $ETS$ (Figure \ref{fig:FIG4}, left), a categorical verification metric that measures forecast skill relative to random chance \cite{schaefer1990critical, mesinger2008bias, mbizvo2024dependence} (Section \ref{M-Categorical-Evaluation}). By correcting for hits expected by chance, $ETS$ penalizes forecasts that predict extreme heat over excessively large areas. This penalizes the hot-biased NCEP and CMA models, although they still perform favourably. ECMWF's IFS and AIFS have substantially different recall, but AIFS also produces smaller sets of hot extremes, which reduces false positives, so that the difference in $ETS$ between both is less pronounced. The overall low but positive $ETS$ means the models retain only marginal but real skill above random chance at these lead times.

The right panels in Figure \ref{fig:FIG4} address the practical question: \textit{"How reliable are forecasts of extreme heat?"} Across all models, once they forecast extreme heat, the odds are high that the true temperature will at least exceed the climatological mean, with probabilities ranging between $\sim70$--$95\%$ across lead times (Figure \ref{fig:FIG4}, top right). Even persistence performs well in this respect. Globally, in about $70\%$ of land that was extremely hot 10 to 15 days earlier, temperatures remain above average, while in $40-35\%$ they remain above one standard deviation (hot), and in $14-11\%$ they remain extreme (Figure \ref{fig:FIG4}, right, top to bottom). For large-scale events such as the ones studied here (Figure \ref{fig:FIG1}), the effect is even more pronounced. Nearly all models surpass this baseline. In affected regions, even with a 15-day lead, at least $\sim1/5$ of the predicted extremes materialize as true extremes for all models (Figure \ref{fig:FIG4}, bottom right, markers without fill), and generally there is at least a one in two chance that it will be hot ($T\geq\mu+\sigma$; middle panel).

Precision is expected to increase with increasing prevalence of extreme events in the reference; a pattern observed comparing global to event-level precision (Figure \ref{fig:FIG4}, left). However, most models also achieve higher recall; their categorical skill here is better for the events than for the full global field. In summary, the precision and recall with which most models identify hot extremes is low. However, once a model predicts extreme heat, a forecaster can generally rely on temperatures being above average, and expect a substantial probability of hot temperatures, although this reliability decreases with lead times from 10 to 15 days. These results are somewhat promising, as the models operate beyond the typical 10-day weather forecasting range, i.e., outside their usual domain of applicability. Additionally, their features lack Earth-system components critical to subseasonal predictability---such as sea-surface temperature, soil moisture, snow cover, vegetation state, and sea ice \cite{van2022using, barriopedro2023heat, prodhomme2021seasonal, miralles2019land, perkins2015review, benson2023soil}---indicating that substantial gains in skill may yet be realized by expanding the feature space.

\subsection{Model behaviour in real-world forecasts}\label{caseStudies}

\begin{figure*} 
    \centering
    \includegraphics[width=\linewidth]{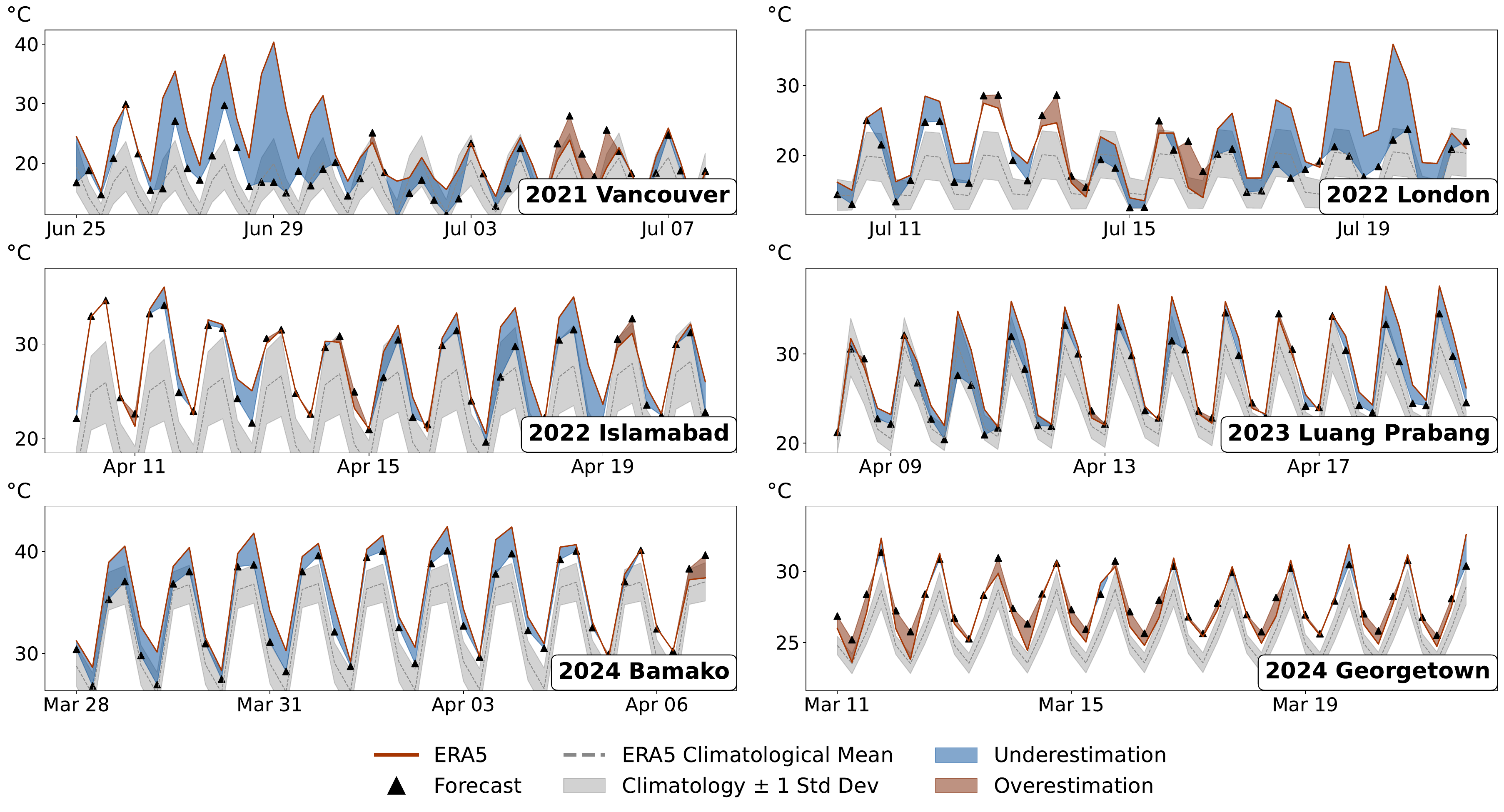}
    \caption{\textbf{AIFS case studies.} AIFS lead day 14 forecasts for the major cities highlighted in Figure \ref{fig:FIG1}, throughout their respective events. Red lines show ERA5 surface temperature at $00h$, $06h$, $12h$ and $18h$ UTC, while black denotes forecasts initialized strictly $14$ days prior. Orange fill denotes temperature overestimation, whereas blue fill denotes underestimation. For reference, a gray line and fill denote the $1990$--$2019$ ERA5 climatological mean and its standard deviation.}
    \label{fig:FIG5}
\end{figure*}

To examine how these results manifest in real-world forecasts, we focus on AIFS, the emulator which provides the most balanced performance across metrics (Sections \ref{section2.2}--\ref{section2.4}), and investigate its 10- to 15-day surface temperature predictions for the six major cities highlighted in Figure \ref{fig:FIG1}. In these case studies, AIFS achieves a 41.5\% reduction in $RMSE$ and 0.62 improvement in $R^2$ relative to climatology (Figure \ref{fig:FIG5}). City-specific $RMSE$ ($R^2$) improvements are 26.1\% (0.73) for Vancouver, 22.8\% (0.45) for London, 55.4\% (1.27) for Islamabad, 44.1\% (0.3) for Luang Prabang, 50.0\% (0.49) for Bamako, and 50.4\% (0.46) for Georgetown (Figure \ref{fig:FIG5}). $RMSE$ is systematically higher for mid-latitude than tropical locations, a pattern driven by background weather variability and observed across all evaluated models (Figure S14).

In all of these case studies, AIFS produces clear departures from climatology, showing variability beyond ordinary diurnal patterns---even at a two-week lead. Although at times it matches peaks in the ERA5 reference, such as the first few days in the 2022 Islamabad time series (Figure \ref{fig:FIG5}, middle left) or several extraordinarily hot nighttime temperatures in 2024 Georgetown (bottom right), it also fails to match the peak temperatures in the 2021 Vancouver and 2022 London time series (upper left and right, respectively). In most cases, extreme temperatures are underestimated. Although AIFS generally under-represents variability across the tropics (Figures S19--S20), the Bamako and Georgetown time series (Figure \ref{fig:FIG5}, bottom panels) show that even in these regions, the model can deviate well beyond two standard deviations from the climatological mean.

Across the six cities, mean $RMSE$ improvement relative to climatology is 41.5\% for AIFS, 35.5\% for FuXi, 33.5\% for Aurora, 30.5\% for ArchesWeather, 30.0\% for NCEP, 28.1\% for GraphCast, 26.4\% for IFS, 17.3\% for persistence, 16.4\% for Pangu-Weather, and -24.1\% for CMA. Not only does AIFS achieve the largest improvements, it also never exceeds climatological $RMSE$ across all lead times and events---the only model to do so (Figure S21). Additionally, it exhibits the lowest variability in $RMSE$ improvement across all lead time--event combinations, indicating robust performance (Figure S21). CMA is the outlier in the case studies. While it achieves comparable scores at other locations, its $RMSE$ inflates substantially in Georgetown (Figure \ref{fig:FIG5}) across all lead times (Figure S21). The case studies confirm that forecasts remain informative beyond climatology. Models can be skillful and successfully predict major heat extremes, but not consistently. Although forecasts are generally valuable during periods of elevated temperatures, extreme heat predictions cannot be reliably expected to precede actual extremes (Figure \ref{fig:FIG5}, right).

None of the evaluated models were explicitly trained for the task of extreme heat forecasting, rather, they were asked to learn the mapping from the current atmospheric reanalysis state to the next, typically six hours ahead. More holistic integration of relevant Earth-system components will likely be essential to unlock the full potential of AI systems for subseasonal heat-risk prediction. A key advantage of deep learning systems lies in their computational efficiency, which enables scaling to large ensembles at relatively low cost. This opens opportunities for probabilistic forecasting of rare extremes, where large sample sizes are essential for robust risk estimation. Several ensemble-based emulators already exist, but were excluded from this study because of accessibility and compute limitations. The deterministic predictors used here can instead be viewed as analogous to control forecasts in NWP ensemble systems, providing a baseline representation of model performance. Continuous research is ongoing regarding pipeline improvements; as well as on potential data corrections and on diverse approaches to ensure physical realism of these models (e.g. \cite{schreck2024community, vonich2025testing, bonev2025fourcastnet, sha2025improving}). More broadly, the emulators are faithfully optimizing the objective they've been given \cite{bi2022pangu, lam2023learning, chen2023fuxi, couairon2024archesweather, lang2024aifs, bodnar2025foundation}. Consequently, improving subseasonal heat-risk prediction may depend not only on model architecture, but also on how the learning problem is formulated through the training data, loss function, target variables, and overall training pipeline.

\section{Conclusion}\label{Conclusion}

This study benchmarks nine models---three physics-based systems and six AI emulators---against climatological and persistence baselines for global surface temperature forecasting and hot extreme prediction at 10–15 day lead times. Results shows that current deterministic weather emulators fail to reliably capture temperature extremes at this frontier between medium-range and subseasonal forecasting. Emulators generally outperform physics-based systems on standard error metrics, but do so by producing overly smooth forecasts that systematically underestimate localized extremes---a critical weakness for heat-risk applications. AIFS stands out as the only emulator combining competitive point prediction scores with good spectral fidelity; while IFS achieves a higher recall of heat extremes than any of the emulators. All models improve on baselines when forecasting within large-scale heat events, suggesting some useful signal persists beyond the typical weather range, although skill degrades substantially with increasing lead time.

These results reflect that none of the evaluated systems were designed or trained for extended lead times or for extreme-heat prediction specifically. Current models offer useful early-warning signals but fall short of reliably anticipating who will be exposed, where, and by how much. Key directions for improvement include purpose-built subseasonal emulators, probabilistic outputs enabling risk-based decision-making, and better integration of Earth-system components critical to extended-lead predictability.

\section{Methods}\label{Methods}
\subsection{Problem statement}\label{M-ProblemStatement}

Practical use of an early warning system for extreme heat necessitates it delivering a high-resolution, reliable product. Therefore, this study focuses on a suite of state-of-the-art deterministic deep learning weather emulators designed to forecast weather, that is, momentary states of the atmosphere at up to 10 days ahead---mostly referred to as the short- to medium-range. Here, we seek to answer to what degree weather models can forecast surface temperature and its hot extremes at the interface of medium-range and subseasonal scales. In recent years, the weather forecasting community has seen a rapid rise of AI emulators; here, we test several of these models against well-established physics-based forecasting systems. As statistical baselines for weather prediction, ERA5 climatology and persistence are used. A threshold of two standard deviations above ERA5 climatological mean is used to define heat extremes. Thresholding based on deviations from climatology is frequently used (e.g. \cite{van2022using, weirich2023subseasonal}), as it is straightforward and avoids the ambiguity of using various heatwave definitions \cite{perkins2013measurement, barriopedro2023heat}. More detail about how climatology is constructed follows in Section \ref{M-Baselines} and Figure S1.

\subsection{Emulators}\label{M-Emulators}

AIFS, GraphCast, Pangu-Weather, FuXi, ArchesWeather and Aurora are evaluated as state-of-the-art deep learning weather emulators; Table S3 for details on model versions. Although several ensemble-based emulators are available (e.g. AIFS-CRPS \cite{lang2026aifs}, GenCast \cite{price2023gencast}, ArchesWeatherGen \cite{couairon2024archesweather}, NVIDIA's FourCastNet family \cite{kurth2023fourcastnet, bonev2025fourcastnet}), these are excluded from this study because of accessibility issues at the time of writing as well as compute limitations. All of the models used here are either fully open-source or at least provide freely available checkpoints.

All of the above methods are used to forecast two-meter surface temperature for the six events mentioned in the introduction, the duration and spatial extent of which can be found in Table S2. With the exception of Aurora, all models (that is, their versions as used in this study---Table S3) are trained exclusively on ERA5 data \cite{hersbach2020era5} or derived products. Aurora, in contrast, is designed as a general-purpose foundation model trained on diverse data sources beyond ERA5; however, the configuration used here is restricted to ERA5 inputs for consistency across models \cite{bodnar2025foundation}. As none of the models were trained on data past $2020$, all events are out-of-sample.

Forecasts are initialized so that, for every day throughout the events (Table S2) and at four distinct times per day ($00h$, $06h$, $12h$ and $18h$ UTC), predictions are available whose valid time is precisely $10$, $11$, $12$, $13$, $14$ and $15$ days after initialization. As a result, in total, this study considers $392$ unique forecast rollouts from each emulator. ArchesWeather produces predictions on a $1.5\degree$ global grid in $24h$ steps, all other models operate on a $0.25\degree$ grid in $6h$ steps.

\subsection{Physics-based Methods}\label{M-PhysicsNumerical}

As golden-standard physics-based counterparts to compare modern emulators against, control forecasts are downloaded from the TIGGE (The International Grand Global Ensemble) database---an initiative of the World Weather Research Programme (WWRP) \cite{bougeault2010tigge, swinbank2016tigge}---for three large weather centers: the European Centre for Medium-Range Weather Forecasts (ECMWF), the China Meteorological Administration (CMA) and the United States National Centers for Environmental Prediction (NCEP). Model versions used are cycle 49r1 from the Integrated Forecasting System (IFS) from ECMWF, the Global and Regional Assimilation and PrEdiction System (GRAPES) from CMA, and the Global Ensemble Forecast System (GEFS) v12 from NCEP. Their forecast data are available on TIGGE at a global $0.5\degree$ grid in $6h$ steps. ECMWF-IFS and CMA models are initialized only at $00h$ and $12h$ UTC, because of which half as many (196) rollouts are included in this evaluation compared to all other models (392).

\subsection{Baseline Methods}\label{M-Baselines}

Throughout the study, climatology and persistence are used as statistical baselines for model evaluation. Here, climatology is fully ERA5-derived and specific to the location, the time of day and the day of the year. It is established at $0.25\degree$ resolution and computed as mean and standard deviation of hourly ERA5 two-meter temperature data over the period $1990$--$2019$. Both mean and standard deviation are then smoothed with a 31-day rolling window, where weights decrease linearly with time-distance from the center day. The result is, for example, climatology specific to 12:00 UTC in Bamako, Mali on the $150^{th}$ day of the year (Figure S1). To compare model forecasts against climatology (e.g. when computing anomalies), the ERA5 climatology is interpolated to the forecast resolution using linear conservative regridding, following Rasp et al. \cite{rasp2024weatherbench}.

Persistence uses the current state of the atmosphere as an estimate for its future state. Here, that means using the ERA5 global temperature field at initialization time as an estimate for global temperatures at the forecast valid time. For example, the 10-day lead persistence forecast for 2021-06-29 18:00 UTC in Vancouver is the ERA5 temperature on 2021-06-19 18:00 UTC at that location. Persistence is at native ERA5 resolution, i.e. hourly and $0.25\degree$ spatial resolution.

\subsection{Evaluation of continuous temperature fields}\label{M-Continuous-Evaluation}

For each lead time and event period, forecasts are evaluated over global land areas---where the ERA5 land-sea mask is greater than or equal to $0.5$---, as well as within the event boxes for heat extremes at locations experiencing three or more days of extreme heat (red-hued regions in Figure \ref{fig:FIG1}). Table S2 details the period and geographic bounds used to study each extreme heat event, and Figure S2 illustrates how, for a given lead time, a forecast time series is evaluated against its reference. For emulators, ERA5 two-meter temperature is considered the reference to which all forecasts are compared. For ArchesWeather, that is ERA5 interpolated to the model ($1.5\degree)$ grid. For physics-based systems, the reference ground truth is their respective initialization state (denoted \textit{fc0}) at the forecast valid time.

$RMSE$, $R^2$, $ACC$, temporal correlation and spectral score are computed for forecast time series; where the first two metrics are based on raw data, the latter three on anomalies. Anomalies are relative to ERA5 climatology, respecting forecast resolution, as described above. To fairly evaluate global equiangular gridded data across a spatial field, area-weighting is used. As all data are on regular latitude-longitude grids, the area \(a_{i,j}\) in \(\mathrm{km}^2\) of a grid cell at latitude $i$ and longitude $j$ on a spherical Earth is given by:

\begin{equation}
\begin{aligned}
a_{i,j} ={}& R^{2}\,\Delta\lambda
\left[
\sin\left(\varphi_i + \frac{\Delta\varphi}{2}\right)
-
\sin\left(\varphi_i - \frac{\Delta\varphi}{2}\right)
\right],\\
\text{where }&
\varphi_i \pm \frac{\Delta\varphi}{2}
\in \left[-\frac{\pi}{2},\frac{\pi}{2}\right].
\end{aligned}
\end{equation}

Here, \(R\) is Earth's radius (approximately \(6371\,\mathrm{km}\)), \(\varphi_{i}\) is the latitude at the centre of the grid cell \(i\), in radians, and \(\Delta\varphi\) and \(\Delta\lambda\) are the latitude and longitude spacing between adjacent grid points, respectively. Simplified, to account for surface area when averaging in space, data at a given latitude $i$ can be weighted:

\begin{equation}
    w_{i} \propto \sin\left(\varphi_{i} + \frac{\Delta\varphi}{2}\right)
             - \sin\left(\varphi_{i} - \frac{\Delta\varphi}{2}\right) = 2\cos\left(\varphi_{i}\right)\sin\left(\frac{\Delta\varphi}{2}\right)\,.
\end{equation}

$RMSE$ and $ACC$ are standard forecast evaluation metrics used in most studies \cite{bi2022pangu, chen2023fuxi, lang2024aifs, lam2023learning, bodnar2025foundation, couairon2024archesweather, persson2007user, rasp2024weatherbench}. $RMSE$ for spatiotemporal data is computed by first taking the root-mean-squared-error in time at each grid point and then applying area weights:

\begin{equation}
\mathrm{RMSE}
= 
\frac{
\sum_{i,j} w_{i,j}\; \mathrm{RMSE}_{i,j}
}{
\sum_{i,j} w_{i,j}
},
\qquad
\mathrm{RMSE}_{i,j}
=
\sqrt{
\frac{1}{T}\sum_{t=1}^T (f_{t,i,j}-o_{t,i,j})^2 
}.
\end{equation}

Here \(f(t,i,j)\) and \(o(t,i,j)\) denote the forecast and reference temperature fields, respectively. The reference for emulators is ERA5, for dynamical models it is their \textit{fc0}. Let $f'(t,i,j)$ denote the forecast anomaly and $o'(t,i,j)$ the reference anomaly:

\begin{equation}
\begin{aligned}
f'(t,i,j) = f(t,i,j) - \overline{o}(t,i,j),\\
o'(t,i,j) = o(t,i,j) - \overline{o}(t,i,j),
\end{aligned}
\end{equation}

where $\overline{o}$ is the location- and time-dependent ERA5 climatological mean. The anomaly correlation coefficient (ACC) for spatiotemporal data is computed by first evaluating a spatially weighted pattern correlation at each verification time and then averaging over time:

\begin{equation}
\mathrm{ACC}
=
\frac{1}{T}
\sum_{t=1}^{T}
\frac{
\sum_{i,j} w_{i,j}\, f'(t,i,j)\, o'(t,i,j)
}{
\sqrt{
\left(
\sum_{i,j} w_{i,j}\, f'(t,i,j)^2
\right)
\left(
\sum_{i,j} w_{i,j}\, o'(t,i,j)^2
\right)
}
}.
\end{equation}

Surface temperature variability on Earth follows a strong latitudinal gradient---the fluctuation of temperatures in time at the poles is several times larger than that in the tropics---driving a corresponding latitudinal dependency in forecast $RMSE$ (Figure S14). As a consequence, a global summary of $RMSE$ will be disproportionately dominated by extratropical error. The use of $R^2$ compensates for this effect by scaling the forecast error relative to the local background variability. $R^2$ is computed gridpoint-wise over time and then
spatially averaged using the area weights:

\begin{equation}
\begin{aligned}
&R^2
=
\frac{
\sum_{i,j} w_{i,j}
\left[
1-\frac{\mathrm{SS}_{\mathrm{res},i,j}}
        {\mathrm{SS}_{\mathrm{tot},i,j}}
\right]
}{
\sum_{i,j} w_{i,j}
},
\\[0.5em]
&\mathrm{SS}_{\mathrm{res},i,j}
=
\sum_{t=1}^{T}
(f_{t,i,j}-o_{t,i,j})^2,
\\
&\mathrm{SS}_{\mathrm{tot},i,j} =
\sum_{t=1}^{T}
\left(o_{t,i,j}-\overline{o}_{i,j}\right)^2.
\end{aligned}
\end{equation}

$\mathrm{SS}_{\mathrm{res}}$ and $\mathrm{SS}_{\mathrm{tot}}$ denote the residual and total sums of squares, respectively, and \(\overline{o}_{i,j}\) the temporal mean of the reference at location $(i,j)$. Temporal correlation ($\mathrm{corr}$), then, is used to gauge whether models are capable of forecasting the evolution of temperature anomalies over time. It is computed at each grid cell as the Pearson
correlation between forecast and reference anomaly over the \(T\) time steps:

\begin{equation}
\mathrm{corr}_{i,j}
=
\rho_{\mathrm{time}}
\!\left(
f'_{:,i,j},\, o'_{:,i,j}
\right).    
\end{equation}

For a given extent, temporal correlation is then the spatially weighted mean:

\begin{equation}
\mathrm{corr}_{\mathrm{event}}
=
\frac{
\sum_{i,j} w_{i,j} \; \mathrm{corr}_{i,j}
}{
\sum_{i,j} w_{i,j}
}.
\end{equation}

As forecast \textit{blurring} upon rollout is a common issue among many emulators, studies often detect this behaviour by decomposing variable fields into power spectra to visually compare if the energy at different wavelengths in the forecast matches that in the reference \cite{lam2023learning, lang2024aifs, rasp2024weatherbench}. This work proposes a method that uses forecast power spectra to formulate a score, which we refer to as the spectral score. Following WeatherBench2 \cite{rasp2024weatherbench}, a discrete Fourier Frequency Transform is applied to derive the data power spectrum at every latitude; the methodology is explained in Section S8. This is the only evaluation for which global temperature fields are not masked to land only, as the transform requires data to be continuous in space. For a given time $t$ and latitude $\varphi_i$, this yields the energy spectrum $S_{t, i}[k]$ along the latitude band at latitude $i$, with k the wavenumber. This wavenumber $k$ corresponds to a certain frequency and wavelength depending on the latitude $\varphi_i$:

\begin{equation}
\nu_{k,i} = \frac{k}{2\pi R\, cos(\varphi_i)}\quad (\text{cycles m}^{-1}),
\qquad
\lambda_{k,i} = \frac{1}{\nu_{k,i}}\quad (\text{m}),
\end{equation}

With R Earth's radius. To obtain wavelength power spectra across latitudes, common logarithmically spaced wavelength bins \(\{\Lambda_m\}_{m=1}^{M}\) are proposed. For each latitude \(i\), interpolate \(\{(\lambda_{k,i}, S_{t,i}[k])\}_k\) onto \(\{\Lambda_m\}\) to obtain \(S_{t,i}^{(\Lambda)}(\Lambda_m)\). For a latitude set \(\mathcal{I}\) (domain of interest), the cosine-weighted average then represents the spatial average of spectral power per wavelength:

\begin{equation}
\overline{S}_t(\Lambda_m) = \frac{\sum_{i\in\mathcal{I}} \cos\varphi_i\,S_{t,i}^{(\Lambda)}(\Lambda_m)}{\sum_{i\in\mathcal{I}}\cos\varphi_i}.    
\end{equation}

Finally, the aim is to assess whether a model's forecasts capture the true spatial variability of surface temperature across scales. For a given wavelength bin $m$, let \(r_t(\Lambda_m)\) be the absolute logarithmic ratio of forecast (\(\overline{S}^{F}_t(\Lambda_m)\)) and true (\(\overline{S}^{T}_t(\Lambda_m)\)) spectral power on a common \(\{\Lambda_m\}\). For a perfect model, \(r_t = 0\). A spectral score quantifying similarity across spatial scales is proposed:

\begin{equation}
\begin{aligned}
&r_t(\Lambda_m) \;=\; \left|\log_{10}\!\left(
\frac{\overline{S}^{F}_t(\Lambda_m)}{\overline{S}^{T}_t(\Lambda_m)}\right)\right|,\\
&\mathrm{Spectral\ score}_t \;=\;
1 - \frac{1}{M}\sum_{m=1}^{M} r_t(\Lambda_m).
\end{aligned}
\end{equation}

The score equals \(1\) when spectra match exactly; and decreases the more they differ. Over a given period, spectral scores at each time step can be averaged.

\subsection{Evaluation of categorical heat extreme fields}\label{M-Categorical-Evaluation}

To further evaluate each model's ability to forecast heat extremes, we cast the forecast problem to a binary classification problem. Classification-based metrics (accuracy, precision, recall, $ETS$, reliability indicators) are computed on threshold-derived binary transformations of the model forecast. With the exception of later reliability indicators, this threshold is set at surface temperatures equal to or exceeding two standard deviations above the climatological mean. For every time step, the reference temperature field identifies grid cells experiencing extreme heat (\(T \geq \mu + 2\sigma\)), and the model forecasts are correspondingly cast into binary format. Let

\begin{equation}
H_{t,i,j} \in \{0,1\}, \qquad
\widehat{H}_{t,i,j} \in \{0,1\}    
\end{equation}

denote the true and forecast heatwave indicators at time \(t\) and grid cell \((i,j)\). Again, all metrics are computed using area weights \(w_{i,j}\) and restricted to land points. For each time step, we form the weighted areas $A$ of true positives (TP), false positives (FP), false negatives (FN), and true negatives (TN):

\begin{equation}
\begin{aligned}
A_{\mathrm{TP}}(t)
&=
\sum_{i,j}
w_{i,j}\,
H_{t,i,j}\widehat{H}_{t,i,j},
\\
A_{\mathrm{FP}}(t)
&=
\sum_{i,j}
w_{i,j}\,
(1-H_{t,i,j})\widehat{H}_{t,i,j},
\\
A_{\mathrm{FN}}(t)
&=
\sum_{i,j}
w_{i,j}\,
H_{t,i,j}(1-\widehat{H}_{t,i,j}),
\\
A_{\mathrm{TN}}(t)
&=
\sum_{i,j}
w_{i,j}\,
(1-H_{t,i,j})(1-\widehat{H}_{t,i,j}).
\end{aligned}
\end{equation}

Then, given the total evaluated area $A_{\mathrm{tot}} = \sum_{i,j} w_{i,j}$, the true affected area $A_{\mathrm{true}}(t)=A_{\mathrm{TP}}(t)+A_{\mathrm{FN}}(t)$, and the forecast area of heat extremes $A_{\mathrm{forecast}}(t)=A_{\mathrm{TP}}(t)+A_{\mathrm{FP}}(t)$:

\begin{equation}
\begin{aligned}
\mathrm{Accuracy}(t)
&=
\frac{A_{\mathrm{TP}}(t)+A_{\mathrm{TN}}(t)}
     {A_{\mathrm{tot}}},
\\
\mathrm{Recall}(t)
&=
\frac{A_{\mathrm{TP}}(t)}
     {A_{\mathrm{TP}}(t)+A_{\mathrm{FN}}(t)},
\\
\mathrm{Precision}(t)
&=
\frac{A_{\mathrm{TP}}(t)}
     {A_{\mathrm{TP}}(t)+A_{\mathrm{FP}}(t)}.
\end{aligned}
\end{equation}

Finally, given that any forecast has a probability of randomly hitting true positives $A_{\mathrm{rand}}(t)=A_{\mathrm{true}}(t) \times A_{\mathrm{forecast}}(t)  / A_{\mathrm{tot}}$, the $ETS$ subtracts the number of hits expected by chance from both the numerator and denominator of the intersection over union:

\begin{equation}
\mathrm{ETS}(t)
=
\frac{
A_{\mathrm{TP}}(t)-A_{\mathrm{rand}}(t)
}{
A_{\mathrm{TP}}(t)
+
A_{\mathrm{FP}}(t)
+
A_{\mathrm{FN}}(t)
-
A_{\mathrm{rand}}(t)
},
\end{equation}

To assess whether a model's forecast of extreme temperatures is trustworthy, we evaluate how often such forecasts coincide with above-average, hot, or extremely hot conditions, observed in the reference. Let \(\mu_{t, i,j}\) denote the location- and time-specific ERA5 climatological mean and \(\sigma_{t, i,j}\) the corresponding standard deviation. A grid cell is classified as

\begin{equation}
\begin{aligned}
&\text{above average: } o_{t,i,j} \geq \mu_{t,i,j},\\
&\text{hot: } o_{t,i,j} \geq \mu_{t,i,j} + \sigma_{t,i,j},\\
&\text{extremely hot: } o_{t,i,j} \geq \mu_{t,i,j} + 2\sigma_{t,i,j}.
\end{aligned}
\end{equation}

\vspace{0.5em}

A forecast is extreme at \((t,i,j)\) when the indicator function $\mathbbm{1}\!(\cdot)$ equals 1:

\begin{equation}
\widehat{H}_{t,i,j} = \mathbbm{1}\!\left(f_{t,i,j} \geq \mu_{t,i,j} + 2\sigma_{t,i,j}\right).    
\end{equation}

\vspace{0.5em}
For each time step, we compute the weighted area of all forecast extreme-heat grid cells,

\begin{equation}
A_{\mathrm{F}}(t) = \sum_{i,j} w_{i,j}\, \widehat{H}_{t,i,j},
\end{equation}

and the weighted areas where this forecast coincides with observed categories:

\begin{equation}
\begin{aligned}
&A_{\mathrm{F\cap N}}(t)   = \sum_{i,j} w_{i,j}\, \widehat{H}_{t,i,j}\, 
                             \mathbbm{1}(o_{t,i,j} \geq \mu_{t,i,j}),\\
&A_{\mathrm{F\cap H}}(t)   = \sum_{i,j} w_{i,j}\, \widehat{H}_{t,i,j}\,
                             \mathbbm{1}(o_{t,i,j} \geq \mu_{t,i,j}+\sigma_{t,i,j}),\\
&A_{\mathrm{F\cap E}}(t)   = \sum_{i,j} w_{i,j}\, \widehat{H}_{t,i,j}\,
                             \mathbbm{1}(o_{t,i,j} \geq \mu_{t,i,j} + 2\sigma_{t,i,j}).
\end{aligned}
\end{equation}

The reliability metrics are the conditional probabilities:

\begin{equation}
\begin{aligned}
&R_{\mathrm{normal}}(t)
  = \frac{A_{\mathrm{F\cap N}}(t)}{A_{\mathrm{F}}(t)},\\
&R_{\mathrm{hot}}(t)
  = \frac{A_{\mathrm{F\cap H}}(t)}{A_{\mathrm{F}}(t)},\\
&R_{\mathrm{extreme}}(t)
  = \frac{A_{\mathrm{F\cap E}}(t)}{A_{\mathrm{F}}(t)}.
\end{aligned}
\end{equation}

These represent the fraction of the forecast extreme heat area that in the reference lies above normal temperatures, above one standard deviation, or above the extreme-heat threshold.

\newpage


\bibliographystyle{naturemag}

\section*{Author contribution}
C.D., T.M. and D.G.M. conceived and designed the study. C.D. conducted the experiments. C.D. analyzed the results and wrote the manuscript. C.D., T.M. and D.G.M. revised the manuscript with inputs from J.K. All authors read and approved the final manuscript.

\section*{Acknowledgements}
This study was funded by the European Research Council (ERC) via the HEAT Consolidator grant (101088405). The funder played no role in study design, data collection, analysis and interpretation of data, or the writing of this manuscript. The computational resources and services used in this work were provided by the VSC (Flemish Supercomputer Center), funded by the Research Foundation, Flanders (FWO), and the Flemish Government.

\section*{Competing interests}
All authors declare no financial or non-financial competing interests. 

\section*{Data availability}
All ERA5 data used is publicly available at \url{https://cds.climate.copernicus.eu/}. In this work, the authors used Analysis Ready, Cloud Optimized (ARCO) ERA5 provided on \url{https://console.cloud.google.com/marketplace/product/bigquery-public-data/arco-era5}. IFS, GRAPES and GEFS forecasts were downloaded from the TIGGE Data Retrieval portal \url{https://apps.ecmwf.int/datasets/data/tigge/}. As of June 2026, these data are now hosted on the ECDS portal \url{https://ecds.ecmwf.int/datasets/tigge-forecasts?tab=overview}. Except for AIFS, which was downloaded via HuggingFace (\url{https://huggingface.co/ecmwf/aifs-single-0.2.1}), all AI emulators were implemented via their public GitHub repositories (\url{https://github.com/198808xc/Pangu-Weather} | \url{https://github.com/tpys/FuXi} | \url{https://github.com/INRIA/geoarches} | \url{https://github.com/google-deepmind/weathernext} | \url{https://github.com/microsoft/aurora}).

\section*{Code availability}
All code used in this study is available on GitHub: \url{https://github.com/Cas-Dec/HEAT-S2S-Benchmarking}.

\clearpage


\section*{Supplementary material}

# S0. DEFINITIONS

***Supplementary Table 1: Definitions.*** *Several terms used throughout the manuscript, their definitions and synonyms.*

| Term | Definition | Synonyms |
|---|---|---|
| Extreme heat | Near-surface temperature exceeding the threshold of ERA5 climatological mean plus two standard deviations. | Hot extreme, heat extreme |
| Dynamical system | A weather model that simulates the evolution of the atmosphere by solving the fundamental equations of fluid dynamics and thermodynamics on a discretized grid. | Physics-based model |
| Emulator | A deep learning weather model that learns statistical relationships between past and future atmospheric states from large historical datasets (here ERA5). Given an initial state, it predicts future fields by applying trained neural network mappings. | AI weather model, deep learning weather model |
| Reference temperature | The "true observed" temperature for a certain model, which is what the model aims to predict. For emulators, this truth is ERA5; for dynamical systems, this is their respective *fc0* field. | Observed temperature |
| Rollout | The temperature timeseries created by a model's autoregressive rollout from a single initialization. | |
| Fixed-lead-time time series | The temperature timeseries created by concatenating the $n^{th}$ step in a rollout from multiple rollouts stemming from subsequent initialization steps. See also Supplementary Figure 2. | X-day lead forecast (X denoting the lead time in days) |
| Global evaluation | Evaluation over all land area on Earth. | |
| Event-level evaluation | Evaluation of forecasts where their reference temperature is extreme, constrained to (1) land area, (2) the event bounds (Supplementary Table 2), and (3) locations with ≥ 3 days of extreme heat in the reference throughout the event. | Evaluation in affected regions |

# S1. EVENT DETAILS

Supplementary Table 2 specifies the details of the studied extreme heat events.

***Supplementary Table 2: Details of studied events and case studies.*** *The period and spatial extent (rows) of the six studied large-scale extreme heat events (columns) are detailed. Time and coordinates follow ERA5 convention, that is, UTC time and regular latitude-longitude (Plate Carrée) coordinate convention, with longitudes ranging 0°-360° (Greenwich).*

| Event Details | 2021 Pacific Heat Dome | 2022 Europe | 2022 India-Pakistan | 2023 Southeast Asia | 2024 Africa | 2024 Latin America |
|---|---|---|---|---|---|---|
| Period | 06/25 – 07/07 | 07/10 – 07/20 | 04/10 – 04/20 | 04/08 – 04/19 | 03/28 – 04/06 | 03/11 – 03/21 |
| Latitude bounds | 65° N – 35°N | 60°N – 35°N | 37.5°N – 20°N | 37.5°N – 5°N | 22°N – 10°S | 12.25°N – 35°S |
| Longitude bounds | 230° – 300° | 348° – 17.5° | 62° – 89.5° | 90.5° – 120° | 340° – 40° | 279° – 320.5° |
| Case study city | Vancouver | London | Islamabad | Luang Prabang | Bamako | Georgetown |
| City location (lat, lon) | 49.25°, 237.25° | 51.5°, 359° | 33.75°, 73° | 19.75°, 102° | 12.5°, 352° | 6.5°, 302° |

# S2. EMULATOR DETAILS

Supplementary Table 3 specifies version and other details of all emulators studied. All models used are publicly available and implemented either using their official GitHub page (Pangu-Weather, FuXi, ArchesWeather, GraphCast), Hugging Face (AIFS) or API (Aurora). Their input typically consists of the current (time *t*) and previous (time *t – Δt*) state of the Earth / atmosphere, represented by surface-level, pressure-level and static variable fields; most of which are default ERA5 data. Some require constructing custom inputs, such as accumulated precipitation.

***Supplementary Table 3: Details of studied emulators.*** *All models and their release version, variable fields count, input states, step size and spatiotemporal resolution are detailed. Variable fields count is the amount of variables the user has to prepare as model input, not considering possible features added in model inference. The state 't' is the set of all input data at the time of model initialization.*

| Model | Release Version | Variable Fields Count | Input States | Step Size | Spatial Resolution | Temporal Resolution | Loss Function |
|---|---|---|---|---|---|---|---|
| Pangu-Weather | v1.0 | 69 | *t* | 24h, 12h, 6h | 0.25° | hourly | $w_\theta \times MSE$ |
| FuXi (3-stage) | FuXi demo | 69 | *t*-6h, *t* | 6h | 0.25° | hourly | $w_\theta \times MAE$ |
| ArchesWeather-Mx4 | v1.0 | 72 | *t*–24h, *t* | 24h | 1. 5° | hourly | $w_\theta \times MSE$ |
| AIFS Single | v0.2.1 | 90 | *t*-6h, *t* | 6h | 0.25° | hourly | $w_\theta \times MSE$ |
| GraphCast | v0.1.1 | 229 | *t*-6h, *t* | 6h | 0.25° | hourly | $w_\theta \times MSE$ |
| Aurora | v1.7.0 | 72 | *t*-6h, *t* | 6h | 0.25° | hourly | $w_\theta \times MSE$ |

# S3. CREATING CLIMATOLOGY

Climatology is created starting from ERA5 two-meter temperature (T) on the default 0.25° latitude x 0.25° longitude grid, years 1990-2019. For every combination of *(location, day of year, time of day)*, with *time of day* in {*00:00 UTC, 06:00 UTC, 12:00 UTC, 18:00 UTC*}, climatology is characterized by the mean ($\mu$) and standard deviation ($\sigma$) of those 30 datapoints. The resulting mean and standard deviation are smoothed along the *day of year* axis by means of a 31-day window that has weights linearly inverse to the time difference between included data and the centre point. The process is illustrated for a combination of (*location*: Bamako, *day of year*: 150) in Supplementary Figure 1.

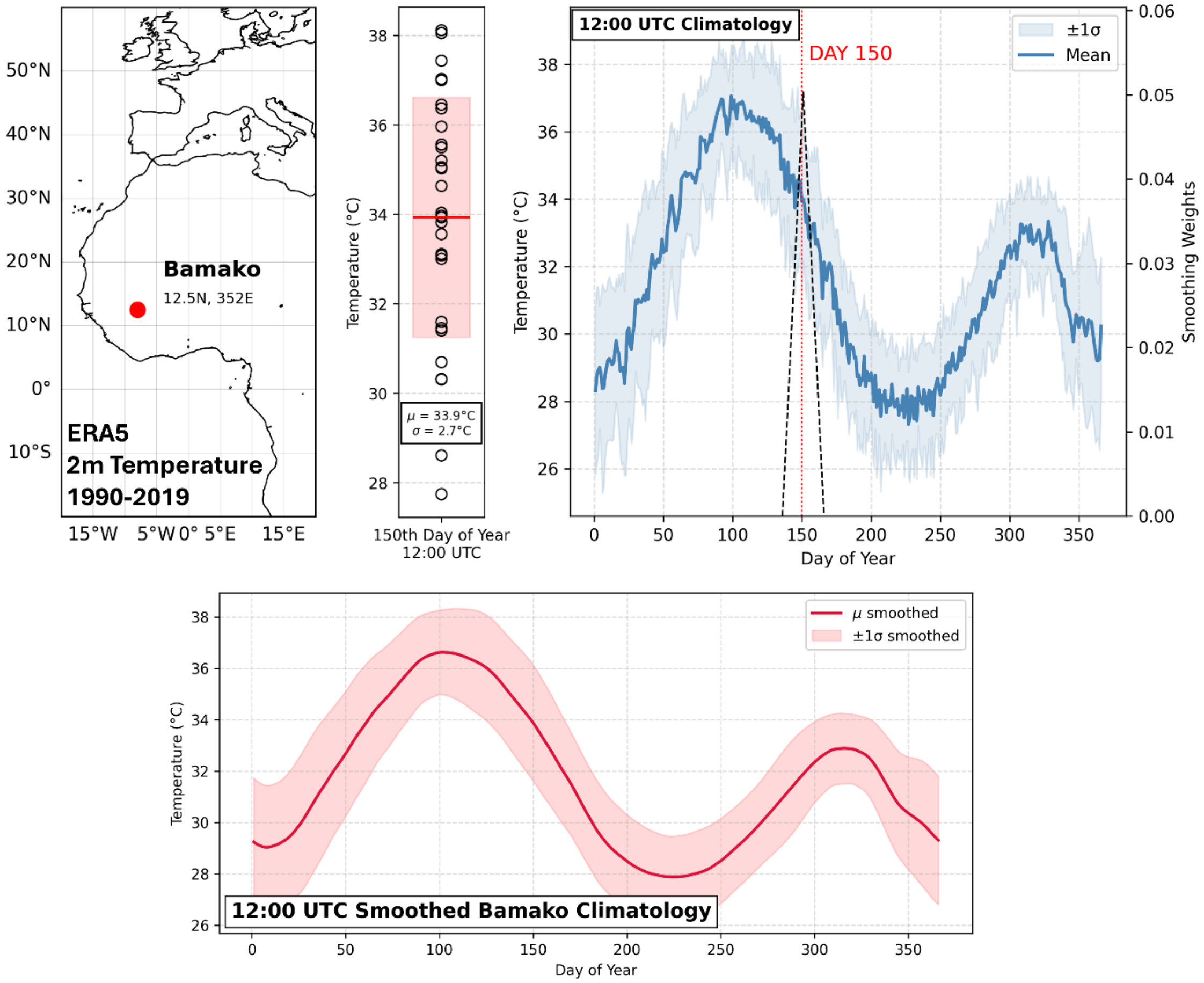


***Supplementary Figure 1: Creating climatology***. *ERA5 two-meter temperature is taken on the default 0.25° latitude x 0.25° longitude grid, years 1990-2019. For every combination of* (location, day of year, time of day), *with* time of day ∈ (00:00 UTC, 06:00 UTC, 12:00 UTC, 18:00 UTC), *climatology is characterized by the mean and standard deviation of those 30 datapoints. The top left and top middle plots illustrate how a point corresponding to the city of Bamako has 30 historical values for the 150$^{th}$ day of the year at 12:00 UTC. Mean and standard deviation are then smoothed for each location along the* day of year *axis by means of a 31-day window that has weights linearly inverse to the time difference between included data and the centre point. The window and its weights are illustrated in the top right panel for smoothing mean and standard deviation for the 150$^{th}$ day of the year. The bottom panel shows the resulting smooth climatology for Bamako, 12:00 UTC.*

# S4. FORECASTS AT FIXED LEAD TIMES

To evaluate weather models as a function of lead time, forecasts from successive rollouts are concatenated to form fixed-lead-time time series. The result is a time series where each reference date-time throughout an event has its own unique forecast at a fixed lead time, with lead times ranging from 10 to 15 days. Supplementary Figure 2 illustrates this for AIFS rollouts at a single location, Vancouver, throughout the 2021 Pacific Heat Dome (2021 June 25th to 2021 July 7th, Supplementary Table 2). In this example, the forecasts at lead day 14 are concatenated from different rollouts to form a single, 14-day lead forecast for the event.

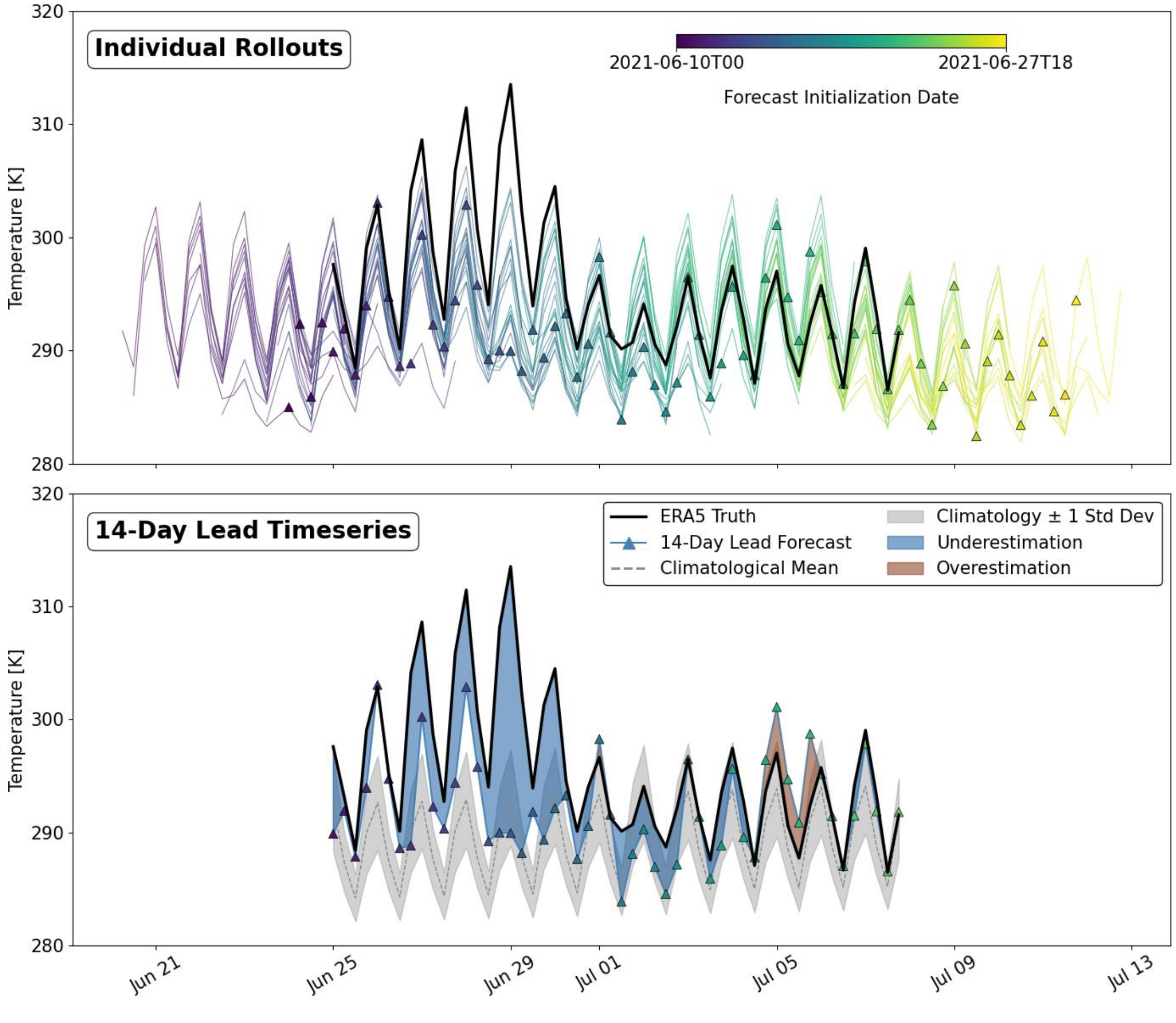


***Supplementary Figure 2: Creating forecasts at fixed lead times.*** *Location: Vancouver, BC, Canada (49.25°, 237.25°). Top panel: AIFS 15-day forecast rollouts from initializations ranging June 10th to June 27th, 2021 and the reference ERA5 two-meter temperature. Bottom panel: concatenation from every 14-day lead prediction into a single timeseries. This is the fixed-lead-time time series used to evaluate AIFS performance at a 14-day lead.*

# S5. DEFINITION OF HEAT EXTREMES

A straightforward threshold of two standard deviations above climatological mean is used to define heat extremes, avoiding the ambiguity of using various heatwave definitions. Using the data standard deviation assumes that the historical 1990-2019 record of surface temperature at a given combination *(location, day of year, time of day)* is normally distributed. Resulting from this, both extreme heat ($T \geq \mu + 2\sigma$) and extreme cold ($T \leq \mu - 2\sigma$) should have about 2.3% probability of occurring at any given time and location. Supplementary Figure 2 shows the observed global frequency of heat and cold extremes throughout the event periods (Table 1). The data shows extreme heat more frequent and extreme cold less frequent than the climatological expectation. This is to be expected, as the study is biased toward large-scale extreme heat events, where especially the 2024 Africa and Latin America events show vast area experiencing extreme heat. Additionally, climate change could also favour frequency of heat over cold relative to a historical baseline.

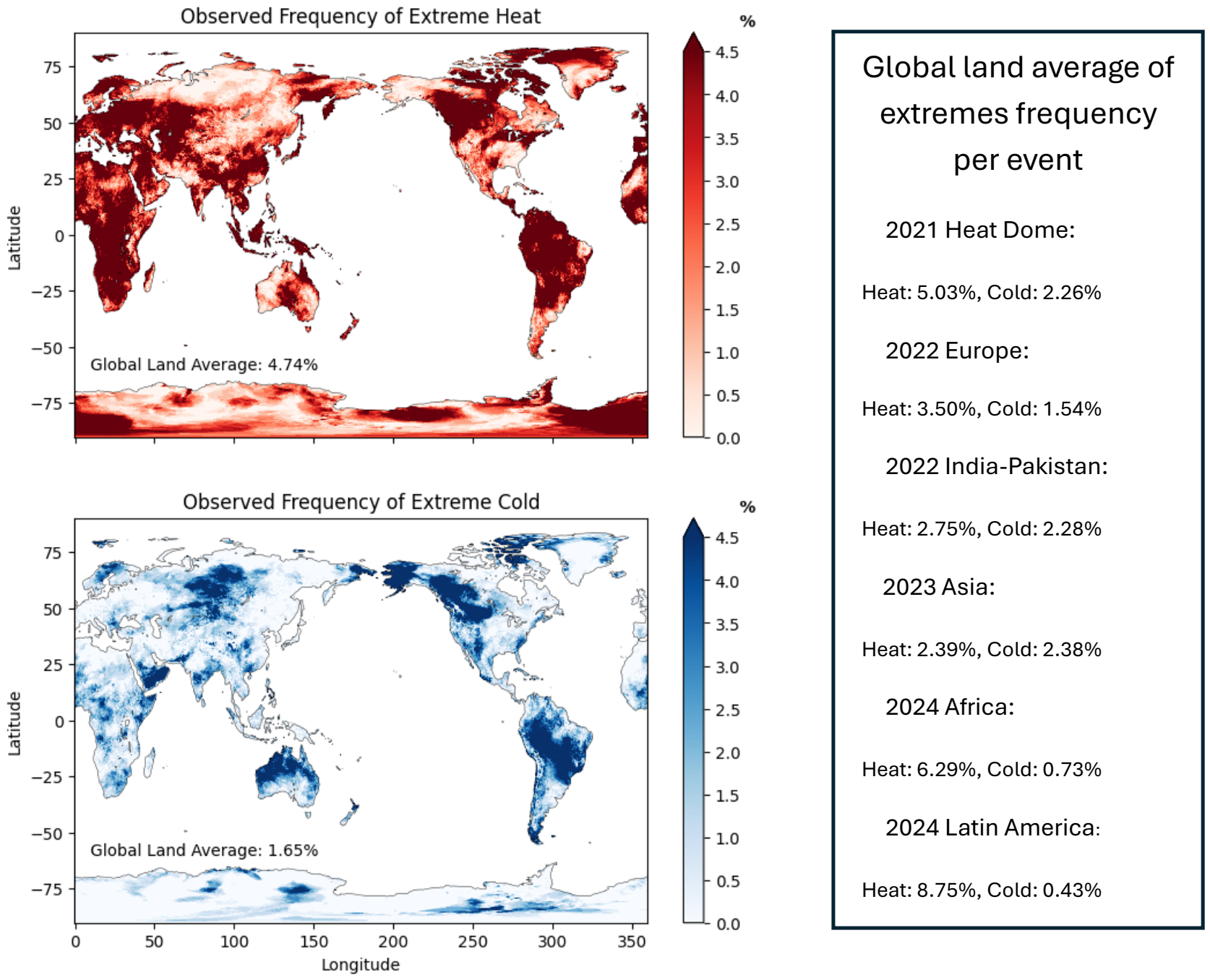


***Supplementary Figure 3: Observed frequency of temperature extremes.*** *Defining a temperature extreme as deviating more than or equal to two standard deviations from climatological mean, the total frequency of global heat extremes (top) and cold extremes (bottom) is shown as observed in the data across studied events (Table 1). The right panel shows this frequency per event. According to climatology, both hot and cold extremes are expected to occur with about 2.3% probability for at any given time and any location on Earth.*

# S6. AREA OF HEAT EXTREMES ON EARTH

Many emulators are prone to blurring, and consequently produce increasingly less surface temperature extremes in their forecasts as lead time increases. Supplementary Figure 4 illustrates how the true global area of extreme heat and cold also differs, depending on whether the reference is ERA5 or the *fc0* states from dynamical systems. Where this figure gives the example for one event period, Supplementary Table 3 reports the average global area of heat extremes in the forecast versus the reference, as a function of lead time. Supplementary Table 4 does the same but for cold extremes.

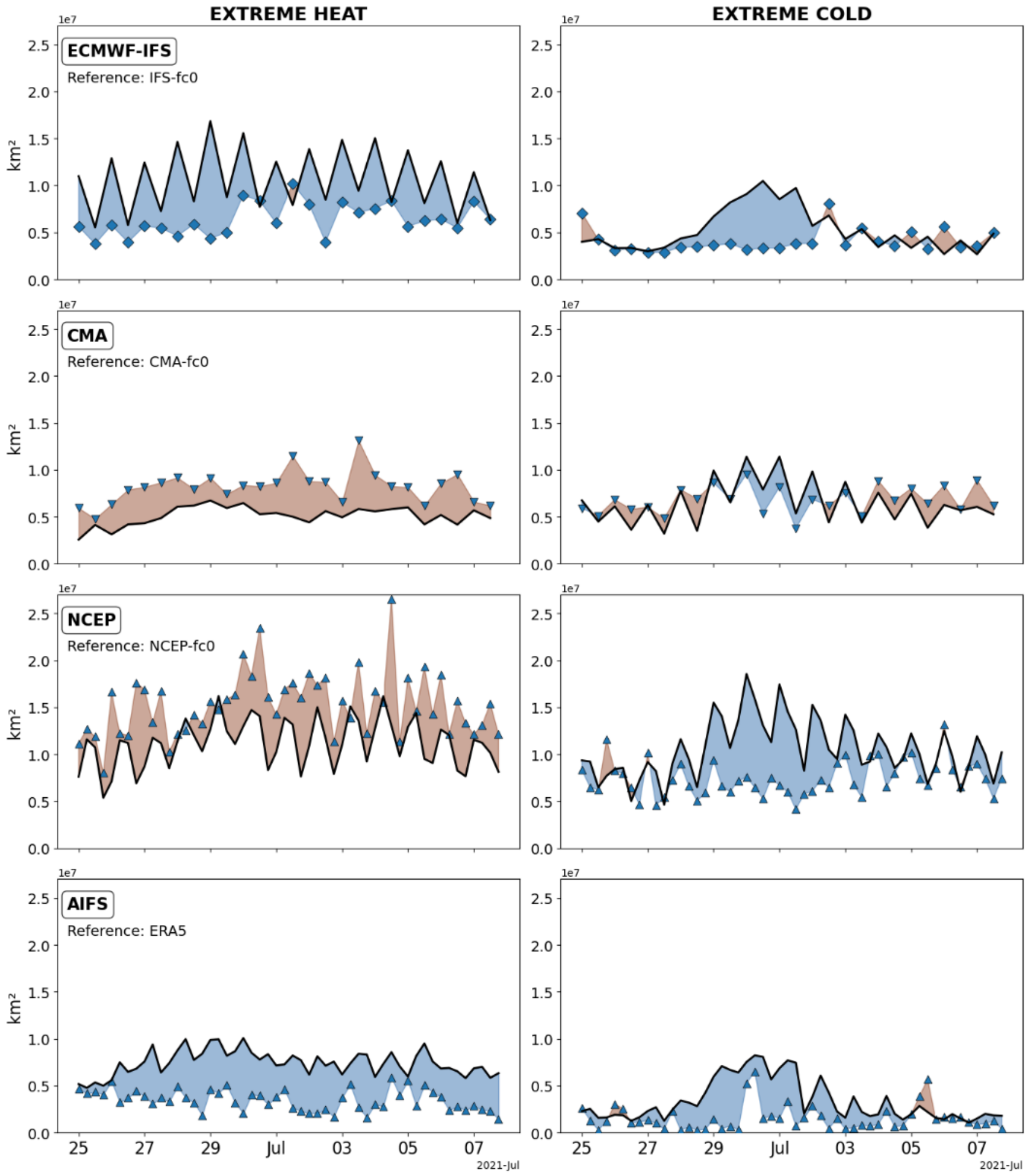


***Supplementary Figure 4: Global hot and cold extremes.*** *For four models (ECMWF-IFS, CMA, NCEP, AIFS), the Supplementary Figure hows the total global area of extreme heat (left) and cold (right) 14-day lead forecast by models (markers) versus their reference (black) throughout the 2021 Heat Dome period (Supplementary Table 2). Each dynamical system has its fc0 reference (initialization state at the forecast valid time), whereas AIFS has 0.25° ERA5 as a reference (0.25°—Supplementary Table 2).*

***Supplementary Table 4: Global hot extremes.*** *Comparison of average predicted and true extent of global hot extremes. Each cell shows area of extreme heat forecast (F, unit: 1e7 km$^2$), area in the "truth" reference (T, unit: 1e7 km$^2$), and their ratio (F/T).*

| Model | 10 days | 11 days | 12 days | 13 days | 14 days | 15 days |
|---|---|---|---|---|---|---|
| Persistence | 5.88 - 7.05 - 0.83 | 5.90 - 7.05 - 0.84 | 6.01 - 7.05 - 0.85 | 6.20 - 7.05 - 0.88 | 6.41 - 7.05 - 0.91 | 6.61 - 7.05 - 0.94 |
| ECMWF-IFS | 6.17 - 7.37 - 0.84 | 6.09 - 7.37 - 0.83 | 6.05 - 7.37 - 0.82 | 6.05 - 7.37 - 0.82 | 5.92 - 7.37 - 0.80 | 5.90 - 7.37 - 0.80 |
| CMA | 9.21 - 8.15 - 1.13 | 9.13 - 8.15 - 1.12 | 8.94 - 8.15 - 1.10 | 9.02 - 8.15 - 1.11 | 9.11 - 8.15 - 1.12 | 9.23 - 8.15 - 1.13 |
| NCEP | 11.39 - 8.20 - 1.39 | 11.50 - 8.20 - 1.40 | 11.51 - 8.20 - 1.40 | 11.40 - 8.20 - 1.39 | 11.26 - 8.20 - 1.37 | 11.38 - 8.20 - 1.39 |
| Pangu-Weather | 3.97 - 7.05 - 0.56 | 3.97 - 7.05 - 0.56 | 3.89 - 7.05 - 0.55 | 3.78 - 7.05 - 0.54 | 3.62 - 7.05 - 0.51 | 3.51 - 7.05 - 0.50 |
| FuXi | 3.33 - 7.05 - 0.47 | 2.04 - 7.05 - 0.29 | 1.75 - 7.05 - 0.25 | 1.65 - 7.05 - 0.23 | 1.64 - 7.05 - 0.23 | 1.69 - 7.05 - 0.24 |
| ArchesWeather | 1.55 - 5.90 - 0.26 | 1.22 - 5.90 - 0.21 | 0.95 - 5.90 - 0.16 | 0.70 - 5.90 - 0.12 | 0.43 - 5.90 - 0.07 | 0.28 - 5.90 - 0.05 |
| AIFS | 4.54 - 7.05 - 0.64 | 4.70 - 7.05 - 0.67 | 4.75 - 7.05 - 0.67 | 4.63 - 7.05 - 0.66 | 4.55 - 7.05 - 0.65 | 4.62 - 7.05 - 0.66 |
| GraphCast | 1.60 - 7.05 - 0.23 | 1.47 - 7.05 - 0.21 | 1.39 - 7.05 - 0.20 | 1.41 - 7.05 - 0.20 | 1.27 - 7.05 - 0.18 | 1.08 - 7.05 - 0.15 |
| Aurora | 1.43 - 7.05 - 0.20 | 3.15 - 7.05 - 0.45 | 3.87 - 7.05 - 0.55 | 4.30 - 7.05 - 0.61 | 4.71 - 7.05 - 0.67 | 5.28 - 7.05 - 0.75 |

***Supplementary Table 5: Global cold extremes.*** *Comparison of average predicted and true extent of cold extremes. Each cell shows area of extreme cold forecast (F, unit: 1e7 km$^2$), area in the "truth" reference (T, unit: 1e7 km$^2$), and their ratio (F/T).*

| Model | 10 days | 11 days | 12 days | 13 days | 14 days | 15 days |
|---|---|---|---|---|---|---|
| Persistence | 4.35 - 2.36 - 1.84 | 4.79 - 2.36 - 2.03 | 5.28 - 2.36 - 2.23 | 5.83 - 2.36 - 2.47 | 6.41 - 2.36 - 2.71 | 6.97 - 2.36 - 2.95 |
| ECMWF-IFS | 4.62 - 4.26 - 1.09 | 4.62 - 4.26 - 1.09 | 4.67 - 4.26 - 1.10 | 4.64 - 4.26 - 1.09 | 4.47 - 4.26 - 1.05 | 4.54 - 4.26 - 1.07 |
| CMA | 5.08 - 4.02 - 1.26 | 5.04 - 4.02 - 1.25 | 4.93 - 4.02 - 1.23 | 4.87 - 4.02 - 1.21 | 4.90 - 4.02 - 1.22 | 5.11 - 4.02 - 1.27 |
| NCEP | 5.54 - 7.30 - 0.76 | 5.63 - 7.30 - 0.77 | 5.65 - 7.30 - 0.77 | 5.73 - 7.30 - 0.78 | 5.75 - 7.30 - 0.79 | 5.79 - 7.30 - 0.79 |
| Pangu-Weather | 2.37 - 2.36 - 1.00 | 2.46 - 2.36 - 1.04 | 2.56 - 2.36 - 1.08 | 2.80 - 2.36 - 1.18 | 2.97 - 2.36 - 1.26 | 2.97 - 2.36 - 1.26 |
| FuXi | 0.64 - 2.36 - 0.27 | 0.50 - 2.36 - 0.21 | 0.42 - 2.36 - 0.18 | 0.40 - 2.36 - 0.17 | 0.41 - 2.36 - 0.17 | 0.45 - 2.36 - 0.19 |
| ArchesWeather | 0.97 - 1.95 - 0.50 | 0.73 - 1.95 - 0.38 | 0.51 - 1.95 - 0.26 | 0.35 - 1.95 - 0.18 | 0.25 - 1.95 - 0.13 | 0.22 - 1.95 - 0.11 |
| AIFS | 1.36 - 2.36 - 0.58 | 1.31 - 2.36 - 0.55 | 1.28 - 2.36 - 0.54 | 1.34 - 2.36 - 0.57 | 1.46 - 2.36 - 0.62 | 1.56 - 2.36 - 0.66 |
| GraphCast | 1.91 - 2.36 - 0.81 | 2.17 - 2.36 - 0.92 | 2.58 - 2.36 - 1.09 | 2.90 - 2.36 - 1.23 | 3.40 - 2.36 - 1.44 | 3.83 - 2.36 - 1.62 |
| Aurora | 1.48 - 2.36 - 0.63 | 1.63 - 2.36 - 0.69 | 1.65 - 2.36 - 0.70 | 1.77 - 2.36 - 0.75 | 1.88 - 2.36 - 0.80 | 2.01 - 2.36 - 0.85 |

Supplementary Figure 5 visualizes Supplementary Table 3. On average, the total extreme heat area on Earth is quite similar for ERA5 and IFS-fc0, and higher for CMA-fc0 and NCEP-fc0. Dynamical systems show that the extent of extreme heat in their forecasts does not depend much on lead time in the 10- to 15-day range, with NCEP forecasts containing millions of square kilometres of extremes in excess relative to its reference. CMA forecasts, too, contain excess extreme heat, but to a lesser degree. ECMWF-IFS extended lead forecasts contain less extremes compared to their reference. Emulators show different results. Here, all areas are too small compared to their references (1.5° ERA5 for ArchesWeather, 0.25° ERA5 for all others—Supplementary Table 2). AIFS, Pangu-Weather and GraphCast extreme heat area appears not to depend on lead time, though there's large area differences between the models. ArchesWeather forecasts increasingly less extremes with increased lead time, and it's extreme heat forecast area is lowest of all models. FuXi shows a marked decrease from lead day 10 to 11, in line with earlier discussion (main body) on the consequences of its three-stage implementation.

Aurora shows a rapid increase of extreme heat area in time, though this study has no data for lead times beyond 15 days.

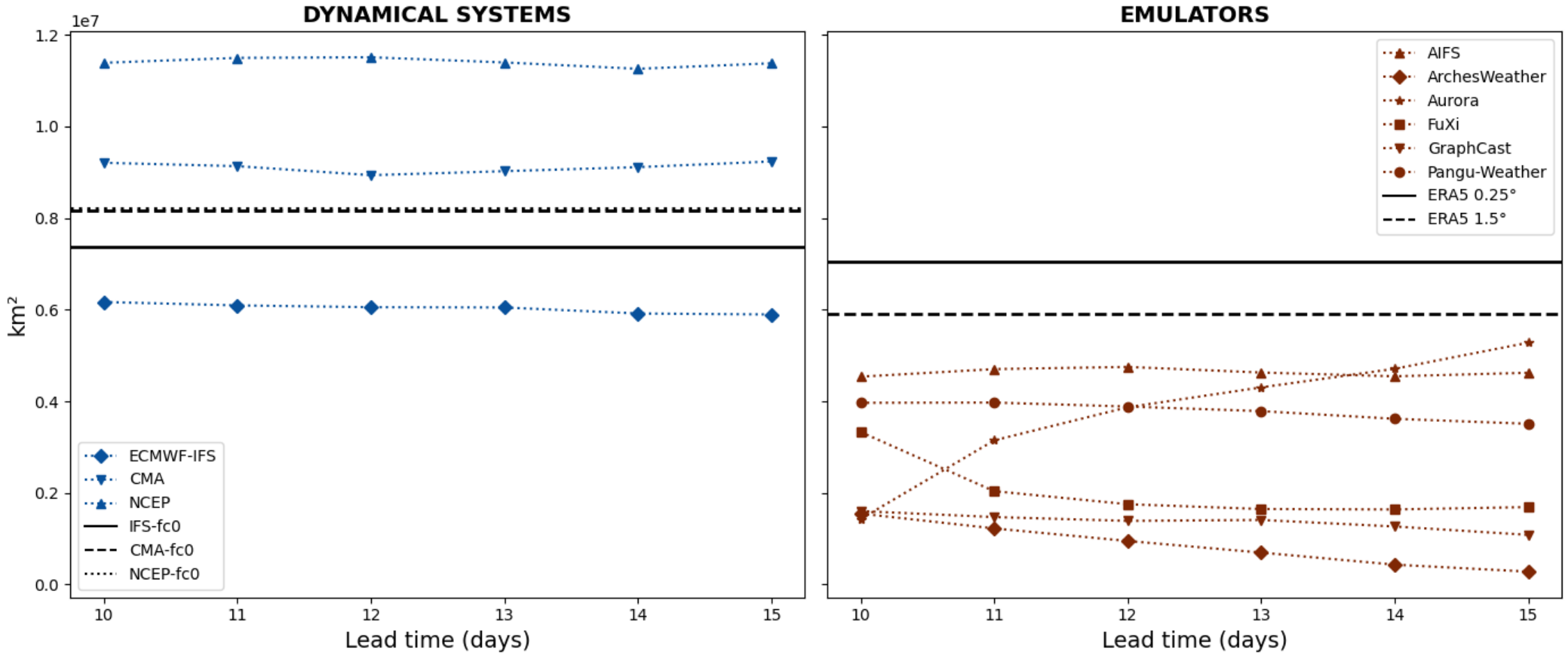


***Supplementary Figure 5: Lead time and extreme heat area forecasts.*** *The average total extreme heat area on Earth is shown for dynamical systems (left, blue), emulators (right, orange), and their references (black), as a function of lead time. Each dynamical system has its fc0 reference, whereas emulators have ERA5 as a reference (1.5° for ArchesWeather, 0.25° for all others—Supplementary Table 2).*

# S7. FORECAST ANOMALY MAPS

To showcase the extent and magnitude of the extreme events studied here, as well as each model's forecast characteristics, the surface temperature anomaly (relative to the ERA5 climatological mean at the forecast resolution) at the peak areal extent of each event is shown in Supplementary Figures 6-11 for the references together with the corresponding 14-day lead forecast from each of the studied models. The global map is cut to the bounding box of each extent (Supplementary Table 2).

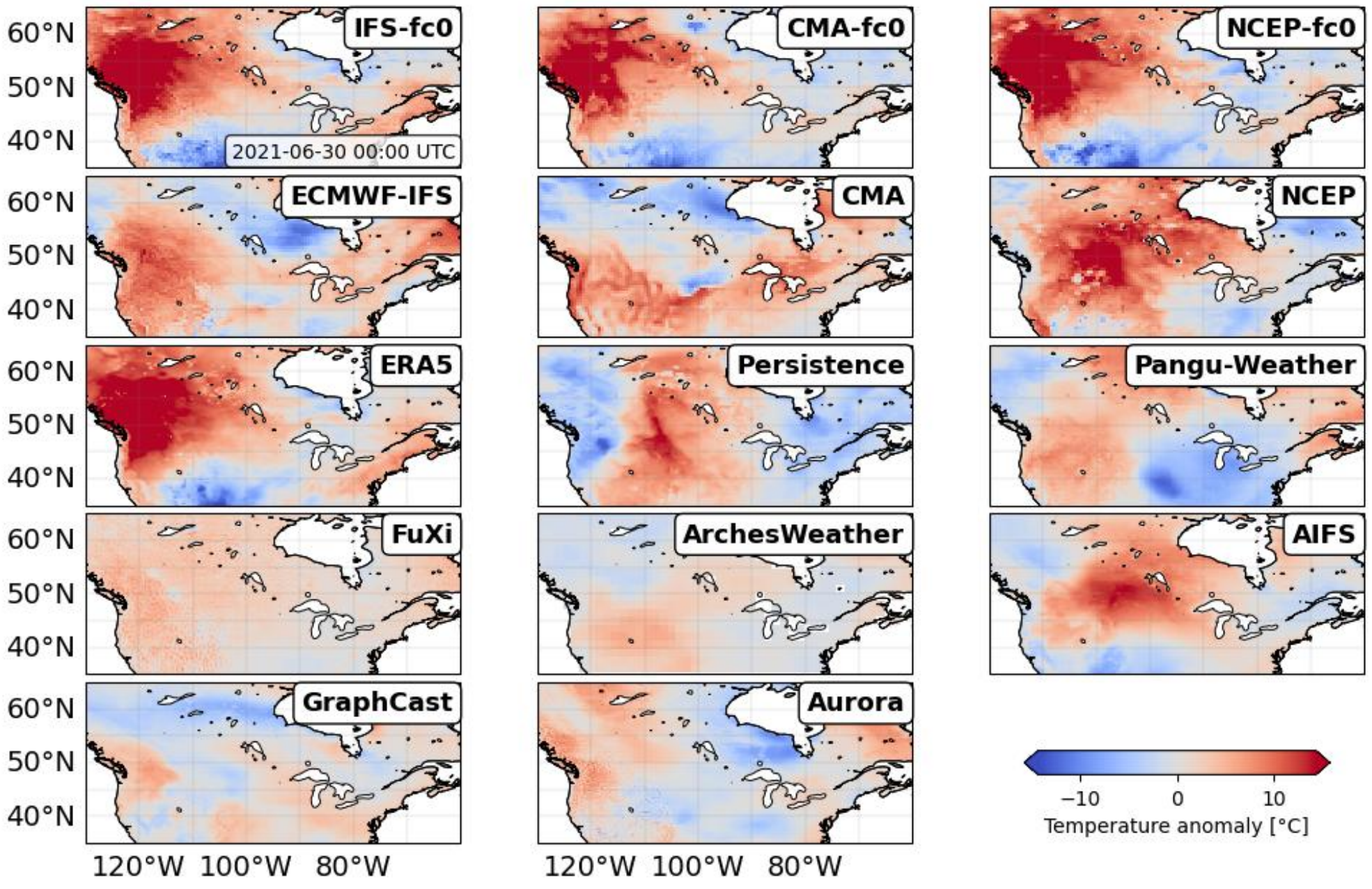


***Supplementary Figure 6: Peak extent of the 2021 Pacific Heat Dome.*** *The Supplementary Figure shows the reference (IFS-fc0, CMA-fc0, NCEP-fc0 and ERA5) and model 14-day forecast surface temperature anomalies at the time the event reached its largest extent.*

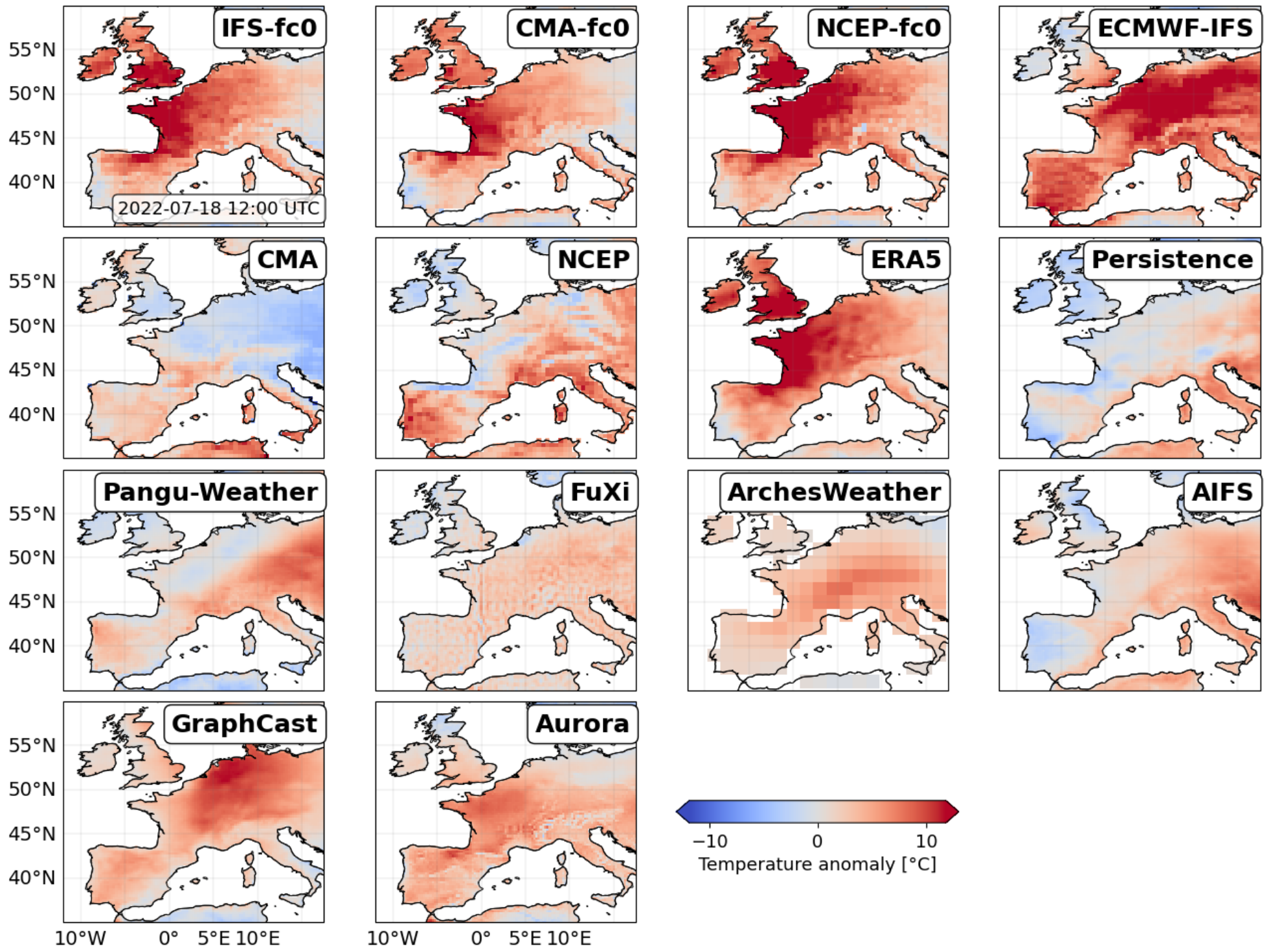


***Supplementary Figure 7: Peak extent of the 2022 Europe event.*** *see caption* ***Supplementary Figure 6***.

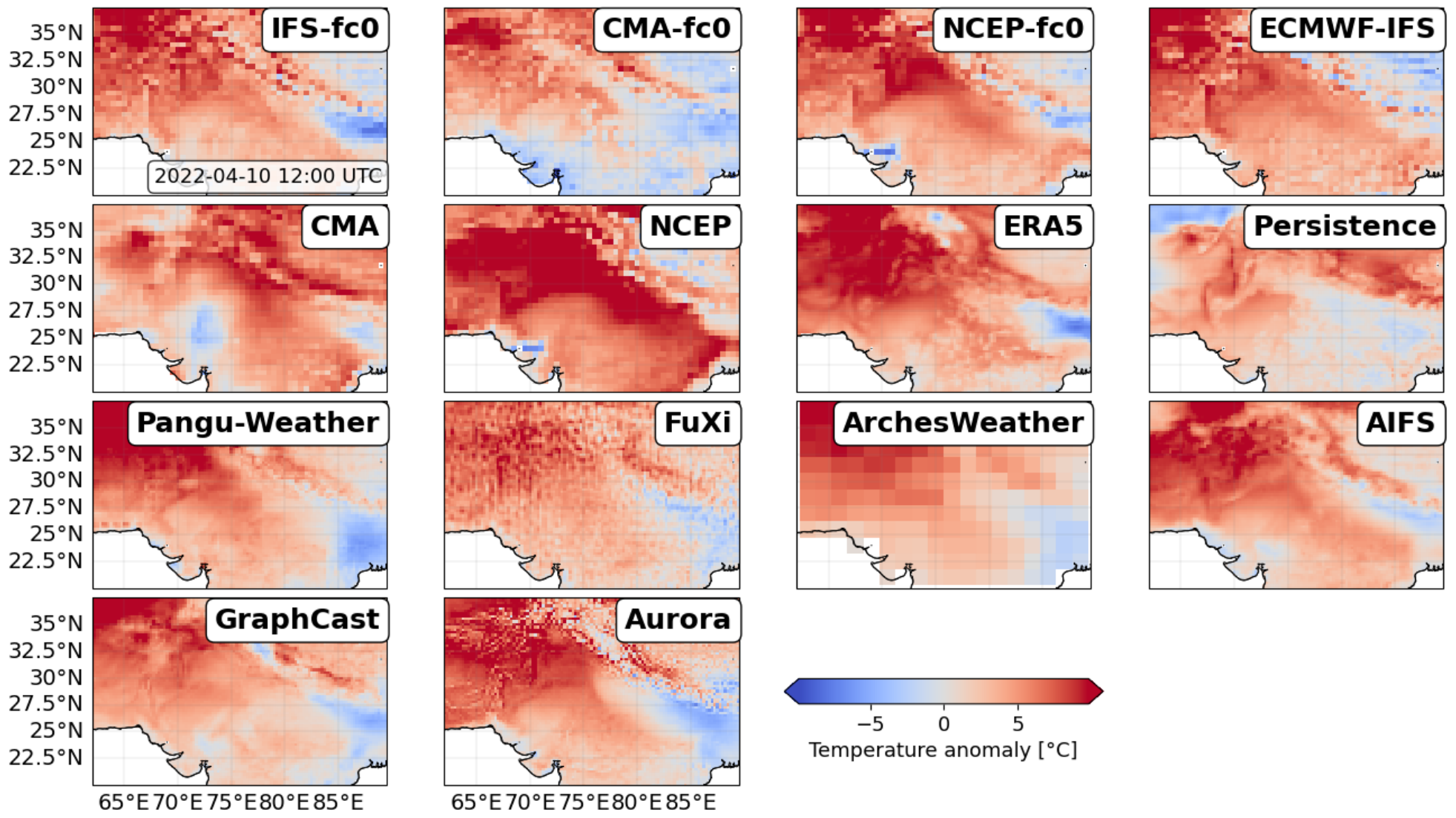


***Supplementary Figure 8: Peak extent of the 2022 India-Pakistan event.*** *See caption* ***Supplementary Figure 6***.

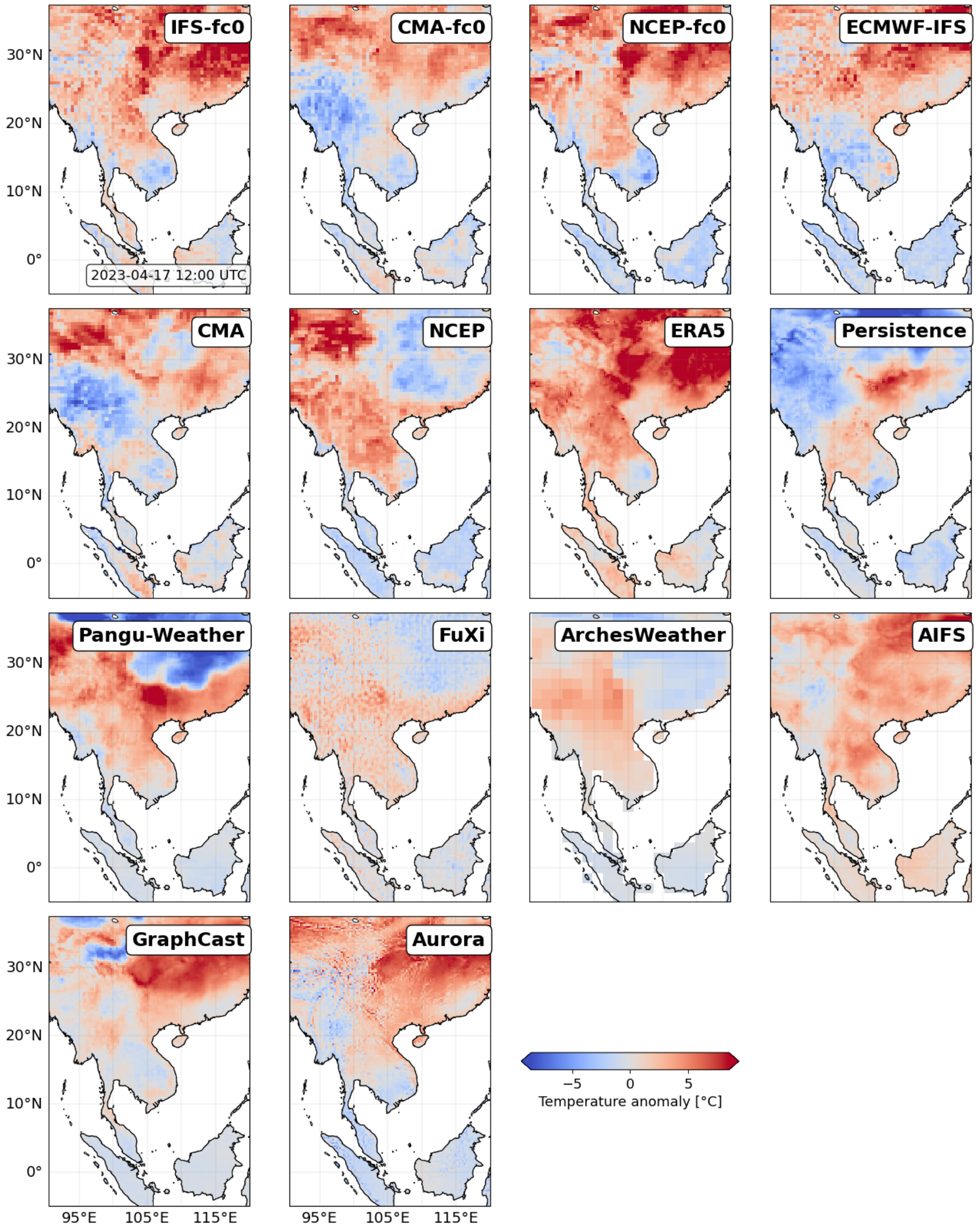


***Supplementary Figure 9: Peak area of the 2023 Southeast Asia event.*** *See caption* ***Supplementary Figure 6***.

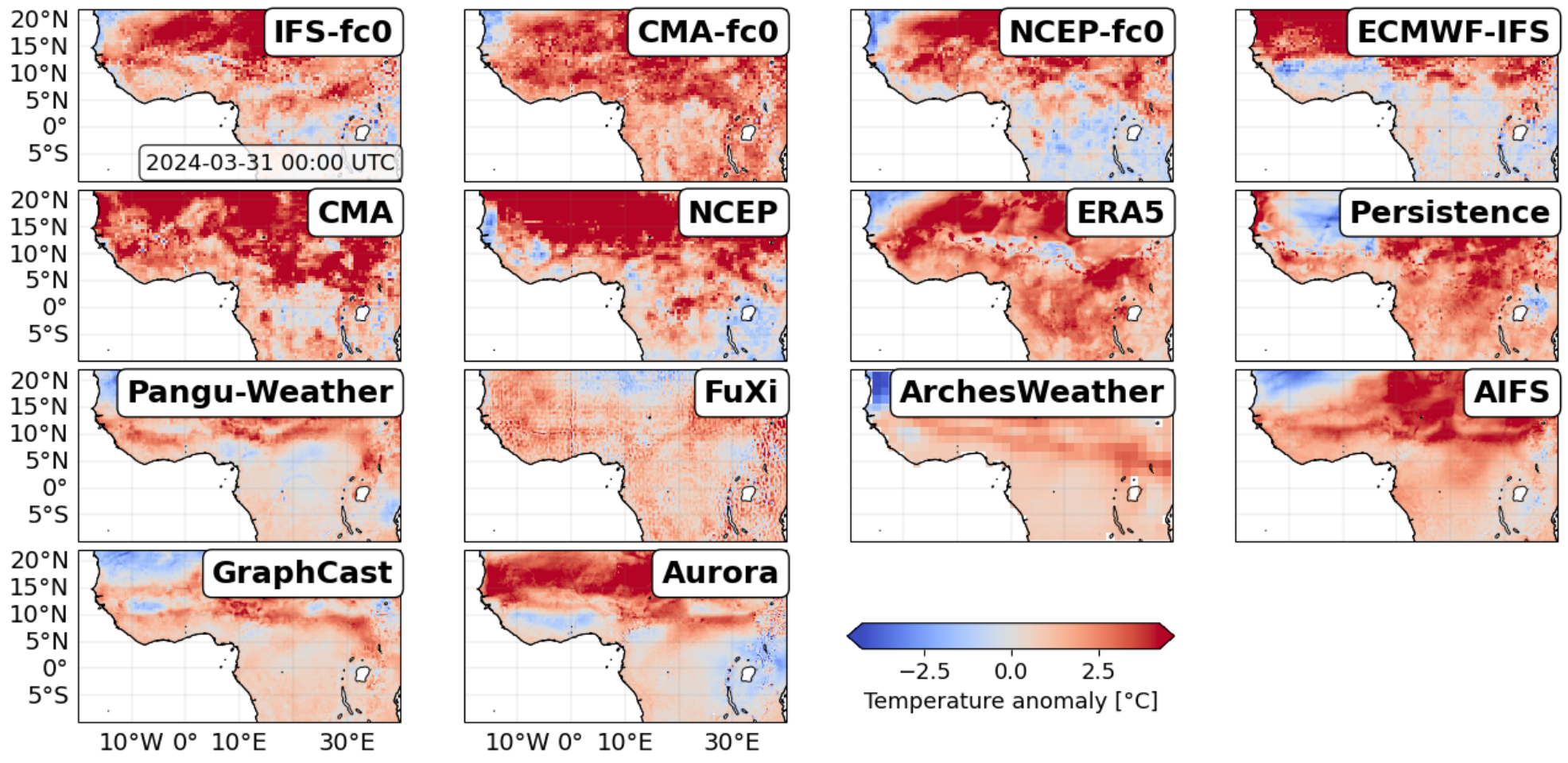


***Supplementary Figure 10: Peak area of the 2024 Africa event.*** *See caption* ***Supplementary Figure 6***.

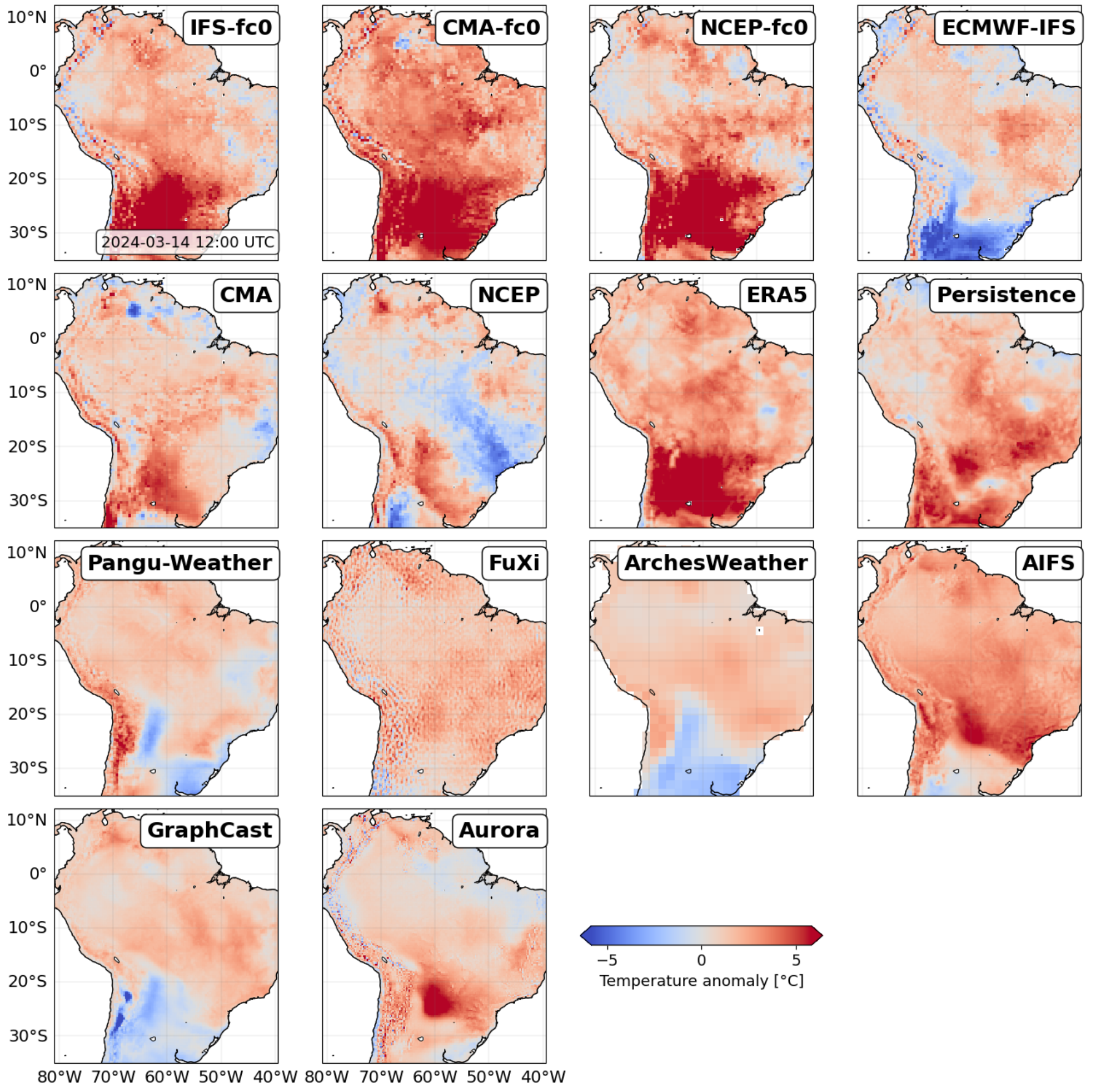


***Supplementary Figure 11: Peak area of the 2024 Latin America event.*** *See caption* ***Supplementary Figure 6***.

# S8. DERIVATION OF THE LINE ENERGY SPECTRUM

**Plain language summary**: a zonal energy spectrum tells us how much of the variability in a geophysical field (such as temperature anomalies) comes from features of different sizes. Large scales (long wavelengths, low wavenumbers) capture broad planetary patterns, while small scales (short wavelengths, high wavenumbers) capture finer details such as regional weather systems. By comparing spectra between forecasts and observations, we can evaluate whether a model produces the correct amount of variability at different spatial scales.

**Background: Discrete Fourier Transform.** The discrete Fourier transform (DFT) decomposes a signal along an axis into a sum of sinusoidal waves of different frequencies (or wavenumbers). Performing this over a latitude circle, the wavenumber $\mathrm{k}$ counts how many complete wave crests fit around the circle. With $\mathrm{k} = 0$ representing the spatial mean, larger $\mathrm{k}$ correspond to shorter wavelengths and finer spatial features.

The DFT outputs complex coefficients $Z[k]$, whose magnitudes squared $Z^2[k]$ represent the power (variance) at each wavenumber. By multiplying by the circle's circumference, we obtain a power spectral density in units that integrate back to the original signal's total variance (Parseval's theorem). Wavelength $\lambda$ is related to wavenumber $\mathrm{k}$ via

$$\lambda_{\mathrm{k,i}} = \frac{\mathrm{C}(\varphi_\mathrm{i})}{\mathrm{k}}$$

where $\mathrm{C}(\varphi_\mathrm{i}) = 2\pi\mathrm{R}\cos\varphi_\mathrm{i}$ is the longitudinal circumference of the globe at latitude $\varphi_\mathrm{i}$.

**The studied signal: temperature anomalies along the longitude band for a fixed latitude.** All spectra are computed from anomalies (deviation from climatology). Doing so isolates variability around the mean seasonal/climatological state, so the spectrum reflects fluctuations rather than the background level. At latitude $\varphi_\mathrm{i}$, longitude $\lambda_\mathrm{i}$, and time $t$, the anomaly is denoted as:

$$\mathrm{z}(\varphi, \lambda, \mathrm{t}).$$

At latitude $\varphi_\mathrm{i}$, data is available at $L_i$ longitudes ($L_i = 1440$ for a 0.25° grid, $L_i = 240$ for a 1.5° grid) with uniform longitude spacing $\Delta\lambda$ (radians):

$$\mathrm{z}_{\mathrm{t,i}}[\mathrm{l}] \equiv \mathrm{z}(\varphi_\mathrm{i}, \lambda_\mathrm{l}, \mathrm{t}), \quad \mathrm{l} = 0, \dots, \mathrm{L_i} - 1,$$

With $R$ as Earth's radius, circle circumference and physical grid spacing are

$$\mathrm{C}(\varphi_\mathrm{i}) = 2\pi\mathrm{R}\cos\varphi_\mathrm{i}\,, \qquad \Delta\mathrm{x_i} = \mathrm{R}\cos\varphi_\mathrm{i}\;\Delta\lambda = \frac{\mathrm{C}(\varphi_\mathrm{i})}{2\pi}\,\Delta\lambda,$$

**Discrete Fourier transform (DFT).** Sticking to the methodology of WeatherBench2 (Rasp et al., 2024), forward normalization is used. A real Fourier Frequency Transform is done by keeping non-negative wavenumbers only:

$$Z_{t,i}[k] = \frac{1}{L_i} \sum_{l=0}^{L_i - 1} z_{t,i}[l] \exp\left(-2\pi i \frac{k\,l}{L_i}\right), \qquad k = 0, \ldots, \left\lfloor \frac{L_i}{2} \right\rfloor.$$

The $1/L_i$ factor normalizes so that energy is handled consistently later.

**Zonal energy spectrum and Parseval's relation.** Define the (line) energy spectrum along the latitude circle by

$$S_{t,i}[k] = \begin{cases} C(\varphi_i)\left|Z_{t,i}[0]\right|^2, & k = 0, \\ 2C(\varphi_i)\left|Z_{t,i}[k]\right|^2, & 1 \le k \le \left\lfloor \frac{L_i}{2} \right\rfloor, \end{cases}$$

so that

$$\int_0^{C(\varphi_i)} |z(l)|^2 \, dl \;\approx\; \sum_k S_{t,i}[k].$$

*Explanation*. Multiplying by $C(\varphi_i)$ converts the spectrum to a line-integral scale (so units are $m \cdot \beta^2$ if $z$ has units $\beta$. The factor 2 for $k \ge 0$ accounts for the omitted negative wavenumbers, ensuring energy conservation (Parseval).

The methods section of the paper explains how this power spectrum is converted to a cosine-weighted average across latitudes in function of wavelength using logarithmically spaced wavelength bins $\{\Lambda_m\}_{m=1}^{M}$. Using the absolute logarithmic ratio of forecast ($\overline{S}_t^F(\Lambda_m)$) and true ($\overline{S}_t^T(\Lambda_m)$) spectral power as a measure of deviation, the spectral score is proposed as the average deviation across wavelength bins:

$$r_t(\Lambda_m) \;=\; \left|\log_{10}\left(\frac{\overline{S}_t^F(\Lambda_m)}{\overline{S}_t^T(\Lambda_m)}\right)\right|, \qquad \text{Spectral Score}_t \;=\; 1 - \frac{1}{M}\sum_{m=1}^{M} r_t(\Lambda_m).$$

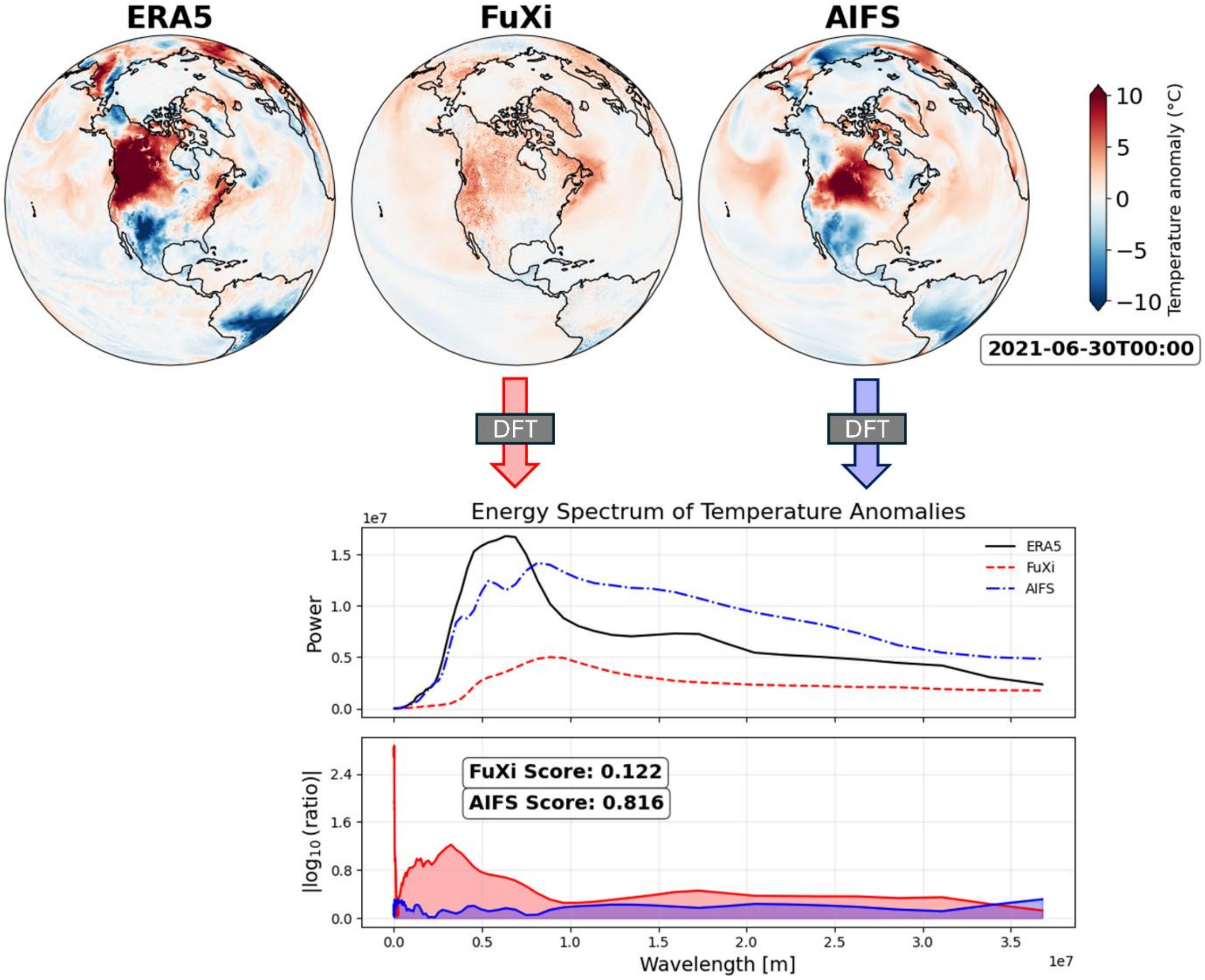


***Supplementary Figure 12: Spectral Scores.*** *Visualization of how forecast surface temperature anomalies translate to a spectral score. For both reference (ERA5) and forecast (model) anomalies, the signal is decomposed at each latitude by means of the discrete Fourier transform. The resulting spectral power is then expressed as a function of wavelength by mapping it onto logarithmically spaced wavelength bins, and averaged with cosine-weights across latitudes; the result is visualized in the middle panel. The logarithmic ratio of the average spectrum between ERA5 and forecast (bottom panel) is averaged across wavelength bins and serves as a measure of deviation between the two spectra. The final spectral score of a forecast is proposed as 1 minus this deviation.*

# S9. METRICS IN TABULAR FORMAT

In the main body of the paper, Figures 3 and 4 summarize—for both global and event-level scales—model performance using several metrics. Raw surface temperature forecasts are scored using RMSE and $R^2$; forecast anomalies relative to ERA5 climatological mean ($\mu$) are scored using ACC, temporal correlation and spectral score. Supplementary Tables 5 (global level) and 6 (event-level) show these metrics and their standard deviation across events for every lead time from 10- to 15-days.

***Supplementary Table 6: Global metric results.*** *Global RMSE, $R^2$, ACC, temporal correlation and the spectral score are reported for each lead time, together with their standard deviation across events**.** From left to right, top to bottom, each cell shows scores for lead days 10, 11, .., to 15.*

<table>
<tr><th>GLOBAL</th><th>RMSE</th><th>R²</th><th>ACC</th><th>CORR</th><th>SPECTRAL</th></tr>
<tr><td rowspan="2">Climatology</td><td>3.56 ± 0.34 | 3.56 ± 0.34 | 3.56 ± 0.34</td><td>0.22 ± 0.12 | 0.22 ± 0.12 | 0.22 ± 0.12</td><td>nan ± nan | nan ± nan | nan ± nan</td><td>nan ± nan | nan ± nan | nan ± nan</td><td>nan ± nan | nan ± nan | nan ± nan</td></tr>
<tr><td>3.56 ± 0.34 | 3.56 ± 0.34 | 3.56 ± 0.34</td><td>0.22 ± 0.12 | 0.22 ± 0.12 | 0.22 ± 0.12</td><td>nan ± nan | nan ± nan | nan ± nan</td><td>nan ± nan | nan ± nan | nan ± nan</td><td>nan ± nan | nan ± nan | nan ± nan</td></tr>
<tr><td rowspan="2">Persistence</td><td>4.92 ± 0.55 | 5.02 ± 0.57 | 5.14 ± 0.56</td><td>-0.58 ± 0.30 | -0.66 ± 0.32 | -0.76 ± 0.33</td><td>0.06 ± 0.04 | 0.03 ± 0.03 | 0.00 ± 0.02</td><td>-0.01 ± 0.02 | -0.00 ± 0.03 | -0.01 ± 0.03</td><td>0.92 ± 0.01 | 0.92 ± 0.01 | 0.91 ± 0.01</td></tr>
<tr><td>5.24 ± 0.57 | 5.30 ± 0.61 | 5.31 ± 0.63</td><td>-0.84 ± 0.36 | -0.89 ± 0.39 | -0.92 ± 0.42</td><td>-0.02 ± 0.03 | -0.02 ± 0.04 | -0.01 ± 0.04</td><td>-0.02 ± 0.02 | -0.01 ± 0.03 | 0.01 ± 0.03</td><td>0.91 ± 0.01 | 0.91 ± 0.01 | 0.91 ± 0.01</td></tr>
<tr><td rowspan="2">ECMWF-IFS</td><td>3.83 ± 0.28 | 4.07 ± 0.33 | 4.23 ± 0.34</td><td>-0.40 ± 0.15 | -0.55 ± 0.14 | -0.65 ± 0.14</td><td>0.42 ± 0.04 | 0.35 ± 0.04 | 0.30 ± 0.04</td><td>0.26 ± 0.05 | 0.20 ± 0.05 | 0.17 ± 0.05</td><td>0.90 ± 0.01 | 0.90 ± 0.01 | 0.90 ± 0.01</td></tr>
<tr><td>4.36 ± 0.34 | 4.47 ± 0.37 | 4.55 ± 0.39</td><td>-0.75 ± 0.13 | -0.83 ± 0.11 | -0.91 ± 0.11</td><td>0.26 ± 0.04 | 0.22 ± 0.04 | 0.20 ± 0.04</td><td>0.13 ± 0.04 | 0.10 ± 0.05 | 0.08 ± 0.04</td><td>0.90 ± 0.01 | 0.89 ± 0.01 | 0.89 ± 0.01</td></tr>
<tr><td rowspan="2">CMA</td><td>3.95 ± 0.29 | 4.11 ± 0.32 | 4.21 ± 0.34</td><td>-1.09 ± 0.21 | -1.23 ± 0.21 | -1.34 ± 0.22</td><td>0.39 ± 0.04 | 0.34 ± 0.05 | 0.31 ± 0.05</td><td>0.26 ± 0.02 | 0.22 ± 0.03 | 0.19 ± 0.04</td><td>0.88 ± 0.02 | 0.88 ± 0.02 | 0.88 ± 0.02</td></tr>
<tr><td>4.28 ± 0.33 | 4.39 ± 0.34 | 4.47 ± 0.35</td><td>-1.40 ± 0.22 | -1.51 ± 0.25 | -1.61 ± 0.29</td><td>0.29 ± 0.05 | 0.25 ± 0.04 | 0.24 ± 0.04</td><td>0.17 ± 0.03 | 0.15 ± 0.03 | 0.14 ± 0.02</td><td>0.88 ± 0.02 | 0.87 ± 0.02 | 0.87 ± 0.02</td></tr>
<tr><td rowspan="2">NCEP</td><td>4.15 ± 0.18 | 4.35 ± 0.22 | 4.51 ± 0.25</td><td>-0.07 ± 0.05 | -0.17 ± 0.07 | -0.26 ± 0.09</td><td>0.43 ± 0.04 | 0.38 ± 0.04 | 0.34 ± 0.04</td><td>0.32 ± 0.02 | 0.27 ± 0.03 | 0.23 ± 0.03</td><td>0.90 ± 0.01 | 0.90 ± 0.01 | 0.90 ± 0.01</td></tr>
<tr><td>4.62 ± 0.27 | 4.70 ± 0.27 | 4.77 ± 0.27</td><td>-0.32 ± 0.11 | -0.37 ± 0.10 | -0.41 ± 0.11</td><td>0.31 ± 0.04 | 0.29 ± 0.04 | 0.27 ± 0.03</td><td>0.21 ± 0.03 | 0.19 ± 0.02 | 0.17 ± 0.02</td><td>0.90 ± 0.01 | 0.90 ± 0.01 | 0.90 ± 0.01</td></tr>
<tr><td rowspan="2">Pangu-Weather</td><td>3.78 ± 0.30 | 4.04 ± 0.34 | 4.23 ± 0.37</td><td>0.07 ± 0.09 | -0.05 ± 0.11 | -0.14 ± 0.13</td><td>0.41 ± 0.06 | 0.34 ± 0.07 | 0.28 ± 0.07</td><td>0.27 ± 0.03 | 0.21 ± 0.03 | 0.17 ± 0.04</td><td>0.71 ± 0.01 | 0.71 ± 0.01 | 0.71 ± 0.01</td></tr>
<tr><td>4.39 ± 0.40 | 4.51 ± 0.39 | 4.59 ± 0.39</td><td>-0.22 ± 0.15 | -0.28 ± 0.15 | -0.31 ± 0.15</td><td>0.23 ± 0.07 | 0.20 ± 0.07 | 0.17 ± 0.07</td><td>0.13 ± 0.04 | 0.10 ± 0.03 | 0.07 ± 0.03</td><td>0.71 ± 0.01 | 0.71 ± 0.01 | 0.71 ± 0.01</td></tr>
<tr><td rowspan="2">FuXi</td><td>3.09 ± 0.25 | 3.14 ± 0.25 | 3.24 ± 0.26</td><td>0.39 ± 0.05 | 0.38 ± 0.05 | 0.34 ± 0.06</td><td>0.53 ± 0.04 | 0.48 ± 0.05 | 0.43 ± 0.06</td><td>0.37 ± 0.03 | 0.31 ± 0.03 | 0.25 ± 0.03</td><td>0.34 ± 0.02 | 0.23 ± 0.02 | 0.18 ± 0.02</td></tr>
<tr><td>3.32 ± 0.26 | 3.38 ± 0.27 | 3.45 ± 0.27</td><td>0.31 ± 0.06 | 0.28 ± 0.06 | 0.26 ± 0.07</td><td>0.39 ± 0.07 | 0.35 ± 0.08 | 0.31 ± 0.07</td><td>0.21 ± 0.03 | 0.17 ± 0.03 | 0.14 ± 0.03</td><td>0.14 ± 0.03 | 0.12 ± 0.04 | 0.09 ± 0.04</td></tr>
<tr><td rowspan="2">ArchesWeather</td><td>3.09 ± 0.29 | 3.23 ± 0.33 | 3.35 ± 0.35</td><td>0.33 ± 0.08 | 0.27 ± 0.09 | 0.22 ± 0.10</td><td>0.51 ± 0.03 | 0.45 ± 0.04 | 0.39 ± 0.04</td><td>0.37 ± 0.02 | 0.30 ± 0.03 | 0.25 ± 0.03</td><td>0.44 ± 0.03 | 0.41 ± 0.03 | 0.39 ± 0.03</td></tr>
<tr><td>3.45 ± 0.34 | 3.52 ± 0.35 | 3.58 ± 0.35</td><td>0.18 ± 0.11 | 0.16 ± 0.11 | 0.13 ± 0.12</td><td>0.33 ± 0.06 | 0.29 ± 0.06 | 0.25 ± 0.06</td><td>0.20 ± 0.04 | 0.15 ± 0.04 | 0.12 ± 0.03</td><td>0.36 ± 0.03 | 0.34 ± 0.03 | 0.32 ± 0.03</td></tr>
<tr><td rowspan="2">AIFS</td><td>3.48 ± 0.32 | 3.70 ± 0.37 | 3.89 ± 0.40</td><td>0.20 ± 0.09 | 0.10 ± 0.12 | -0.00 ± 0.14</td><td>0.46 ± 0.04 | 0.40 ± 0.05 | 0.34 ± 0.05</td><td>0.34 ± 0.03 | 0.27 ± 0.03 | 0.22 ± 0.03</td><td>0.81 ± 0.01 | 0.81 ± 0.01 | 0.81 ± 0.01</td></tr>
<tr><td>4.06 ± 0.42 | 4.20 ± 0.42 | 4.32 ± 0.42</td><td>-0.09 ± 0.15 | -0.16 ± 0.16 | -0.21 ± 0.16</td><td>0.28 ± 0.05 | 0.24 ± 0.05 | 0.20 ± 0.05</td><td>0.18 ± 0.03 | 0.15 ± 0.02 | 0.12 ± 0.02</td><td>0.81 ± 0.01 | 0.81 ± 0.00 | 0.81 ± 0.00</td></tr>
<tr><td rowspan="2">GraphCast</td><td>3.64 ± 0.32 | 3.88 ± 0.40 | 4.08 ± 0.41</td><td>0.10 ± 0.11 | -0.03 ± 0.15 | -0.14 ± 0.16</td><td>0.40 ± 0.04 | 0.33 ± 0.04 | 0.27 ± 0.05</td><td>0.30 ± 0.02 | 0.24 ± 0.02 | 0.19 ± 0.02</td><td>0.76 ± 0.01 | 0.76 ± 0.01 | 0.75 ± 0.01</td></tr>
<tr><td>4.25 ± 0.40 | 4.40 ± 0.42 | 4.54 ± 0.45</td><td>-0.24 ± 0.16 | -0.33 ± 0.17 | -0.43 ± 0.20</td><td>0.21 ± 0.04 | 0.17 ± 0.04 | 0.13 ± 0.03</td><td>0.16 ± 0.03 | 0.13 ± 0.03 | 0.09 ± 0.02</td><td>0.75 ± 0.01 | 0.75 ± 0.01 | 0.74 ± 0.01</td></tr>
<tr><td rowspan="2">Aurora</td><td>3.42 ± 0.30 | 3.72 ± 0.36 | 3.92 ± 0.41</td><td>0.23 ± 0.08 | 0.09 ± 0.10 | -0.02 ± 0.13</td><td>0.40 ± 0.03 | 0.35 ± 0.05 | 0.30 ± 0.06</td><td>0.26 ± 0.03 | 0.20 ± 0.04 | 0.17 ± 0.04</td><td>0.52 ± 0.01 | 0.59 ± 0.01 | 0.62 ± 0.01</td></tr>
<tr><td>4.09 ± 0.43 | 4.23 ± 0.43 | 4.36 ± 0.42</td><td>-0.11 ± 0.15 | -0.20 ± 0.15 | -0.27 ± 0.16</td><td>0.26 ± 0.06 | 0.24 ± 0.06 | 0.21 ± 0.05</td><td>0.14 ± 0.04 | 0.12 ± 0.03 | 0.10 ± 0.03</td><td>0.64 ± 0.00 | 0.65 ± 0.00 | 0.65 ± 0.01</td></tr>
</table>

***Supplementary Table 7: Event-level metric results.*** *Event-level RMSE, $R^2$, ACC, temporal correlation and the spectral score are reported for each lead time, together with their standard deviation across events. From left to right, top to bottom, each cell shows scores for lead days 10, 11, .., to 15.*

| EVENTS | RMSE | ACC | CORR |
|---|---|---|---|
| Climatology | 7.01 ± 2.48 \| 6.88 ± 2.53 \| 6.88 ± 2.53 | nan ± nan \| nan ± nan \| nan ± nan | nan ± nan \| nan ± nan \| nan ± nan |
| | 6.88 ± 2.53 \| 6.88 ± 2.53 \| 6.88 ± 2.53 | nan ± nan \| nan ± nan \| nan ± nan | nan ± nan \| nan ± nan \| nan ± nan |
| Persistence | 6.01 ± 3.22 \| 6.14 ± 3.60 \| 6.35 ± 3.72 | 0.39 ± 0.39 \| 0.42 ± 0.40 \| 0.35 ± 0.44 | 0.16 ± 0.09 \| 0.13 ± 0.14 \| 0.13 ± 0.20 |
| | 6.36 ± 3.63 \| 6.33 ± 3.52 \| 6.19 ± 3.30 | 0.31 ± 0.46 \| 0.29 ± 0.45 \| 0.27 ± 0.45 | 0.11 ± 0.16 \| 0.10 ± 0.13 \| 0.20 ± 0.09 |
| ECMWF-IFS | 4.37 ± 1.63 \| 4.57 ± 1.87 \| 4.79 ± 1.96 | 0.74 ± 0.14 \| 0.70 ± 0.15 \| 0.69 ± 0.15 | 0.15 ± 0.06 \| 0.14 ± 0.08 \| 0.12 ± 0.06 |
| | 4.99 ± 2.21 \| 5.26 ± 2.25 \| 5.41 ± 2.25 | 0.67 ± 0.14 \| 0.60 ± 0.16 \| 0.56 ± 0.14 | 0.13 ± 0.07 \| 0.06 ± 0.13 \| 0.12 ± 0.13 |
| CMA | 4.53 ± 3.28 \| 4.46 ± 3.31 \| 4.36 ± 2.95 | 0.69 ± 0.24 \| 0.64 ± 0.28 \| 0.67 ± 0.21 | 0.11 ± 0.16 \| 0.13 ± 0.13 \| 0.13 ± 0.13 |
| | 4.32 ± 2.80 \| 4.39 ± 2.64 \| 4.48 ± 2.57 | 0.68 ± 0.21 \| 0.63 ± 0.25 \| 0.61 ± 0.25 | 0.12 ± 0.15 \| 0.11 ± 0.13 \| 0.05 ± 0.11 |
| NCEP | 4.20 ± 2.51 \| 4.36 ± 2.88 \| 4.67 ± 3.01 | 0.79 ± 0.15 \| 0.78 ± 0.19 \| 0.72 ± 0.22 | 0.32 ± 0.16 \| 0.35 ± 0.18 \| 0.31 ± 0.16 |
| | 4.82 ± 2.99 \| 4.89 ± 2.83 \| 4.96 ± 2.69 | 0.71 ± 0.23 \| 0.70 ± 0.20 \| 0.69 ± 0.20 | 0.31 ± 0.13 \| 0.29 ± 0.12 \| 0.28 ± 0.10 |
| Pangu-Weather | 4.29 ± 1.99 \| 4.67 ± 2.28 \| 5.03 ± 2.25 | 0.72 ± 0.13 \| 0.65 ± 0.19 \| 0.59 ± 0.20 | 0.01 ± 0.24 \| -0.03 ± 0.25 \| -0.03 ± 0.23 |
| | 5.18 ± 2.24 \| 5.34 ± 2.10 \| 5.62 ± 2.13 | 0.50 ± 0.21 \| 0.45 ± 0.24 \| 0.43 ± 0.26 | -0.00 ± 0.21 \| -0.02 ± 0.21 \| -0.07 ± 0.14 |
| FuXi | 3.57 ± 1.58 \| 3.96 ± 1.87 \| 4.29 ± 1.96 | 0.86 ± 0.07 \| 0.85 ± 0.10 \| 0.83 ± 0.13 | 0.31 ± 0.17 \| 0.24 ± 0.23 \| 0.20 ± 0.23 |
| | 4.53 ± 1.98 \| 4.73 ± 1.99 \| 4.96 ± 2.05 | 0.79 ± 0.16 \| 0.74 ± 0.21 \| 0.70 ± 0.24 | 0.16 ± 0.21 \| 0.17 ± 0.19 \| 0.15 ± 0.16 |
| ArchesWeather | 3.83 ± 1.94 \| 4.27 ± 2.10 \| 4.79 ± 2.35 | 0.88 ± 0.10 \| 0.80 ± 0.14 \| 0.73 ± 0.15 | 0.15 ± 0.16 \| 0.08 ± 0.15 \| 0.10 ± 0.15 |
| | 4.98 ± 2.59 \| 5.34 ± 2.67 \| 5.53 ± 2.49 | 0.70 ± 0.16 \| 0.59 ± 0.21 \| 0.54 ± 0.21 | 0.14 ± 0.22 \| 0.10 ± 0.19 \| 0.06 ± 0.16 |
| AIFS | 3.52 ± 1.31 \| 3.65 ± 1.68 \| 3.83 ± 1.93 | 0.84 ± 0.12 \| 0.85 ± 0.14 \| 0.84 ± 0.14 | 0.29 ± 0.08 \| 0.27 ± 0.09 \| 0.20 ± 0.13 |
| | 4.09 ± 2.10 \| 4.45 ± 2.42 \| 4.70 ± 2.69 | 0.82 ± 0.12 \| 0.76 ± 0.16 \| 0.71 ± 0.18 | 0.15 ± 0.17 \| 0.13 ± 0.21 \| 0.11 ± 0.19 |
| GraphCast | 4.18 ± 1.65 \| 4.41 ± 1.90 \| 4.74 ± 1.92 | 0.76 ± 0.10 \| 0.70 ± 0.13 \| 0.64 ± 0.17 | 0.22 ± 0.15 \| 0.19 ± 0.17 \| 0.11 ± 0.19 |
| | 4.91 ± 1.79 \| 5.18 ± 1.73 \| 5.48 ± 1.87 | 0.58 ± 0.17 \| 0.50 ± 0.21 \| 0.47 ± 0.23 | 0.09 ± 0.16 \| 0.09 ± 0.16 \| 0.06 ± 0.12 |
| Aurora | 4.04 ± 0.92 \| 3.79 ± 0.86 \| 3.99 ± 1.00 | 0.78 ± 0.17 \| 0.75 ± 0.21 \| 0.69 ± 0.26 | 0.27 ± 0.16 \| 0.16 ± 0.21 \| 0.10 ± 0.25 |
| | 4.23 ± 1.06 \| 4.61 ± 1.49 \| 4.88 ± 2.10 | 0.66 ± 0.25 \| 0.60 ± 0.23 \| 0.55 ± 0.28 | 0.08 ± 0.16 \| 0.11 ± 0.14 \| 0.15 ± 0.15 |

Categorical evaluation is done by casting forecasts to a binary format, becoming 1 when the forecast temperature exceeds the extreme threshold (indicator $\mathbb{1}(T_{forecast} \geq \mu + 2\sigma)$, with $\sigma$ the ERA5 climatological standard deviation), and zero otherwise. The categorical forecast is then evaluated against a similar transformation of the reference ($\mathbb{1}(T_{reference} \geq \mu + 2\sigma)$), using accuracy (not in figures), recall, precision and Equitable Threat Score (ETS). For each lead time, these and their standard deviation across events are reported in Supplementary Figures 7 (global) and 8 (event-scale).

***Supplementary Table 8: Global classification results***. *Global accuracy, recall, precision and the ETS are reported for each lead time, together with their standard deviation across events. From left to right, top to bottom, each cell shows scores for lead days 10, 11, ..., to 15.*

| GLOBAL | ACCURACY | RECALL | PRECISION | ETS |
|---|---|---|---|---|
| Climatology | 95.2 ± 2.27 \| 95.2 ± 2.27 \| 95.2 ± 2.27 | 0 ± 0 \| 0 ± 0 \| 0 ± 0 | nan ± nan \| nan ± nan \| nan ± nan | 0 ± 0 \| 0 ± 0 \| 0 ± 0 |
| | 95.2 ± 2.27 \| 95.2 ± 2.27 \| 95.2 ± 2.27 | 0 ± 0 \| 0 ± 0 \| 0 ± 0 | nan ± nan \| nan ± nan \| nan ± nan | 0 ± 0 \| 0 ± 0 \| 0 ± 0 |
| Persistence | 92.7 ± 2.56 \| 92.5 ± 2.75 \| 92.5 ± 2.75 | 11.5 ± 6.1 \| 11.3 ± 6.52 \| 11 ± 6.14 | 14 ± 9 \| 13.8 ± 8.9 \| 13 ± 8.85 | 4.33 ± 3.54 \| 4.67 ± 3.3 \| 4.17 ± 3.24 |
| | 92.2 ± 2.79 \| 92 ± 3.06 \| 92 ± 3.06 | 10.3 ± 5.85 \| 10.7 ± 6.05 \| 10.7 ± 6.57 | 12.3 ± 8.03 \| 12 ± 8.12 \| 11.2 ± 7.4 | 3.83 ± 2.79 \| 3.67 ± 2.69 \| 3.33 ± 2.56 |
| ECMWF-IFS | 93.2 ± 1.67 \| 93.2 ± 1.67 \| 92.8 ± 1.57 | 23.3 ± 9.91 \| 21.2 ± 8.88 \| 19.5 ± 8.02 | 27.2 ± 6.41 \| 24.5 ± 6.55 \| 22.5 ± 5.38 | 11.7 ± 4.82 \| 10.5 ± 4.54 \| 9 ± 3.65 |
| | 92.5 ± 1.89 \| 92.5 ± 1.89 \| 92.3 ± 1.89 | 17.2 ± 7.24 \| 15.3 ± 6.18 \| 14 ± 6.06 | 20.2 ± 5.08 \| 18.7 ± 4.27 \| 17.3 ± 4.85 | 7.5 ± 3.3 \| 6.5 ± 2.57 \| 6 ± 2.65 |
| CMA | 91.5 ± 2.36 \| 91.5 ± 2.36 \| 91.3 ± 2.36 | 28.7 ± 6.16 \| 26.5 ± 6.24 \| 25.3 ± 5.99 | 24.5 ± 8.4 \| 23 ± 8.62 \| 22 ± 8.54 | 12.2 ± 3.39 \| 11 ± 3.46 \| 10 ± 3.37 |
| | 91.2 ± 2.19 \| 91 ± 2.08 \| 90.8 ± 2.19 | 24.8 ± 6.26 \| 24.5 ± 6.26 \| 23.8 ± 6.04 | 21.5 ± 8.18 \| 20.8 ± 8.35 \| 20.2 ± 8.41 | 10 ± 3.06 \| 9.83 ± 3.24 \| 9 ± 3.32 |
| NCEP | 91 ± 1.73 \| 90.3 ± 1.97 \| 90.2 ± 1.95 | 36.7 ± 2.49 \| 34.3 ± 2.29 \| 32 ± 2.58 | 26.5 ± 3.59 \| 24.2 ± 3.53 \| 22.5 ± 3.5 | 14.7 ± 1.8 \| 13.5 ± 1.5 \| 11.7 ± 1.49 |
| | 90 ± 1.73 \| 90 ± 1.73 \| 89.8 ± 1.86 | 29.5 ± 2.57 \| 28 ± 2.52 \| 26.7 ± 3.64 | 21.3 ± 3.3 \| 20.5 ± 3.3 \| 19.5 ± 3.45 | 11.2 ± 1.21 \| 10.2 ± 1.21 \| 9.17 ± 1.34 |
| Pangu-Weather | 93.7 ± 2.49 \| 93.5 ± 2.36 \| 93.5 ± 2.36 | 12.3 ± 2.43 \| 10 ± 1.63 \| 8.33 ± 1.7 | 22.3 ± 6.34 \| 18 ± 4.58 \| 14.5 ± 4.86 | 7 ± 1.73 \| 5.33 ± 0.471 \| 3.83 ± 0.687 |
| | 93.5 ± 2.36 \| 93.5 ± 2.36 \| 93.5 ± 2.36 | 7 ± 1.91 \| 6 ± 1.53 \| 5.33 ± 0.943 | 13 ± 4.2 \| 12 ± 4.04 \| 10.8 ± 3.98 | 3 ± 0.816 \| 2.33 ± 0.745 \| 2 ± 0.577 |
| FuXi | 94.2 ± 2.48 \| 94.8 ± 2.19 \| 94.8 ± 2.19 | 12.7 ± 6.34 \| 7 ± 4.86 \| 5.17 ± 4.37 | 28 ± 7.26 \| 23.8 ± 8.97 \| 21.3 ± 9.12 | 8 ± 3.51 \| 4.67 ± 3.09 \| 3.67 ± 2.98 |
| | 94.8 ± 2.19 \| 94.8 ± 2.19 \| 94.7 ± 2.49 | 4.5 ± 4.11 \| 4.33 ± 3.77 \| 4 ± 3.56 | 19.5 ± 9.55 \| 17.7 ± 9.05 \| 16 ± 9.29 | 3 ± 2.45 \| 2.83 ± 2.54 \| 2.5 ± 2.14 |
| ArchesWeather | 95.8 ± 2.19 \| 95.8 ± 2.19 \| 95.8 ± 2.19 | 8.83 ± 4.6 \| 6.33 ± 3.25 \| 4.5 ± 2.22 | 35.5 ± 9.84 \| 32 ± 9.02 \| 29.2 ± 11.1 | 6.67 ± 3.73 \| 4.67 ± 2.81 \| 3.33 ± 1.97 |
| | 95.8 ± 2.19 \| 95.8 ± 2.19 \| 96 ± 2.08 | 3.17 ± 1.77 \| 1.83 ± 0.898 \| 1.17 ± 0.687 | 27 ± 11.5 \| 29.3 ± 13.8 \| 25.8 ± 13.2 | 2.17 ± 1.34 \| 1.5 ± 1.12 \| 1 ± 0.816 |
| AIFS | 94 ± 2.38 \| 93.8 ± 2.19 \| 93.7 ± 2.36 | 17.7 ± 7.45 \| 16.2 ± 7.65 \| 15.2 ± 7.45 | 27.7 ± 9.72 \| 24.8 ± 8.93 \| 23.2 ± 8.63 | 10.5 ± 4.39 \| 9.17 ± 4.02 \| 8.33 ± 3.86 |
| | 93.7 ± 2.36 \| 93.7 ± 2.36 \| 93.5 ± 2.63 | 13.2 ± 6.84 \| 12 ± 7 \| 11.2 ± 7.03 | 20.7 ± 9.03 \| 18.8 ± 9.26 \| 17.5 ± 8.75 | 7 ± 3.51 \| 6.33 ± 3.9 \| 5.67 ± 3.64 |
| GraphCast | 94.8 ± 2.19 \| 94.8 ± 2.19 \| 94.7 ± 2.21 | 7 ± 1.53 \| 5.5 ± 0.957 \| 4.67 ± 0.471 | 29 ± 7.39 \| 25 ± 6.73 \| 23.3 ± 3.77 | 5.17 ± 1.21 \| 3.67 ± 0.943 \| 3.33 ± 0.471 |
| | 94.7 ± 2.21 \| 94.8 ± 2.19 \| 94.8 ± 2.19 | 4.17 ± 1.34 \| 3.17 ± 1.07 \| 2.5 ± 0.957 | 20.7 ± 6.05 \| 18.3 ± 5.82 \| 16.3 ± 5.71 | 2.83 ± 0.687 \| 2 ± 1 \| 1.67 ± 0.745 |
| Aurora | 94.8 ± 2.19 \| 94 ± 2.38 \| 93.7 ± 2.29 | 5.67 ± 2.56 \| 9.17 ± 4.02 \| 9.5 ± 4.31 | 24.3 ± 8.12 \| 19.2 ± 7.01 \| 16 ± 6.11 | 3.83 ± 1.95 \| 5 ± 2.45 \| 4.5 ± 2.22 |
| | 93.2 ± 2.27 \| 92.7 ± 2.13 \| 92.5 ± 2.22 | 9 ± 4.4 \| 9 ± 3.92 \| 9.67 ± 2.75 | 14 ± 5 \| 13.2 ± 4.45 \| 12 ± 3.92 | 3.67 ± 2.21 \| 3.5 ± 1.71 \| 3.5 ± 1.26 |

***Supplementary Table 9: Event-level classification results.*** *Event-level accuracy, recall, precision and the ETS are reported for each lead time, together with their standard deviation across events**.** From left to right, top to bottom, each cell shows scores for lead days 10, 11, .., to 15.*

| EVENTS | ACCURACY | RECALL | PRECISION | JACCARD |
|---|---|---|---|---|
| Climatology | 86.5 ± 8.54 \| 86.5 ± 8.54 \| 86.5 ± 8.54 | 0 ± 0 \| 0 ± 0 \| 0 ± 0 | nan ± nan \| nan ± nan \| nan ± nan | 0 ± 0 \| 0 ± 0 \| 0 ± 0 |
| | 86.5 ± 8.54 \| 86.5 ± 8.54 \| 86.5 ± 8.54 | 0 ± 0 \| 0 ± 0 \| 0 ± 0 | nan ± nan \| nan ± nan \| nan ± nan | 0 ± 0 \| 0 ± 0 \| 0 ± 0 |
| Persistence | 84 ± 8.56 \| 84.2 ± 8.27 \| 83.7 ± 8.48 | 14.2 ± 11.9 \| 13.8 ± 12.7 \| 11.7 ± 12.1 | 22.2 ± 13.8 \| 21.5 ± 13.6 \| 19.2 ± 13.9 | 8.67 ± 7.27 \| 8.17 ± 7.29 \| 6.83 ± 6.87 |
| | 83 ± 8.54 \| 82.8 ± 9.1 \| 82.8 ± 9.26 | 10.8 ± 11.2 \| 11 ± 11.9 \| 11.2 ± 13.2 | 19.7 ± 12.9 \| 18.7 ± 12.1 \| 18.3 ± 11.4 | 6.67 ± 6.07 \| 6.17 ± 5.81 \| 6 ± 5.69 |
| ECMWF-IFS | 84.5 ± 5.5 \| 83.8 ± 5.4 \| 83.7 ± 4.92 | 28.8 ± 10.7 \| 25 ± 10.1 \| 24.3 ± 9.53 | 35.7 ± 12.4 \| 33.2 ± 13.7 \| 30.5 ± 12.8 | 15.8 ± 7.86 \| 13.5 ± 7.37 \| 12.5 ± 6.42 |
| | 83.7 ± 4.75 \| 83.3 ± 4.71 \| 83.5 ± 4.68 | 22.7 ± 11.1 \| 21.2 ± 11 \| 19.3 ± 9.76 | 28.7 ± 13.2 \| 28 ± 11.9 \| 25.2 ± 12.4 | 11 ± 6.78 \| 10.5 ± 5.97 \| 9.67 ± 5.76 |
| CMA | 84.5 ± 9.23 \| 84.7 ± 9.69 \| 84.3 ± 9.64 | 35.7 ± 14 \| 31.5 ± 14.6 \| 28.5 ± 13.8 | 28 ± 14 \| 27.3 ± 14.8 \| 26 ± 14.7 | 15.7 ± 6.94 \| 14.2 ± 6.69 \| 12.5 ± 5.91 |
| | 84.5 ± 10.1 \| 84.5 ± 10.2 \| 84.7 ± 9.84 | 27.7 ± 12.7 \| 28 ± 10.6 \| 27.2 ± 10.1 | 27.3 ± 13.9 \| 27.3 ± 15.4 \| 27.8 ± 14.4 | 12.5 ± 5.28 \| 12.2 ± 6.54 \| 12.3 ± 5.09 |
| NCEP | 82.7 ± 4.07 \| 82.2 ± 4.45 \| 82.2 ± 4.37 | 44.7 ± 13.9 \| 40.3 ± 13.9 \| 37.2 ± 13.7 | 31.7 ± 8.12 \| 29.7 ± 8.96 \| 27.7 ± 9.21 | 20.7 ± 5.47 \| 18.5 ± 5.62 \| 17.2 ± 5.98 |
| | 82.7 ± 4.19 \| 82.8 ± 4.18 \| 82.5 ± 4.11 | 36 ± 12.3 \| 34.7 ± 12.1 \| 31.8 ± 11.1 | 27.7 ± 8.69 \| 28.5 ± 8.04 \| 27.7 ± 7.06 | 16.8 ± 5.11 \| 16.3 ± 4.53 \| 15 ± 4.43 |
| Pangu-Weather | 84.7 ± 7.43 \| 84.8 ± 7.45 \| 84.7 ± 7.65 | 15.2 ± 5.96 \| 11.5 ± 5.16 \| 10.2 ± 4.49 | 30.3 ± 12.2 \| 25.5 ± 13.1 \| 25 ± 12.4 | 10.2 ± 3.85 \| 8 ± 3.7 \| 7 ± 3.16 |
| | 84.8 ± 7.73 \| 84.8 ± 7.69 \| 85 ± 7.83 | 8.83 ± 6.39 \| 8.67 ± 6.8 \| 8 ± 6.08 | 22.8 ± 13.3 \| 24 ± 11.2 \| 23.5 ± 11 | 6.17 ± 4.06 \| 5.67 ± 4.07 \| 5.5 ± 3.4 |
| FuXi | 85.2 ± 7.65 \| 85.5 ± 8.18 \| 85.8 ± 8.31 | 18.7 ± 7.39 \| 10.3 ± 7.16 \| 8.17 ± 6.89 | 32.3 ± 9.81 \| 29.2 ± 10.7 \| 25.8 ± 10.8 | 12.8 ± 5.73 \| 8 ± 5.35 \| 5.83 ± 5.46 |
| | 85.8 ± 8.31 \| 85.7 ± 8.67 \| 85.7 ± 8.67 | 6.83 ± 6.34 \| 6.33 ± 6.24 \| 5.5 ± 5.85 | 24.3 ± 11.4 \| 22.8 ± 11.6 \| 22.3 ± 12.1 | 5.33 ± 4.92 \| 4.67 ± 4.71 \| 3.83 ± 4.37 |
| ArchesWeather | 87.3 ± 7.23 \| 87.7 ± 7.2 \| 87.7 ± 7.56 | 10.8 ± 6.91 \| 8.5 ± 6.1 \| 6 ± 4.65 | 41 ± 13.2 \| 37.3 ± 14.7 \| 31.2 ± 17 | 7.5 ± 5.28 \| 6.5 ± 4.99 \| 4.5 ± 3.82 |
| | 87.7 ± 7.95 \| 87.8 ± 7.95 \| 87.5 ± 8.18 | 4.33 ± 3.94 \| 3.67 ± 3.3 \| 2.17 ± 2.34 | 31.5 ± 21.4 \| 36 ± 21.5 \| 44.6 ± 25.8 | 3.33 ± 2.92 \| 3 ± 2.16 \| 1.67 ± 1.7 |
| AIFS | 85.3 ± 7.3 \| 85 ± 7.26 \| 84.2 ± 7.56 | 19.8 ± 9.56 \| 19.5 ± 7.07 \| 18.5 ± 6.95 | 32.7 ± 11.2 \| 32.8 ± 10 \| 30 ± 10.2 | 13.8 ± 7.65 \| 12.8 ± 5.7 \| 11.7 ± 5.5 |
| | 84 ± 7.33 \| 84.5 ± 7.8 \| 84.7 ± 7.8 | 16.5 ± 5.99 \| 14.8 ± 7.22 \| 14.2 ± 7.54 | 27.5 ± 9.98 \| 25.8 ± 8.78 \| 26.5 ± 9.54 | 10.5 ± 4.99 \| 9.83 ± 5.21 \| 9.33 ± 5.09 |
| GraphCast | 86.2 ± 8.43 \| 85.8 ± 8.29 \| 86 ± 8.31 | 10.5 ± 6.63 \| 8.5 ± 5.5 \| 7.33 ± 4.19 | 32.8 ± 12.7 \| 28.3 ± 12.6 \| 25.5 ± 10.1 | 7.83 ± 4.74 \| 6.33 ± 3.99 \| 5.67 ± 2.81 |
| | 86 ± 8.29 \| 86 ± 8.33 \| 85.8 ± 8.25 | 6.17 ± 4.45 \| 4.83 ± 3.29 \| 3.67 ± 2.92 | 24.5 ± 10.5 \| 23 ± 9.59 \| 21.5 ± 9.6 | 4.67 ± 3.5 \| 3.67 ± 2.56 \| 2.83 ± 2.41 |
| Aurora | 85.7 ± 7.97 \| 83.8 ± 6.79 \| 83 ± 6.51 | 12 ± 7.44 \| 15.8 ± 8.74 \| 16 ± 9.52 | 30.7 ± 14 \| 28.3 ± 15.1 \| 25.5 ± 16.1 | 8.5 ± 5.65 \| 10.2 ± 6.15 \| 9.5 ± 5.5 |
| | 82.3 ± 6.72 \| 82.5 ± 6.95 \| 82.8 ± 7.08 | 14.5 ± 9.07 \| 12.7 ± 9.21 \| 11.5 ± 9.22 | 24.3 ± 15.3 \| 21.7 ± 14.3 \| 21.2 ± 13.2 | 8.33 ± 4.99 \| 7 ± 4.55 \| 6.17 ± 3.85 |

Finally, reference temperatures are mapped to two additional thresholds corresponding to $\mathbb{1}(T_{reference} \geq \mu)$, and $\mathbb{1}(T_{reference} \geq \mu + \sigma)$, referred to as above-normal and above-hot temperatures, respectively. Reliability is tested, by computing the observed probability of an extreme heat forecast preceding above-normal, above-hot and extreme heat temperatures in the reference. The results and their standard deviation across events are reported in Supplementary Tables 9 (global) and 10 (event-level), for every lead time.

***Supplementary Table 10: Global reliability results.*** *Global observed probability of extreme heat forecasts preceding above-normal, above-hot and extreme heat temperatures is reported for each lead time, together with standard deviation across events. From left to right, top to bottom, each cell shows scores for lead days 10, 11, .., to 15.*

<table>
<tr><th>GLOBAL</th><th>ABOVE NORMAL</th><th>ABOVE HOT</th><th>ABOVE EXTREME</th></tr>
<tr><td rowspan="2">Persistence</td><td>70.8 ± 8.93 | 71.7 ± 8.86 | 71 ± 9.06</td><td>40.3 ± 12.3 | 40.7 ± 12.1 | 39 ± 12.1</td><td>14 ± 9 | 13.8 ± 8.9 | 13 ± 8.85</td></tr>
<tr><td>69.7 ± 9.59 | 68.3 ± 10.1 | 68 ± 9.06</td><td>37.5 ± 12.2 | 37.2 ± 12 | 36.3 ± 12.1</td><td>12.3 ± 8.03 | 12 ± 8.12 | 11.2 ± 7.4</td></tr>
<tr><td rowspan="2">ECMWF-IFS</td><td>85.7 ± 4.31 | 83.8 ± 5.55 | 81.8 ± 5.05</td><td>61.3 ± 8.73 | 57 ± 10.1 | 54.3 ± 8.58</td><td>27.2 ± 6.41 | 24.5 ± 6.55 | 22.5 ± 5.38</td></tr>
<tr><td>80.8 ± 4.74 | 78.7 ± 4.96 | 76.2 ± 4.98</td><td>51.8 ± 7.24 | 47.8 ± 7.24 | 46.3 ± 7.11</td><td>20.2 ± 5.08 | 18.7 ± 4.27 | 17.3 ± 4.85</td></tr>
<tr><td rowspan="2">CMA</td><td>85.8 ± 3.67 | 84 ± 4.58 | 82.5 ± 5.5</td><td>59 ± 7.72 | 55.8 ± 7.9 | 53.5 ± 9.03</td><td>24.5 ± 8.4 | 23 ± 8.62 | 22 ± 8.54</td></tr>
<tr><td>81.3 ± 5.56 | 79.7 ± 5.56 | 78.8 ± 5.58</td><td>52.5 ± 8.64 | 50.7 ± 9.32 | 49.2 ± 9.19</td><td>21.5 ± 8.18 | 20.8 ± 8.35 | 20.2 ± 8.41</td></tr>
<tr><td rowspan="2">NCEP</td><td>85.2 ± 2.41 | 83.3 ± 2.92 | 81 ± 3.16</td><td>61 ± 4.43 | 57.7 ± 4.78 | 54.8 ± 4.49</td><td>26.5 ± 3.59 | 24.2 ± 3.53 | 22.5 ± 3.5</td></tr>
<tr><td>79.5 ± 3.2 | 78 ± 3.65 | 76.5 ± 3.64</td><td>52.3 ± 4.5 | 50.8 ± 4.67 | 48.7 ± 4.64</td><td>21.3 ± 3.3 | 20.5 ± 3.3 | 19.5 ± 3.45</td></tr>
<tr><td rowspan="2">Pangu-Weather</td><td>85.7 ± 4.85 | 81 ± 6.35 | 77.3 ± 6.6</td><td>58 ± 7.48 | 50.2 ± 8.23 | 45.3 ± 8.77</td><td>22.3 ± 6.34 | 18 ± 4.58 | 14.5 ± 4.86</td></tr>
<tr><td>75.5 ± 6.85 | 73.7 ± 6.87 | 71.8 ± 6.36</td><td>41.7 ± 7.2 | 39.8 ± 6.36 | 36.7 ± 6.97</td><td>13 ± 4.2 | 12 ± 4.04 | 10.8 ± 3.98</td></tr>
<tr><td rowspan="2">FuXi</td><td>90.7 ± 3.94 | 87.7 ± 5.44 | 85.5 ± 6.24</td><td>66.7 ± 8.01 | 60.3 ± 10.7 | 56.3 ± 11.7</td><td>28 ± 7.26 | 23.8 ± 8.97 | 21.3 ± 9.12</td></tr>
<tr><td>83.3 ± 7.04 | 82.3 ± 7.78 | 81.2 ± 8.13</td><td>52.5 ± 12.7 | 50 ± 13.2 | 47.8 ± 13.6</td><td>19.5 ± 9.55 | 17.7 ± 9.05 | 16 ± 9.29</td></tr>
<tr><td rowspan="2">ArchesWeather</td><td>94.5 ± 2.22 | 91.5 ± 3.5 | 87.8 ± 6.62</td><td>73.8 ± 7.49 | 70.2 ± 8.21 | 63.2 ± 13.2</td><td>35.5 ± 9.84 | 32 ± 9.02 | 29.2 ± 11.1</td></tr>
<tr><td>85.7 ± 6.02 | 85.3 ± 6.47 | 85.8 ± 5.08</td><td>59 ± 12.6 | 58.5 ± 12.1 | 54.5 ± 12.4</td><td>27 ± 11.5 | 29.3 ± 13.8 | 25.8 ± 13.2</td></tr>
<tr><td rowspan="2">AIFS</td><td>90.2 ± 4.02 | 87.2 ± 4.71 | 84 ± 5.97</td><td>65.5 ± 8.88 | 61.5 ± 9.45 | 57.7 ± 10</td><td>27.7 ± 9.72 | 24.8 ± 8.93 | 23.2 ± 8.63</td></tr>
<tr><td>81.2 ± 7.24 | 80.3 ± 7.76 | 77.8 ± 8.74</td><td>54 ± 11.5 | 51.2 ± 12.3 | 48.2 ± 12.4</td><td>20.7 ± 9.03 | 18.8 ± 9.26 | 17.5 ± 8.75</td></tr>
<tr><td rowspan="2">GraphCast</td><td>87.5 ± 6.02 | 83.8 ± 6.62 | 81.3 ± 6.26</td><td>62.8 ± 10.2 | 57.7 ± 10.1 | 54.5 ± 7.21</td><td>29 ± 7.39 | 25 ± 6.73 | 23.3 ± 3.77</td></tr>
<tr><td>78.8 ± 7.69 | 76 ± 8.02 | 74 ± 8.16</td><td>50 ± 8.5 | 46.3 ± 9.98 | 43.3 ± 8.67</td><td>20.7 ± 6.05 | 18.3 ± 5.82 | 16.3 ± 5.71</td></tr>
<tr><td rowspan="2">Aurora</td><td>87.5 ± 2.93 | 83.7 ± 2.92 | 80.5 ± 2.99</td><td>61.3 ± 7.36 | 54.3 ± 6.45 | 49.2 ± 5.58</td><td>24.3 ± 8.12 | 19.2 ± 7.01 | 16 ± 6.11</td></tr>
<tr><td>76.7 ± 3.45 | 75 ± 3.87 | 74.7 ± 5.76</td><td>44 ± 5.48 | 42.3 ± 5.22 | 41.2 ± 5.34</td><td>14 ± 5 | 13.2 ± 4.45 | 12 ± 3.92</td></tr>
</table>

***Supplementary Table 11: Event-level reliability results.*** *Event-level observed probability of extreme heat forecasts preceding above-normal, above-hot and extreme heat temperatures is reported for each lead time, together with standard deviation across events. From left to right, top to bottom, each cell shows scores for lead days 10, 11, .., to 15.*

<table>
<tr><th>EVENTS</th><th>ABOVE NORMAL</th><th>ABOVE HOT</th><th>ABOVE EXTREME</th></tr>
<tr><td rowspan="2">Persistence</td><td>82 ± 10.1 | 83.2 ± 9.75 | 81.5 ± 9.34</td><td>55.5 ± 15.4 | 55.2 ± 15.6 | 50.7 ± 15.4</td><td>22.2 ± 13.8 | 21.5 ± 13.6 | 19.2 ± 13.9</td></tr>
<tr><td>80 ± 11.7 | 81 ± 11.4 | 81.7 ± 11.1</td><td>49.5 ± 16.5 | 50 ± 17.1 | 50.2 ± 16.5</td><td>19.7 ± 12.9 | 18.7 ± 12.1 | 18.3 ± 11.4</td></tr>
<tr><td rowspan="2">ECMWF-IFS</td><td>92.5 ± 4.79 | 91 ± 7.07 | 90.7 ± 7.34</td><td>71.8 ± 9.65 | 69 ± 12.6 | 68 ± 13.7</td><td>35.7 ± 12.4 | 33.2 ± 13.7 | 30.5 ± 12.8</td></tr>
<tr><td>88.2 ± 9.56 | 86 ± 10.3 | 84.3 ± 9.83</td><td>63.2 ± 15.7 | 60 ± 15.2 | 56.2 ± 15.3</td><td>28.7 ± 13.2 | 28 ± 11.9 | 25.2 ± 12.4</td></tr>
<tr><td rowspan="2">CMA</td><td>85.3 ± 13.1 | 85.2 ± 12.5 | 82.3 ± 15.7</td><td>62.2 ± 21.5 | 62.3 ± 20.5 | 58.7 ± 22.3</td><td>28 ± 14 | 27.3 ± 14.8 | 26 ± 14.7</td></tr>
<tr><td>81.5 ± 15 | 82.8 ± 13.4 | 83 ± 12</td><td>58.8 ± 20.8 | 58 ± 21.4 | 58.2 ± 19.1</td><td>27.3 ± 13.9 | 27.3 ± 15.4 | 27.8 ± 14.4</td></tr>
<tr><td rowspan="2">NCEP</td><td>87.7 ± 9.09 | 86.3 ± 10.3 | 84.7 ± 11.3</td><td>69 ± 13 | 65.5 ± 13.9 | 62.5 ± 15</td><td>31.7 ± 8.12 | 29.7 ± 8.96 | 27.7 ± 9.21</td></tr>
<tr><td>84.3 ± 10.6 | 84.2 ± 11.6 | 83.7 ± 10.5</td><td>62.5 ± 15.2 | 62.2 ± 14.2 | 61.5 ± 13.9</td><td>27.7 ± 8.69 | 28.5 ± 8.04 | 27.7 ± 7.06</td></tr>
<tr><td rowspan="2">Pangu-Weather</td><td>91.8 ± 6.64 | 89.3 ± 8.06 | 89.3 ± 7.27</td><td>69.2 ± 10.1 | 64.5 ± 10.7 | 64.8 ± 10.8</td><td>30.3 ± 12.2 | 25.5 ± 13.1 | 25 ± 12.4</td></tr>
<tr><td>88.3 ± 7.34 | 89.5 ± 7.14 | 89 ± 6.66</td><td>61.3 ± 12.1 | 61.3 ± 11.6 | 60 ± 11.8</td><td>22.8 ± 13.3 | 24 ± 11.2 | 23.5 ± 11</td></tr>
<tr><td rowspan="2">FuXi</td><td>93.3 ± 4.31 | 92.8 ± 5.15 | 89.8 ± 6.2</td><td>74.2 ± 8.33 | 70.7 ± 10.4 | 67.2 ± 12.2</td><td>32.3 ± 9.81 | 29.2 ± 10.7 | 25.8 ± 10.8</td></tr>
<tr><td>88.7 ± 6.13 | 86.5 ± 8.4 | 85.5 ± 7.76</td><td>62.5 ± 13.4 | 59 ± 14.1 | 53.7 ± 15.2</td><td>24.3 ± 11.4 | 22.8 ± 11.6 | 22.3 ± 12.1</td></tr>
<tr><td rowspan="2">ArchesWeather</td><td>95.3 ± 4.5 | 95 ± 7.7 | 93.8 ± 9.19</td><td>80 ± 12.1 | 75.2 ± 14.5 | 70.3 ± 20.7</td><td>41 ± 13.2 | 37.3 ± 14.7 | 31.2 ± 17</td></tr>
<tr><td>95.7 ± 5.71 | 91.2 ± 12 | 97 ± 4.52</td><td>66.5 ± 31.8 | 71 ± 29.5 | 67 ± 23.8</td><td>31.5 ± 21.4 | 36 ± 21.5 | 44.6 ± 25.8</td></tr>
<tr><td rowspan="2">AIFS</td><td>93.7 ± 4.57 | 93.3 ± 5.76 | 92.5 ± 5.62</td><td>73.5 ± 7.21 | 72.7 ± 7.97 | 70.7 ± 8.46</td><td>32.7 ± 11.2 | 32.8 ± 10 | 30 ± 10.2</td></tr>
<tr><td>90.2 ± 4.84 | 89.7 ± 5.31 | 87.8 ± 7.36</td><td>65.3 ± 10.5 | 64 ± 9.33 | 63 ± 11.1</td><td>27.5 ± 9.98 | 25.8 ± 8.78 | 26.5 ± 9.54</td></tr>
<tr><td rowspan="2">GraphCast</td><td>94 ± 6.11 | 90.8 ± 7.49 | 89 ± 8</td><td>74.2 ± 10.8 | 65.8 ± 11.4 | 62.8 ± 12.1</td><td>32.8 ± 12.7 | 28.3 ± 12.6 | 25.5 ± 10.1</td></tr>
<tr><td>86.2 ± 12.2 | 87.5 ± 10.2 | 82.8 ± 12</td><td>60 ± 11.8 | 59.3 ± 12.7 | 56.8 ± 13.8</td><td>24.5 ± 10.5 | 23 ± 9.59 | 21.5 ± 9.6</td></tr>
<tr><td rowspan="2">Aurora</td><td>92.8 ± 4.41 | 90.5 ± 5.47 | 89.5 ± 5.59</td><td>67.3 ± 12.1 | 63.2 ± 12.8 | 61 ± 12.7</td><td>30.7 ± 14 | 28.3 ± 15.1 | 25.5 ± 16.1</td></tr>
<tr><td>85.7 ± 8.88 | 82.3 ± 13 | 81.8 ± 13.2</td><td>56.8 ± 14.4 | 53.2 ± 17.1 | 52.8 ± 17.1</td><td>24.3 ± 15.3 | 21.7 ± 14.3 | 21.2 ± 13.2</td></tr>
</table>

## S11. METRIC VISUALS

Supplementary Figure 13 shows global (top) and event-level (bottom) forecast RMSE as a function of lead time, visualizing mean results from Supplementary Tables 5 and 6. Few models are capable of scoring RMSE below climatological baseline at a 10-day lead, and only FuXi manages to retain this feat until the 15-day lead time. At event-level, the forecasts for heat extremes all significantly improve on both the persistence and climatological baselines—some halving the climatological RMSE. Though performance degrades with further lead times, every model scores better than climatology even at 15 days out.

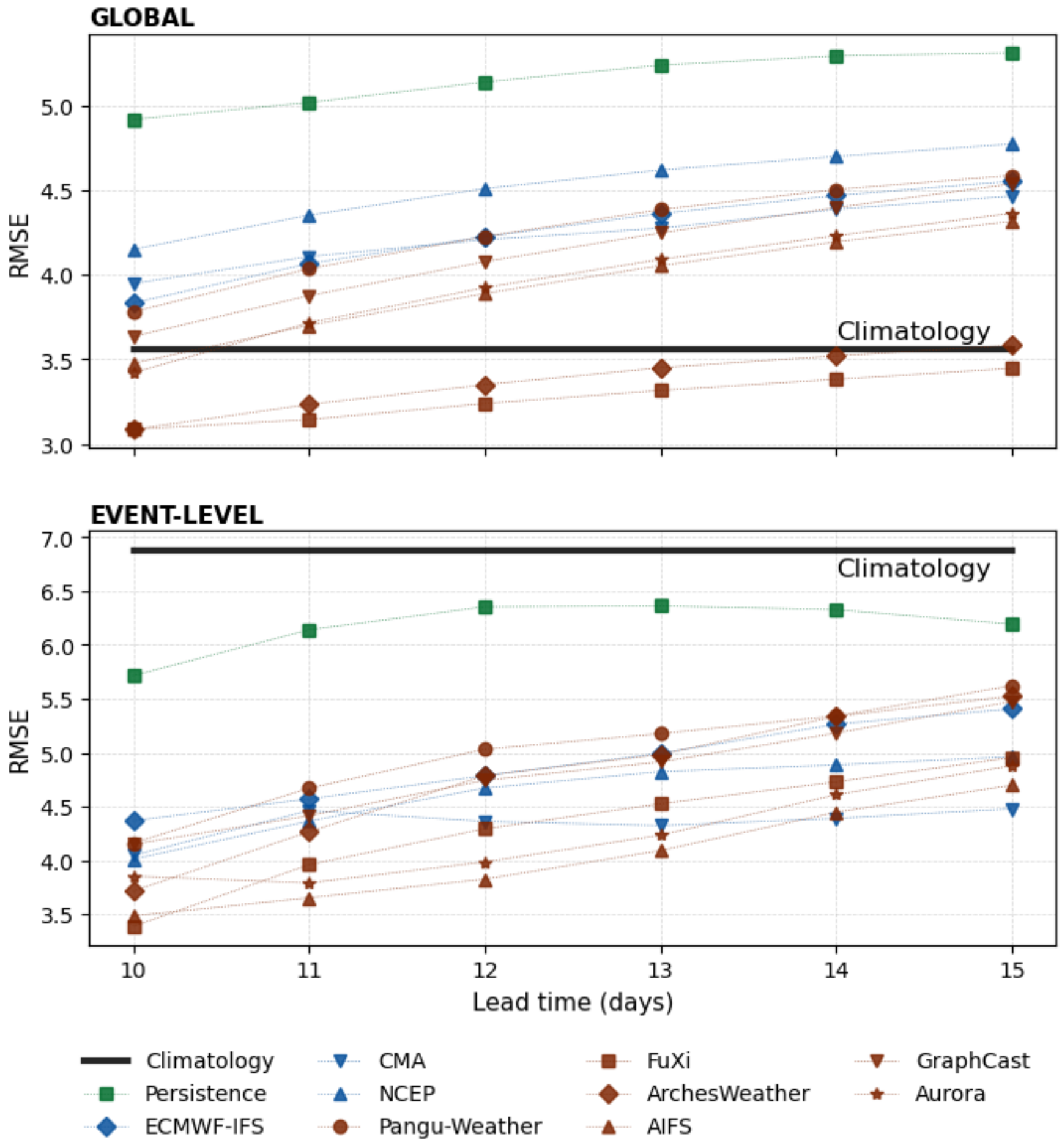


***Supplementary Figure 13: RMSE in function of lead time.*** *For global (top) and event-level (bottom) scales, average RMSE across events is shown as a function of lead time.*

Latitudinal patterns of 14-day lead RMSE, $R^2$, correlation and anomaly correlation are shown in Supplementary Figures 14—17, respectively. Temperature variability is lowest in the tropics and increases towards the Earths poles, which is reflected in the RMSE patterns—forecast errors generally increase with increasing latitude. There remains a latitudinal dependency in the $R^2$—scaling error with respect to the background signal—

indicating that the forecast error increases more with latitude than does the variability of the background signal.

For each of the metrics shown here, emulators average better scores than dynamical systems. CMA and ECMWF-IFS systems in this study show poor $R^2$ relative to the other models in the Southern Hemisphere specifically; GraphCast shows a discontinuity at the South Pole. Supplementary Figure 16—S17 show that while most forecasts are fairly correlated with their reference, this is largely driven by diurnal variability. Subtract this, and models show poor ability to forecast the temporal evolution of temperature anomalies.

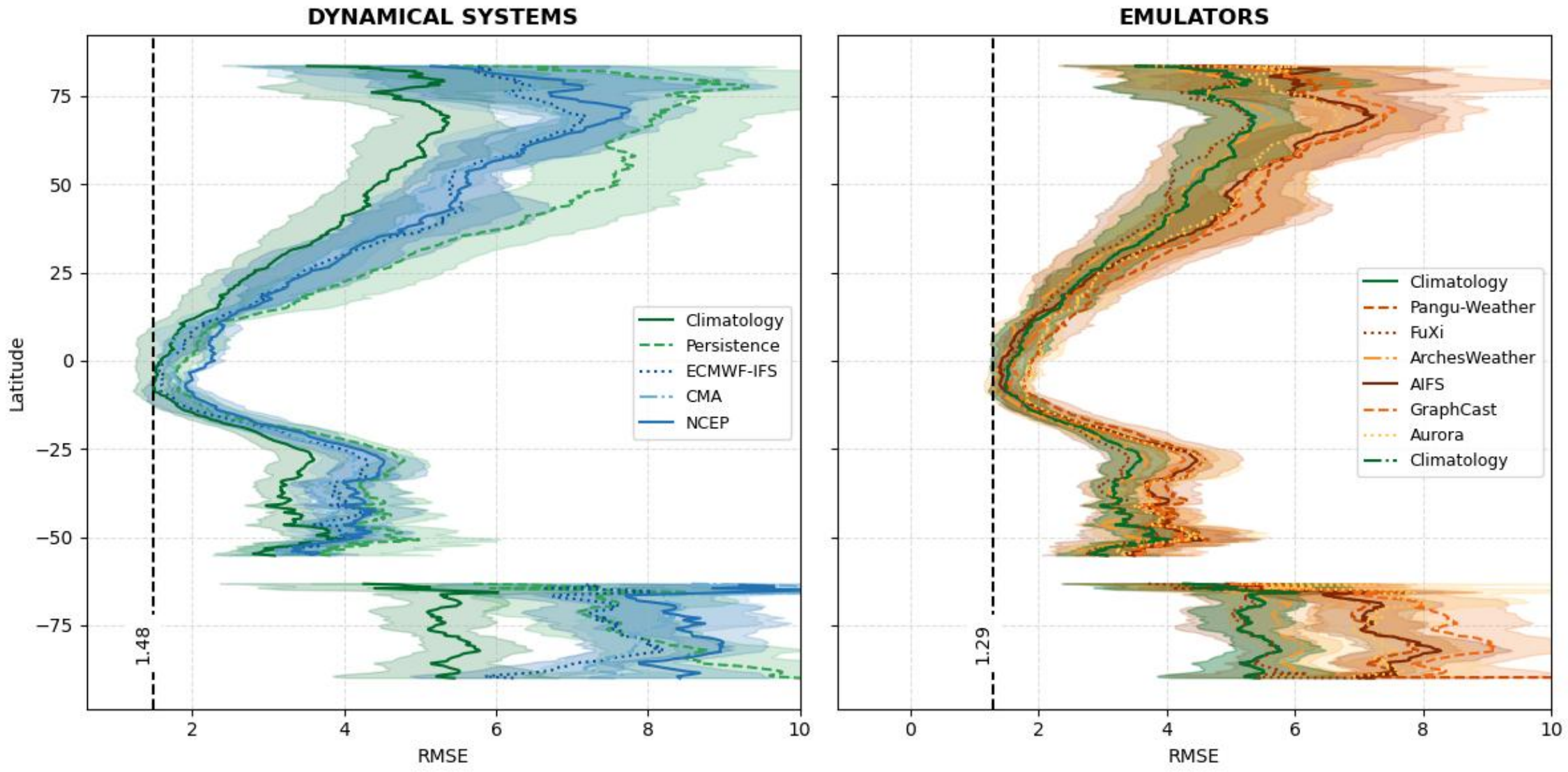


***Supplementary Figure 14: Latitudinal RMSE.*** *Global gridded forecast RMSE of land surface temperature is computed for each event period (Supplementary Table 2); subsequently, mean and standard deviation are taken across events and averaged over the longitude coordinate. The left plot shows results for dynamical systems and both baselines, the right plot for emulators and climatology. A horizontal dashed line shows the best observed mean score. Shaded bands show standard deviation across events.*

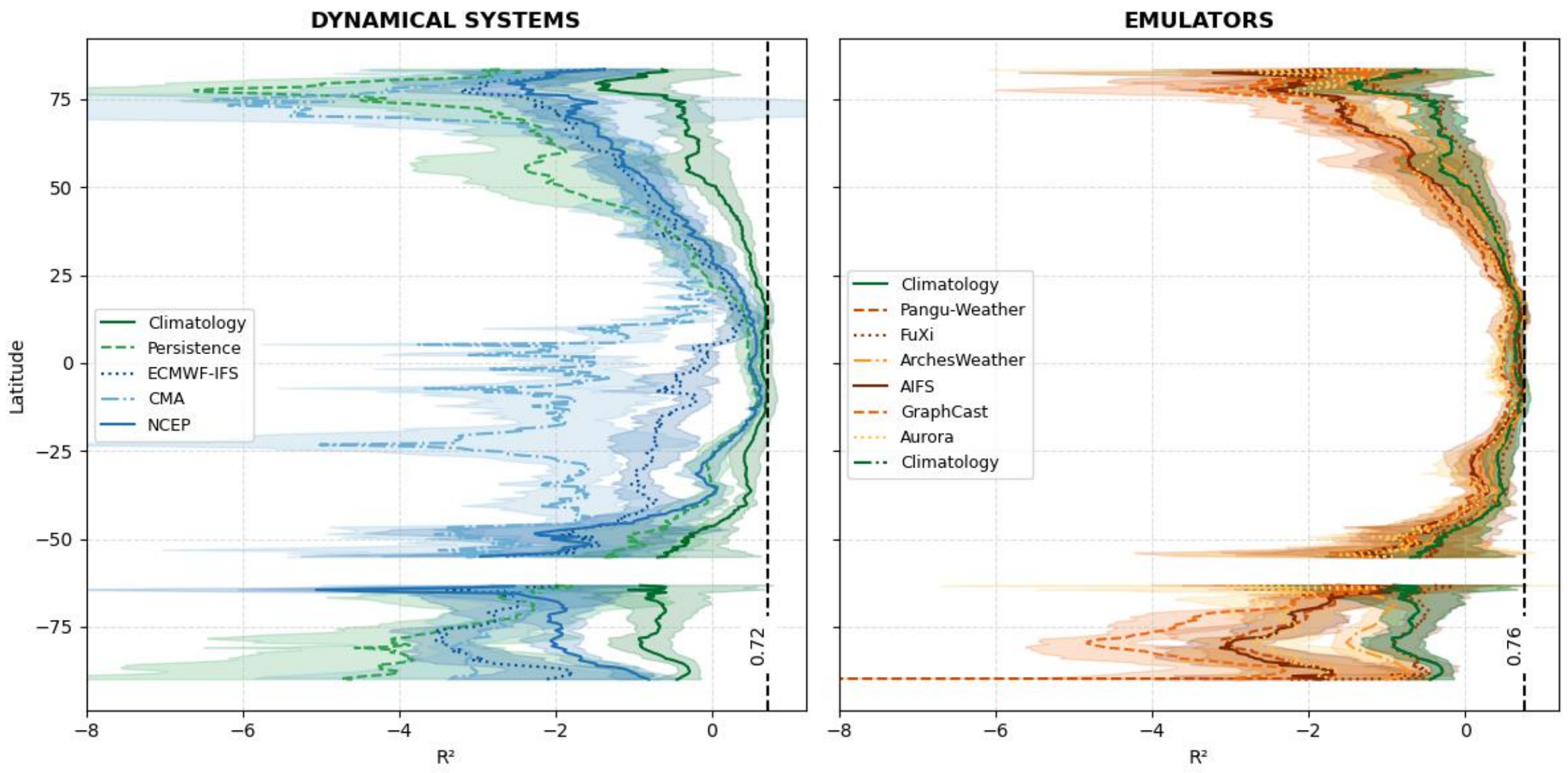


***Supplementary Figure 15: Latitudinal $R^2$.*** *Global gridded forecast $R^2$ of land surface temperature is computed for each event period (Supplementary Table 2); subsequently, mean and standard deviation are taken across events and averaged over the longitude coordinate. The left plot shows results for dynamical systems and both baselines, the right plot for emulators and climatology. A horizontal dashed line shows the best observed mean score. Shaded bands show standard deviation across events.*

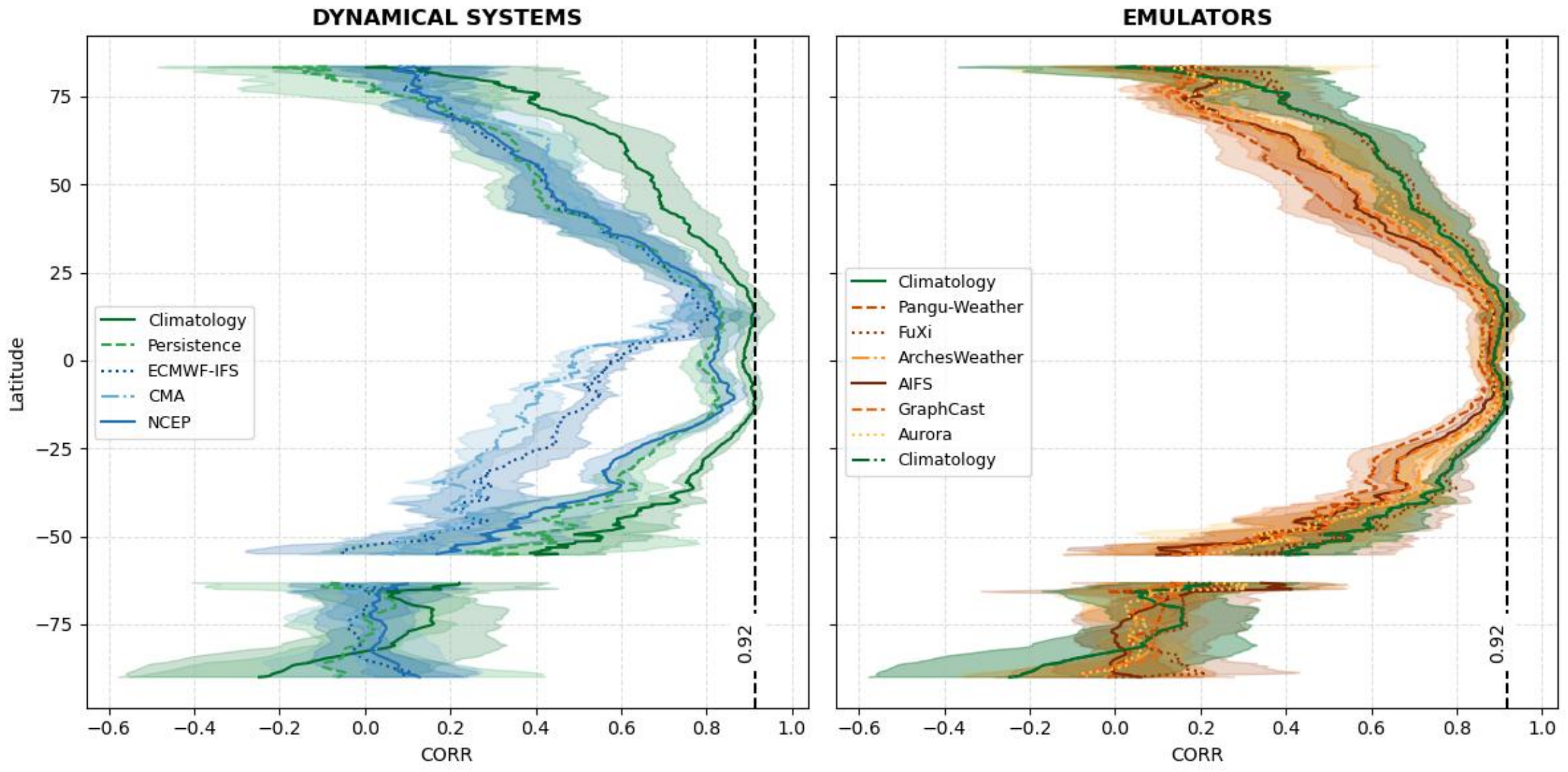


***Supplementary Figure 16: Latitudinal correlation.*** *Global gridded forecast correlation of land surface temperature is computed for each event period (Supplementary Table 2); subsequently, mean and standard deviation are taken across events and averaged over the longitude coordinate. The left plot shows results for dynamical systems and both baselines, the right plot for emulators and climatology. A horizontal dashed line shows the best observed mean score. Shaded bands show standard deviation across events. Note that this is correlation between forecast and reference temperature, not anomalies (as in main body)—those results are shown in Supplementary Figure 17.*

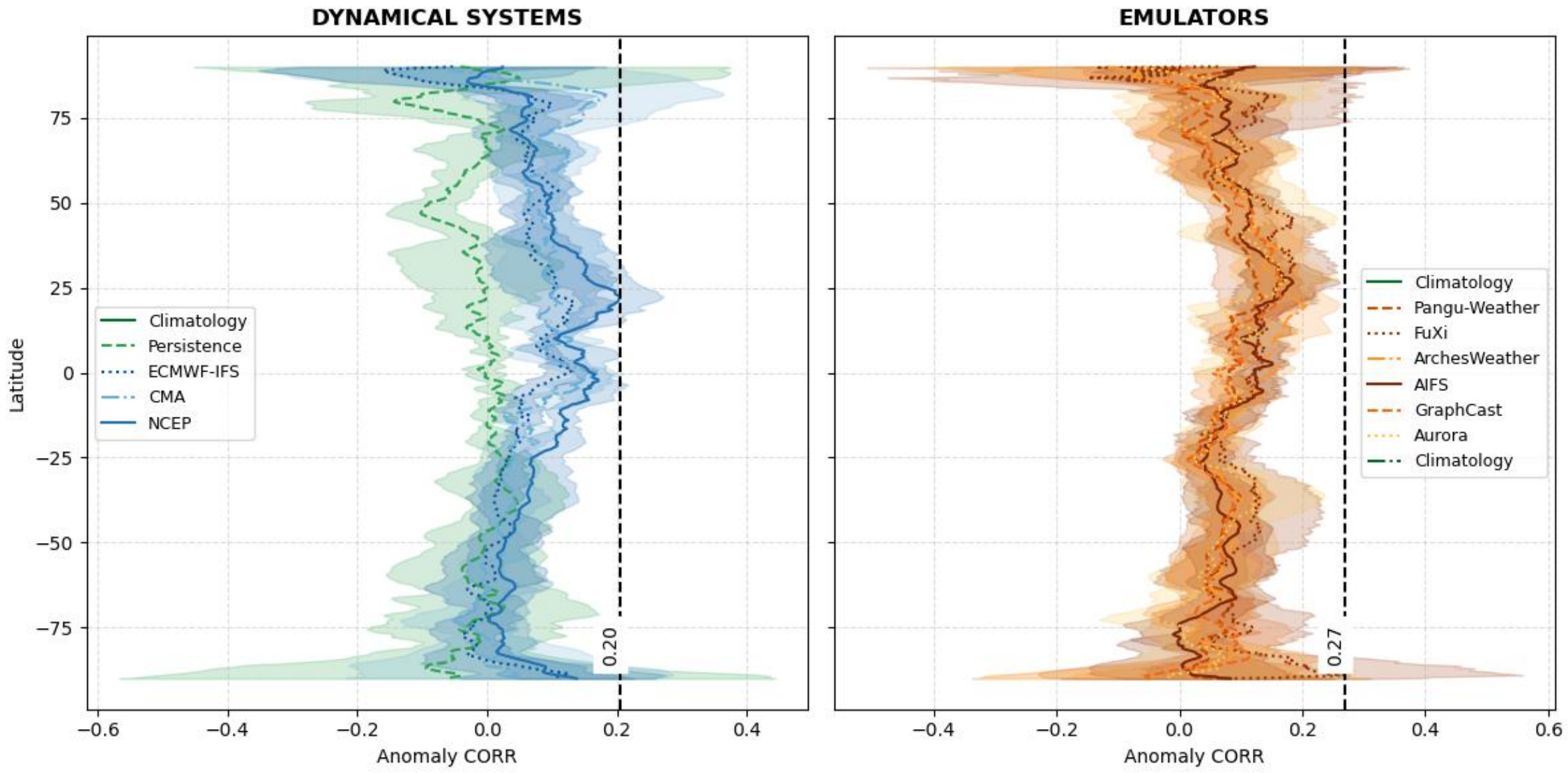


***Supplementary Figure 17: Latitudinal anomaly correlation.*** *Global gridded forecast anomaly correlation of land surface temperature is computed for each event period (Supplementary Table 2); subsequently, mean and standard deviation are taken across events and averaged over the longitude coordinate. The left plot shows results for dynamical systems and both baselines, the right plot for emulators and climatology. A horizontal dashed line shows the best observed mean score. Shaded bands show standard deviation across events.*

In weather forecasting, the standard deviation of a field in either space or time can be referred to as the forecast activity (ECMWF, 2024). Additional to spectral decomposition (Supplementary Figure 12), comparing forecast to reference activity can be a means of quantifying the degree to which a model produces overly smooth (too little variability) or even sharp (too much variability) fields. The spectral analysis in the main body evaluated the spatial characteristics of temperature forecasts; here, we use the ratio of forecast to reference temperature activity as a measure for the spatial smoothness/sharpness of models at extended lead times. Supplementary Figure 18 does so for rollouts from lead day 10 to 15, Supplementary Figure 19 for fixed-lead-time 14-day lead timeseries.

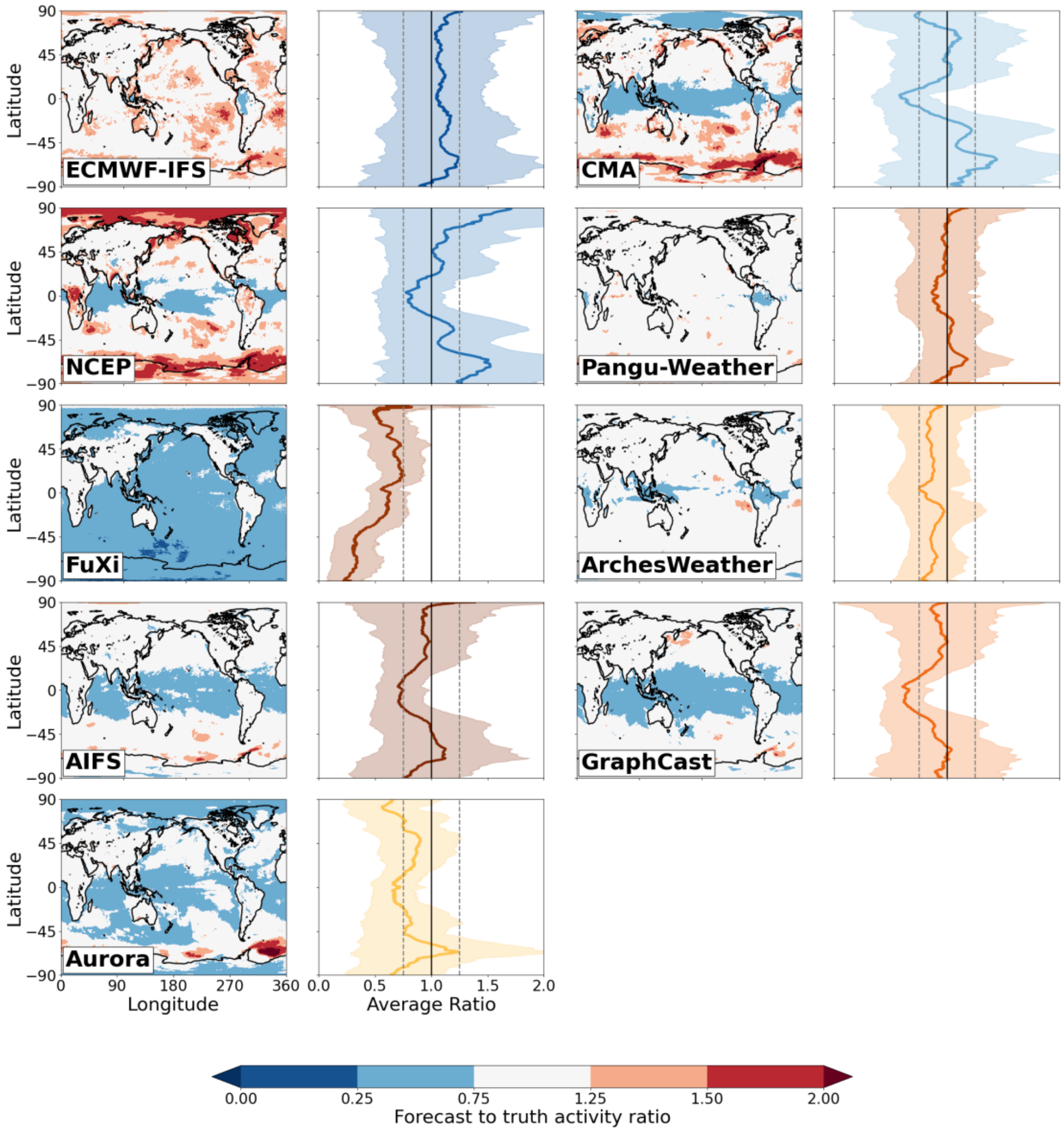


***Supplementary Figure 18: Activity of rollouts days 10-15.*** *Global maps show the average ratio of forecast to reference activity across events, to their right is the corresponding latitudinal profile (averaged across longitudes). Shaded bands show the latitudinal profile of activity standard deviation across events.*

Supplementary Figure 18 shows that many models produce similar temperature variability in time relative to their reference. Pangu-Weather, ArchesWeather and ECMWF-IFS are well-balanced. Many models show a distinct low activity in the tropical

belt, which for emulators is likely a consequence of MSE optimization (main body). FuXi shows the lowest activity of all, though its forecasts are less smooth over land than over oceans.

As the atmosphere is a chaotic deterministic system, forecasts initialized at different times are expected to lead to vastly different trajectories with increasing lead times [1-3] (Lorenz, 1969; Judt, 2018; Zhang et al., 2019). Supplementary Figure 19 shows the activity ratio but for the 14-day lead concatenated time series rather than rollouts, illustrating that the dynamical systems showcase this expected behaviour. For most emulators, activity increases; yet FuXi and ArchesWeather—the latter quite balanced in Supplementary Figure 18—show reduced activity in their concatenated forecasts compared to their rollouts.

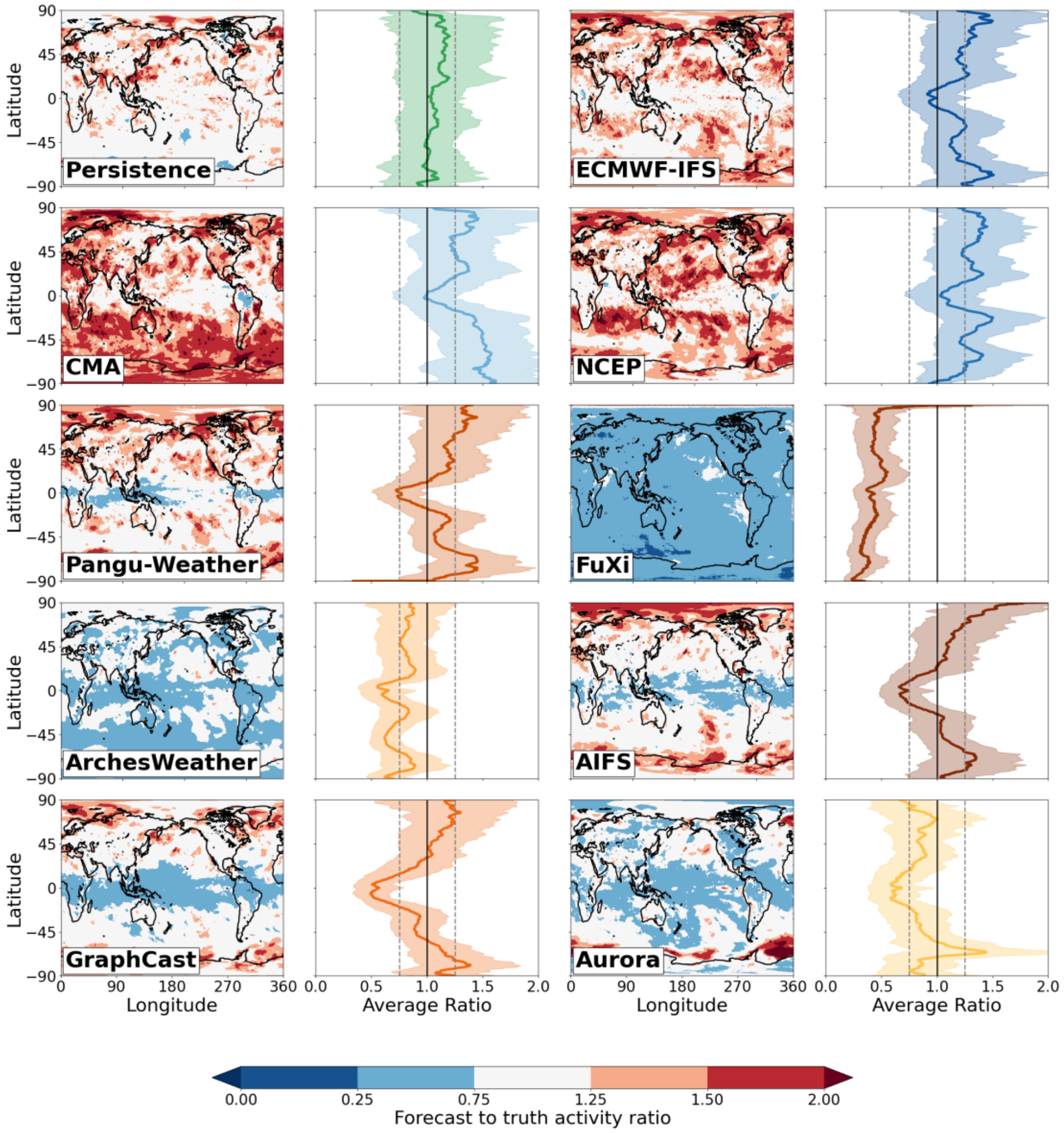


***Supplementary Figure 19: Activity of fixed-lead-time time series.*** *Global maps show the average ratio of forecast to reference activity across events, to their right is the corresponding latitudinal profile (averaged across longitudes). Shaded bands show the latitudinal profile of activity standard deviation across events.*

# S15. CASE STUDIES

Supplementary Figure 20 shows each model's 14-day lead forecast for Vancouver throughout the 2021 “Heat Dome” (Supplementary Table 2). The top-left panel illustrates the importance of the reference, as the “true” temperature timeseries is different for ERA5 and dynamical model *fc0* states. The concatenated 14-day forecast from CMA is highly variable, but its reference is also more variable compared to the other references. In Vancouver, the model that best captures the heat extremes is GraphCast (lower middle panel). Some characteristic model behaviour is visible, such as FuXi's showing a near-climatological time series, yet still picking up some of the heat signal 14 days in advance.

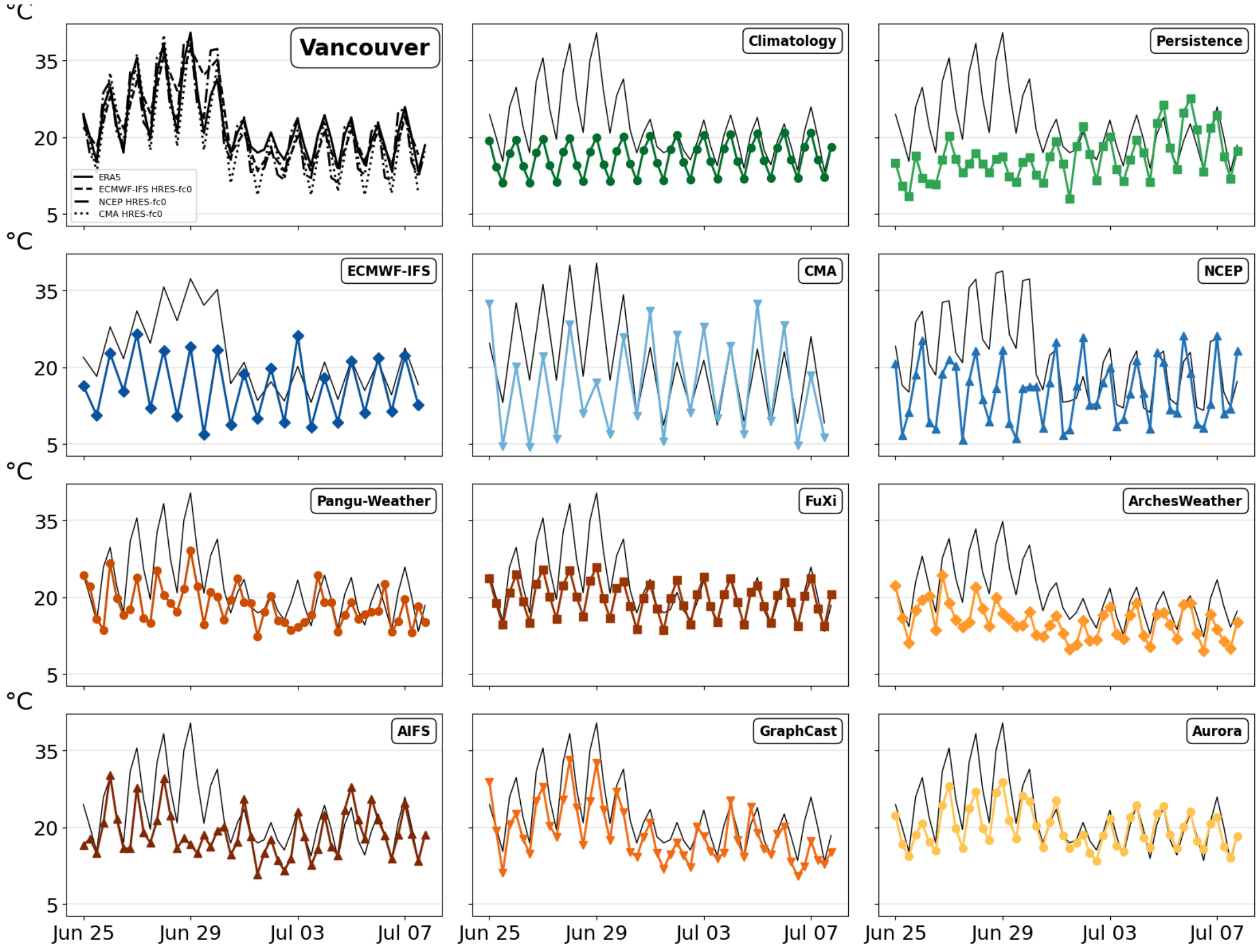


***Supplementary Figure 20: 2021 "Heat Dome" lead day 14 forecasts in Vancouver.*** *The top-left panel shows the “true” temperature timeseries observed in Vancouver (Supplementary Table 2) according to the four different references used. All other panels then show each baseline or model 14-day lead concatenated forecast (Supplementary Figure 2) together with its reference.*

For all case studies (Supplementary Table 2) and lead times 10 to 15 days, Supplementary Figure 21 shows to what degree models were able to improve upon climatological RMSE. All models manage to improve upon the baseline, except for CMA—which scores a poor RMSE for all lead times in Georgetown (Supplementary Table 2). AIFS has the largest average improvement, lowest standard deviation of this improvement, and does not score worse than climatology for any of the lead time—event combinations (n=36).

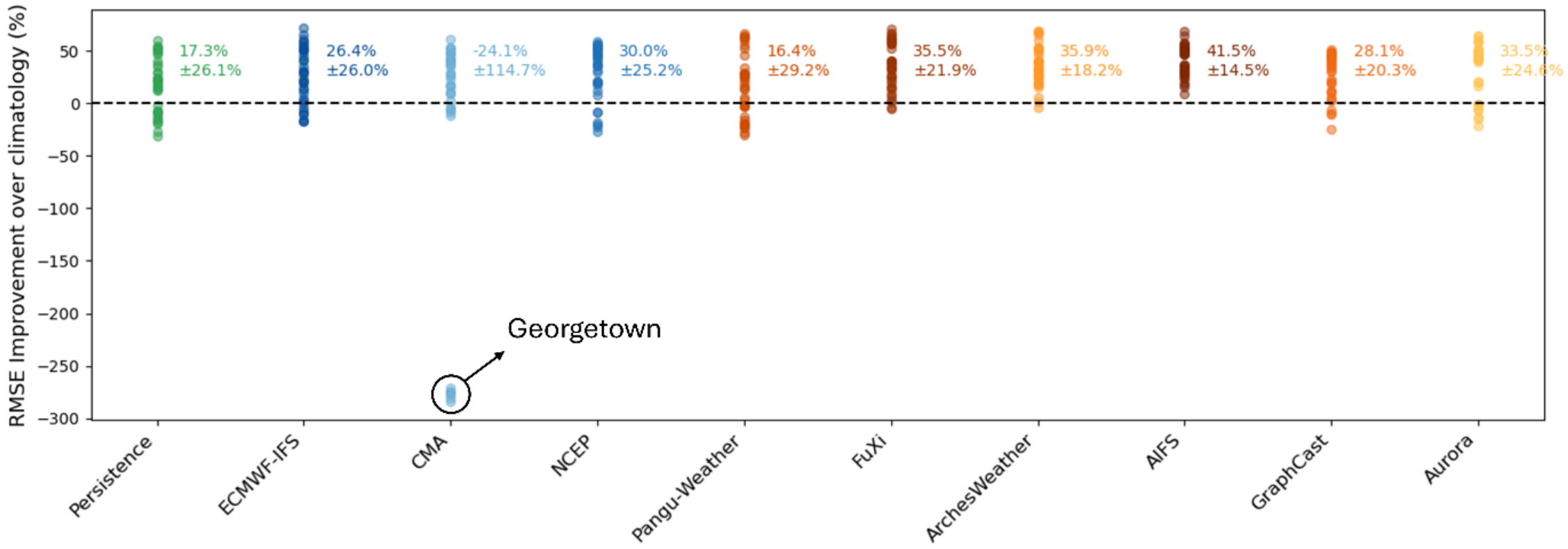


***Supplementary Figure 21: Case study RMSE relative to climatology.*** *For each lead time–event combination (n=36), the forecast RMSE at the event's case study location relative to that of the climatological forecast. Computed as* $(RMSE_{forecast} - RMSE_{climatology}) / RMSE_{climatology}$.



\end{document}